\documentclass[twocolumn]{aastex62}
\usepackage{graphicx, subfigure, amsmath, amsfonts, amssymb, footnote, color}
\usepackage{captcont}
\usepackage[bookmarks=false]{hyperref}
\usepackage{times}
\hypersetup{
  colorlinks= true, %Colours links instead of ugly boxes
  urlcolor  = blue, %Colour for external hyperlinks
  linkcolor = blue, %Colour of internal links
  citecolor = blue %Colour of citations
} % makes stupid green boxes go away, paper now looks more like it will when published!
\submitjournal{ApJ}
\shorttitle{Newly radio-loud quasars}
\shortauthors{Kristina Nyland et al.}

\begin{document}

%\title{Variable AGN Identified in a Search for Transients in VLASS Epoch 1 and FIRST}
%\title{Newly Radio-loud AGN Identified in a Search for Transients between VLASS and FIRST}
%\title{Radio-quiet AGN that have Transitioned to Radio-loud on Decadal Timescales Revealed by VLASS and FIRST}
%\title{Powerful AGN that have Transitioned from Radio-quiet to Radio-loud on Decadal Timescales Revealed by VLASS and FIRST}
\title{Quasars That Have Transitioned from Radio-quiet to Radio-loud on Decadal Timescales Revealed by VLASS and FIRST}
%\title{Quasars with Young Radio Jets Revealed by VLASS and FIRST: Radio-loudness Can Change on Human Timescales}
%\title{From Radio-quiet to Radio-loud in 10 Years: Transitioning Quasars Revealed by VLASS and FIRST}
%\title{Newly Radio-loud AGN Identified in a Search for Transients between VLASS and FIRST}

\correspondingauthor{Kristina Nyland}
\email{kristina.nyland.ctr@nrl.navy.mil}

\author{Kristina Nyland}
\affiliation{National Research Council, resident at the U.S. Naval Research Laboratory, 4555 Overlook Ave SW, Washington, DC 20375, USA}

\author{Dillon Z. Dong}
\affiliation{California Institute of Technology, 1200 E. California Blvd. MC 249-17, Pasadena, CA 91125, USA}

\author{Pallavi Patil}
\affiliation{National Radio Astronomy Observatory, Charlottesville, VA 22903, USA}
\affiliation{University of Virginia, Department of Astronomy, Charlottesville, VA 22903, USA}

\author{Mark Lacy}
\affiliation{National Radio Astronomy Observatory, Charlottesville, VA 22903, USA}

\author{Sjoert van Velzen} %[0000-0002-3859-8074]
\affiliation{Center for Cosmology and Particle Physics, New York University, New York, NY 10003}
\affiliation{Department of Astronomy, University of Maryland, College Park, MD 20742}

\author{Amy E. Kimball}
\affiliation{National Radio Astronomy Observatory, 1003 Lopezville Rd, Socorro, NM 87801, USA}

\author{Sumit K. Sarbadhicary}
\affiliation{Department of Physics and Astronomy, Michigan State University, East Lansing, MI 48824, USA}

\author{Gregg Hallinan}
\affiliation{California Institute of Technology, 1200 E. California Blvd. MC 249-17, Pasadena, CA 91125, USA}

\author{Vivienne Baldassare}
\altaffiliation{Einstein Fellow}
\affiliation{Yale University, Department of Astronomy, 52 Hillhouse Avenue, New Haven, CT 06511, USA}

\author{Tracy E. Clarke}
\affiliation{U.S. Naval Research Laboatory, 4555 Overlook Ave SW, Washington, DC 20375, USA}

\author{Andy D. Goulding}
\affiliation{Department of Astrophysical Sciences, Princeton University, Princeton, NJ 08540, USA}

\author{Jenny Greene}
\affiliation{Department of Astrophysical Sciences, Princeton University, Princeton, NJ 08540, USA}

\author{Andrew Hughes}
\affiliation{Department of Physics, University of Alberta, CCIS 4-181, Edmonton, AB T6G 2E1, Canada}

\author{Namir Kassim}
\affiliation{U.S. Naval Research Laboatory, 4555 Overlook Ave SW, Washington, DC 20375, USA}

\author{Magdalena Kunert-Bajraszewska}
\affiliation{Institute of Astronomy, Faculty of Physics, Astronomy and Informatics, NCU, Grudziadzka 5/7, 87-100, Torun, Poland}

\author{Thomas J. Maccarone}
\affiliation{Department of Physics \& Astronomy, Texas Tech University, Box 41051, Lubbock, TX 79409-1051, USA}

\author{Kunal Mooley}
\affiliation{Cahill Center for Astronomy, MC 249-17, California Institute of Technology, Pasadena, CA 91125, USA}

\author{Dipanjan Mukherjee} 
\affiliation{Inter-University Centre for Astronomy and Astrophysics, Post Bag 4, Ganeshkhind, Pune, Maharashtra 411007, India}

\author{Wendy Peters}
\affiliation{U.S. Naval Research Laboatory, 4555 Overlook Ave SW, Washington, DC 20375, USA}

\author{Leonid Petrov}
\affiliation{NASA GSFC, 8800 Greenbelt Rd, Greenbelt, MD 20771 USA}

\author{Emil Polisensky}
\affiliation{U.S. Naval Research Laboatory, 4555 Overlook Ave SW, Washington, DC 20375, USA}

\author{Wiphu Rujopakarn}
\affiliation{Department of Physics, Faculty of Science, Chulalongkorn University, 254 Phayathai Road,
Pathumwan, Bangkok 10330, Thailand}
\affiliation{National Astronomical Research Institute of Thailand (Public Organization), Don Kaeo, Mae
Rim, Chiang Mai 50180, Thailand}

\author{Mark Whittle}
\affiliation{University of Virginia, Department of Astronomy, Charlottesville, VA 22903, USA}

\author{Mattia Vaccari}
\affiliation{Inter-university Institute for Data Intensive Astronomy, Department of Physics and Astronomy, University of the Western Cape, 7535 Bellville, Cape Town, South Africa}
\affiliation{INAF - Istituto di Radioastronomia, via Gobetti 101, 40129 Bologna, Italy}

%%%%%%%%%%%%%%%%%%%%%%%%%%%%%%%%%%%%%%%%%%%%%%%
\begin{abstract}
We have performed a search over 3440 deg$^2$ of Epoch 1 (2017--2019) of the Very Large Array Sky Survey to identify unobscured quasars in the optical ($0.2 < z < 3.2$) and obscured active galactic nuclei (AGN) in the infrared that have brightened dramatically in the radio over the past one to two decades.  These sources would have been previously classified as ``radio-quiet" quasars based on upper limits from the Faint Images of the Radio Sky at Twenty Centimeters survey (1993--2011), but they are now consistent with ``radio-loud" quasars ($L_{\rm 3\,GHz} = 10^{40 - 42} \,\, {\rm erg} \,{\rm s}^{-1}$).  A quasi-simultaneous, multiband ($\sim1-18$ GHz) follow-up study of 14 sources with the VLA has revealed compact sources ($<0.1^{\prime \prime}$ or $<$1 kpc) with peaked radio spectral shapes.  The high-amplitude variability over decadal timescales at 1.5 GHz (100\% to $>$2500\%), but roughly steady fluxes over a few months at 3 GHz, are inconsistent with extrinsic variability due to propagation effects, thus favoring an intrinsic origin.  We conclude that our sources are powerful quasars hosting compact/young jets.  This challenges the generally accepted idea that ``radio-loudness" is a property of the quasar/AGN population that remains fixed on human timescales.  Our study suggests that frequent episodes of short-lived AGN jets that do not  necessarily grow to large scales may be common at high redshift.  We speculate that intermittent but powerful jets on subgalactic scales could interact with the interstellar medium, possibly driving feedback capable of influencing galaxy evolution.
\end{abstract}
%%%%%%%%%%%%%%%%%%%%%%%%%%%%%%%%%%%%%%%%%%%%%%%

%%%%%%%%%%%%%%%%%%%%%%%%%%%%%%%%%%%%%%%%%%%%%%%
%%%%%%%%%%%%%%%%%%%%%%%%%%%%%%%%%%%%%%%%%%%%%%%

%%%%%%%%%%%%%%%%%%%%%%%%%%%%%%%%%%%%%%%%%%%%%%%
%%%%%%%%%%%%%%%%%%%%%%%%%%%%%%%%%%%%%%%%%%%%%%%
%%%%%%%%%%%%%%%%%%%%%%%%%%%%%%%%%%%%%%%%%%%%%%%
\section{Introduction}
%Until recently, systematic surveys of the dynamic sky have been dominated by optical, X-ray, and gamma-ray observing campaigns, with radio studies of transient and variable sources often performed as follow-up in response to a shorter wavelength trigger.  Thus, the transient/variable radio sky remains largely unexplored, particularly among source populations invisible to synoptic surveys conducted at shorter wavelengths (e.g., due to obscuration).  
Until recently, synoptic surveys of the dynamic sky have been dominated by optical, X-ray, and gamma-ray observing campaigns, with systematic radio studies of transient and variable sources, in particular those at high redshift, often performed as follow-up in response to a shorter wavelength trigger.  Thus, the slow transient (or slowly varying) extragalactic radio source population remains largely unexplored, particularly among sources invisible to synoptic surveys conducted at shorter wavelengths (e.g., due to obscuration by dense columns of gas and dust). Large-scale synoptic radio surveys are thus a key way forward for both current radio telescopes, such as the Karl G. Jansky Very Large Array (VLA), and prospective instruments under development over the next decade, such as the next-generation Very Large Array\footnote{https://ngvla.nrao.edu} (ngVLA; \citealt{murphy+18,selina+18}) and Square Kilometre Array\footnote{https://www.skatelescope.org} (SKA; \citealt{dewdney+09, SKA2015, SKA2019}).   
  
%Although the dynamic radio source population is extraordinarily diverse in nature (Galactic vs.\ extragalactic sources, coherent vs.\ incoherent emission, etc.), recent studies (e.g., \citealt{mooley+16}) have demonstrated that the slow transient/variable radio sky is dominated by emission from active galactic nuclei (AGN) associated with accreting supermassive black holes (SMBHs). While a subset of AGN identified in synoptic radio studies represent bonafide transients (e.g., tidal disruption events), the majority are likely associated with variability \citep{bannister+11}, a well-known multiwavelength hallmark of AGN \citep{hovatta+07}.  
% from Bannister et al. 2011: "We conclude that many variable sources can be explained as scintillating AGN"

%kunal.mooley71: This is a reasonable opening paragraph, but I recommend rewording it as follows. 
Active Galactic Nuclei (AGN) represent the most dominant source of variability and transient activity in the radio sky (e.g. \citealt{carilli+03, thyagarajan+11, bannister+11, frail+12, bell+15, mooley+16}). The majority of AGN exhibit variability (a well-known multiwavelength hallmark of an AGN \citealt{hovatta+07}) in the radio regime at the few tens of percent level over a wide range of timescales, between a few days and a few decades (e.g. \citealt{barvainis+05, thyagarajan+11, gralla+20, sarbadhicary+20}). This variability is often attributed to extrinsic effects related to propagation (e.g. interstellar scattering) or intrinsic effects directly related to AGN itself (\citealt{bignall+15}, and references therein) such as the propagation of shocks along the jet (e.g. \citealt{marscher+85}).   

More extreme radio AGN variability (with a variability amplitude   $\gtrsim100$\%) occurring over longer timescales of years to decades has also been observed (e.g. \citealt{barvainis+05}), and may be caused by jets that have reoriented themselves toward our line of sight or young, expanding jets that have been recently triggered.  
In the case of jet reorientation, precession due to perturbations from the presence of a binary SMBH or stochastic effects related to accretion \citep{an+13} may alter the jet inclination angle and degree of Doppler boosting, thereby leading to high-amplitude, long-term variability.  Bonafide candidates for new jet activity have also been identified  \citep{mooley+16, kunert-bajraszewska+20}, suggesting  that short-lived, intermittent AGN jet activity recurring  on timescales of $\sim 10^4-10^5$~years could be common, consistent with predictions (e.g. \citealt{reynolds+97, czerny+09}).

%AGN variability may arise from extrinsic effects related to propagation (e.g., interstellar scattering) or intrinsic effects directly related to AGN itself \citep{thyagarajan+11,bignall+15}. Common intrinsic mechanismss include accretion disk instabilities and ``shock-in-jet models" \citep{marscher+85}.  
%Newborn radio jets that have been recently triggered, or jets that have recently reoriented themselves toward our line of sight may also exhibit variability.  In the case of jet reorientation, precession due to perturbations from the presence of a binary SMBH or stochastic effects related to accretion \citep{an+13} may alter the jet inclination angle and degree of Doppler boosting, thereby leading to variability.  

%Whether driven by extrinsic or intrinsic effects, the majority of variable radio AGN identified beyond the low-$z$ Universe are compact in nature due to line-of-sight effects, youth/confinement by a dense medium, or a combination of these effects.  
Compact radio AGN with subgalactic extents residing in gas-rich host galaxies, especially those at redshifts of 1$\lesssim z \lesssim 3$, are an important, yet still poorly studied class of objects for understanding the role of jet-ISM feedback in influencing galaxy growth and evolution (\citealt{mukherjee+16,nyland+18,jarvis+19}).    
Specifically, the prevalence of radio AGN variability driven by different intrinsic effects, as well as the magnitude and impact feedback from subgalactic jets on SMBH-galaxy co-evolution, remain uncertain.  
%The prevalence remains unknown and just to complete the thought, that presumably therefore means the magnitude of the impact remains uncertain as well.

%One particularly interesting possibility is that large changes in radio flux may be associated with AGN transitioning between a radio-quiet state to one in which synchrotron-emitting radio jets are present. Extreme radio variability is typical of Galactic radio sources such as X-ray binaries \citep[e.g.][]{mirabel+99}. In these sources, the short timescales associated with black holes with masses $\sim 1-10\, M_{\odot}$ mean that the change in state from a radio-quiet mode associated with soft X-ray emission, and a mode with a hard X-ray spectrum and radio jets can happen on timescales of minutes. Given the $\sim 10^7-10^9\, M_{\odot}$ SMBH masses that are typical of AGN, similar transitions may occur \citep{maccarone+03,falcke+04,nipoti+05,kording+06}, but the corresponding timescales may be longer. Attempts have been made to identify radio sources whose jets may have recently switched off \citep[e.g.][]{marecki+11}, but such objects still have radio emission from their lobes. A less ambiguous approach is to try to find young sources whose jets have recently switched on. By their very nature these objects will be rare, but by surveying the radio emission from a large number of AGN in two or more well-separated epochs, it may be possible to find objects that are candidates for AGN undergoing this transition.

Here, we present follow-up, quasi-simultaneous, multiband VLA observations of a subset of AGN with high-amplitude (100\% to $>$2500\%) radio variability identified in a search for transients on decades-long timescales between the {Faint Images of the Sky at Twenty centimeters (FIRST; \citealt{Becker+95})} and $\sim$3,440 deg$^2$ of Epoch~1 of the Very Large Array Sky Survey (VLASS; \citealt{lacy+20}).  
We describe our sample and selection criteria in Section~\ref{sec:sample}.  In Section~\ref{sec:data}, we describe the observing and data reduction heuristics for our new multiband VLA observations.  An assessment of the radio continuum properties of our sources, variability, and multiband spectral energy distributions (SEDs) is provided in Section~\ref{sec:results}.  We consider a variety of possibilities for the origin of the radio variability in our sources in Section~\ref{sec:origin}, and discuss implications for the life cycles of radio AGN and their connection to galaxy evolution in Section~\ref{sec:discussion}.  We summarize our results and assess prospects for further insights through future follow-up observations in Section~\ref{sec:summary}.  We adopt a standard $\Lambda$CDM cosmology with $H_{0}$ = 67.7  km s$^{-1}$ Mpc$^{-1}$, $\Omega_{\Lambda}$ = 0.691 and $\Omega_{\textrm{\small{M}}}$ = 0.307 \citep{planck+15} throughout our study.  Errors shown are 1$\sigma$ unless otherwise stated.

%%%%%%%%%%%%%%%%%%%%%%%%%%%%%%%%%%%%%%%%%%%%%%%%
%%%%%%%%%%%%%%%%%%%%%%%%%%%%%%%%%%%%%%%%%%%%%%%
%%%%%%%%%%%%%%%%%%%%%%%%%%%%%%%%%%%%%%%%%%%%%%%
%%%%%%%%%%%%%%%%%%%%%%%%%%%%%%%%%%%%%%%%%%%%%%%
\section{Sample}
\label{sec:sample}
As part of our ongoing compilation of radio transients observed at 3~GHz in 
Epoch~1 of VLASS, %the Very Large Array Sky Survey (VLASS; \citealt{lacy+20}), 
which will be presented in its entirety in Dong et al.\ (in prep.), we have selected a preliminary sample of radio AGN exhibiting extreme variability on timescales of 
%several years to about two decades.  
$\sim$1--2 decades. 
%In the following subsections, 
In the remainder of this section, we provide details on the radio surveys used to conduct our search for variable AGN and describe our sample selection criteria.

%%%%%%%%%%%%%%%%%%%%%%%%%%%%%%%%%%%%%%%%%%%%%%%
\begin{figure*}[t!]
\centering
\includegraphics[clip=true, trim=0.15cm 0.3cm 0.2cm 0cm, width=\textwidth]{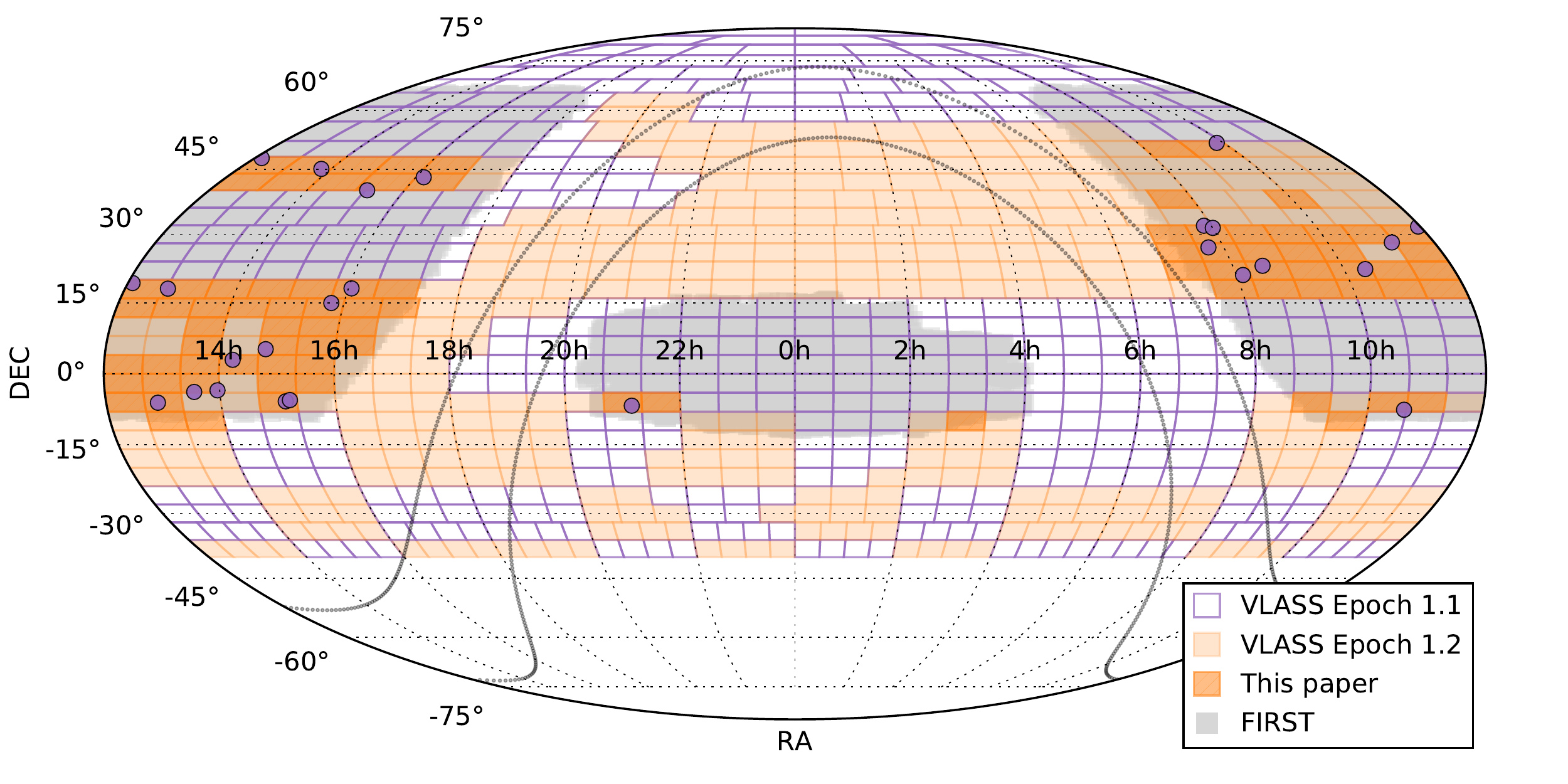}
\caption{Relative sky footprints of the VLASS and FIRST surveys.  VLASS tiles from Epochs 1.1 and 1.2 are shown in purple and light orange, respectively.  The large gray shaded regions illustrate the coverage of the FIRST survey.  The 86 tiles from VLASS Epoch~1.2 analyzed in this paper are indicated by the dark orange regions.  The gray curves denote coordinates within 10 degrees of the Galactic plane.  
The positions of our 26 candidate radio-variable AGN identified via comparison with data from FIRST are shown by the purple circles.    
\\}
\label{fig:footprint}
\end{figure*}
%%%%%%%%%%%%%%%%%%%%%%%%%%%%%%%%%%%%%%%%%%%%%%

%%%%%%%%%%%%%%%%%%%%%%%%%%%%%%%%%%%%%%%%%%%%%%%
\begin{figure*}
\centering
\includegraphics[clip=true, trim=1cm 0cm 1cm 0cm, width=0.7\textwidth]{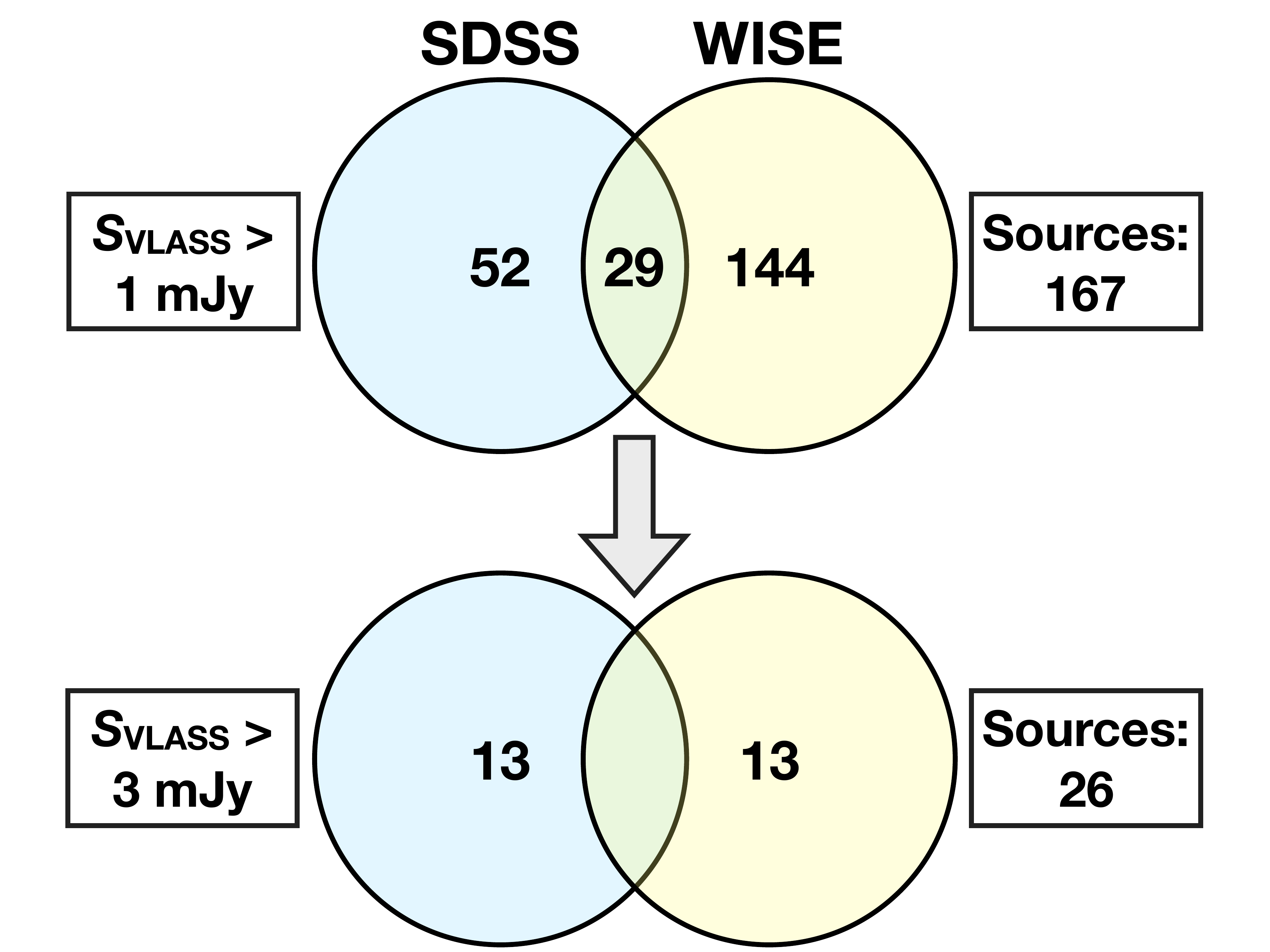}
\caption{Venn diagram illustrating the selection criteria for our sample of AGN that have recently brightened in the radio in the last 1--2 decades.  Previously radio-quiet AGN are selected based on both optical and infrared AGN diagnostics in SDSS and {\it WISE}, respectively.  Further details are provided in Section~\ref{sec:selection_criteria}.\\}
\label{fig:venn_diagram}
\end{figure*}
%%%%%%%%%%%%%%%%%%%%%%%%%%%%%%%%%%%%%%%%%%%%%%

%%%%%%%%%%%%%%%%%%%%%%%%%%%%%%%%%%%%%%%%%%%%%%%
\subsection{Radio Surveys}

%%%%%%%%%%%%%%%%%%%%%%%%%%%%%%%%%%%%%%%%%%%%%%%
\subsubsection{FIRST}
FIRST is a 20~cm (1.4~GHz) survey conducted with the pre-upgrade VLA covering a total sky footprint of 10,575~deg$^2$ \citep{Becker+95,helfand+15}. FIRST observed approximately 9,000~deg$^2$ of the north Galactic cap from 1993--2004, and a smaller $\sim2.5\degr$-wide strip of the south Galactic cap along the Celestial equator in 2009--2011. The VLA was undergoing the upgrade of its receivers during the latter time range, leading to slightly different image characteristics in the different regions of sky covered by FIRST, but the quality of the survey's sky coverage in the northern and southern regions are functionally equivalent.\footnote{\url{http://sundog.stsci.edu/}} FIRST images have a resolution of approximately 5\arcsec, rms sensitivity of approximately 0.15~mJy beam$^{-1}$, and astrometric uncertainty of $\lesssim1\arcsec$. The published catalog of emission components from FIRST is 95\% complete to 2 mJy and 80\% complete to the catalog limit of 1 mJy. Peak flux densities and integrated flux densities in the published catalog were determined from two-dimensional Gaussian fits to each co-added FIRST image.

%%%%%%%%%%%%%%%%%%%%%%%%%%%%%%%%%%%%%%%%%%%%%%%
\subsubsection{VLASS}
VLASS is a wide-area (33,885~deg$^2$), high-resolution (2.5$^{\prime \prime}$), full-polarization, broadband radio continuum survey currently being carried out by NRAO at $S$-band (2--4~GHz). The relative sky footprints of FIRST and VLASS are illustrated in Figure~\ref{fig:footprint}.  In order to enable transient science, a key science goal of the survey, VLASS observations are being taken over a series of 3 epochs with a cadence of 32 months.  Each VLASS epoch reaches a $1\sigma$ sensitivity of 0.12~mJy beam$^{-1}$, comparable to the depth of FIRST.  

VLASS Epoch~1 observations have recently been completed (2017--2019), and preliminary ``QuickLook" images have been made publicly available on the NRAO archive\footnote{\url{https://archive-new.nrao.edu/vlass/quicklook/}}.  The VLASS QuickLook images have flux uncertainties up to  $\sim20$\% and are not final survey data products (see \citealt{lacy+20} for details).  Despite the limitations of the QuickLook images, they are valuable for identifying transient and variable sources that have experienced substantial changes in flux (larger than a factor of 2) in the time period between FIRST and VLASS.

%%%%%%%%%%%%%%%%%%%%%%%%%%%%%%%%%%%%%%%%%%%%%%
\subsection{Variable AGN Selection Criteria}
\label{sec:selection_criteria}
Currently, no ``official" source catalog for VLASS has been made publicly available.  We therefore produced our own source catalog over a subset of VLASS Epoch~1.2 covering an area of $\sim$3,440~deg$^2$.  We used the source finding algorithm PyBDSF to compile a raw VLASS catalog based on the QuickLook images of sources brighter than 1~mJy.  Although our goal is to identify sources that have recently brightened in the radio, we note that similar studies of {\it reverse} transients (i.e.\ sources that were previously detected in FIRST but discovered to have declined drastically in flux when re-observed in VLASS 1-2 decades later) are also currently in progress (e.g.\ \citealt{law+18,cendes+20}). 

We performed a positional cross-match (within a radius of $1^{\prime \prime}$) with the FIRST catalog.  
The high prevalence of artifacts (uncleaned sidelobes) in the VLASS QuickLook images poses a challenge for reliable source extraction.  We manually vetted our preliminary catalog of VLASS-FIRST transients by eye to remove spurious VLASS sources associated with image artifacts or diffuse emission (radio lobes) with inherently ill-defined positions.  Such spurious sources account for the vast majority ($\sim$80--90\%) of our preliminary transient catalog. 
Our manually-vetted catalog covering the 3,440~deg$^2$ of VLASS Epoch~1 analyzed thus far  contains $\sim$2000 candidate transients that are compact and brighter than 1~mJy in VLASS but below the formal detection threshold ($1$~mJy) of FIRST.  Additional details will be presented in Dong et al. (in prep.). 

%%%%%%%%%%%%%%%%%%%%%%%%%%%%%%%%%%%%%%%%%%%%%%%
\begin{deluxetable*}{ccccccccccccc}[t!]
\tablecaption{Sample \label{tab:sample}}
\tablecolumns{12}
%\tablenum{1}
\tablewidth{0pt}
\tablehead{
\colhead{Source} & \colhead{RA} & \colhead{Dec} & \colhead{$z$} & \colhead{$L_{\rm Bol}$} & \colhead{$\log(M_{\rm SMBH})$}  & \colhead{$L_{\rm Bol}/L_{\rm Edd}$} & \colhead{Type} & \\
\colhead{} & \colhead{(J2000)} & \colhead{(J2000)} & \colhead{} & \colhead{(erg~s$^{-1}$)} & \colhead{(M$_{\odot}$)} & \colhead{} & \colhead{}\\
\colhead{(1)} & \colhead{(2)} & \colhead{(3)} & \colhead{(4)} & \colhead{(5)} & \colhead{(6)} & \colhead{(7)} & \colhead{(8)} 
}
\startdata
J0742+2704  & 115.701796  &  27.070113  &   0.6264  &    45.57  &     8.67  &     -1.20  &  SDSS, WISE\\ 
J0751+3154  & 117.877557  &  31.904037  &   1.8640  &    46.57  &     9.47  &     -1.00  &  SDSS, WISE\\ 
J0800+3124  & 120.044683  &  31.415871  &   1.9368  &    46.93  &     9.56  &     -0.73  &  SDSS, WISE\\ 
J0807+2102  & 121.767970  &  21.035786  &   1.5588  &    46.31  &     9.34  &     -1.13  &  SDSS, WISE\\ 
J0832+2302  & 128.198513  &  23.042816  &   0.9430  &  \nodata  &  \nodata  &   \nodata  &  SDSS, WISE\\ 
J0950+5128  & 147.653173  &  51.477212  &   0.2142  &    45.18  &     7.98  &     -0.90  &        SDSS\\ 
J1023+2219  & 155.842720  &  22.321296  &  \nodata  &  \nodata  &  \nodata  &   \nodata  &        WISE\\ 
J1037-0736  & 159.491631  &  -7.607362  &  \nodata  &  \nodata  &  \nodata  &   \nodata  &        WISE\\ 
J1112+2809  & 168.055854  &  28.165137  &  \nodata  &  \nodata  &  \nodata  &   \nodata  &        WISE\\ 
J1157+3142  & 179.388528  &  31.707514  &   0.8910  &  \nodata  &  \nodata  &   \nodata  &  SDSS, WISE\\ 
J1204+1918  & 181.218737  &  19.306199  &   2.3440  &  \nodata  &  \nodata  &   \nodata  &  SDSS, WISE\\ 
J1208+4741  & 182.241352  &  47.698944  &   1.0915  &  \nodata  &  \nodata  &   \nodata  &  SDSS, WISE\\ 
J1246+1805  & 191.518270  &  18.089036  &  \nodata  &  \nodata  &  \nodata  &   \nodata  &        WISE\\ 
J1254-0606  & 193.521438  &  -6.109209  &  \nodata  &  \nodata  &  \nodata  &   \nodata  &        WISE\\ 
J1333-0349  & 203.334269  &  -3.832307  &  \nodata  &  \nodata  &  \nodata  &   \nodata  &        WISE\\ 
J1347+4505  & 206.836689  &  45.098706  &  \nodata  &  \nodata  &  \nodata  &   \nodata  &        WISE\\ 
J1357-0329  & 209.443830  &  -3.484817  &  \nodata  &  \nodata  &  \nodata  &   \nodata  &        WISE\\ 
J1413+0257  & 213.454848  &   2.953648  &  \nodata  &  \nodata  &  \nodata  &   \nodata  &        WISE\\ 
J1447+0512  & 221.817580  &   5.207433  &   1.7475  &    46.36  &     9.02  &     -0.76  &  SDSS, WISE\\ 
J1507-0549  & 226.962335  &  -5.819718  &  \nodata  &  \nodata  &  \nodata  &   \nodata  &        WISE\\ 
J1512-0533  & 228.096364  &  -5.550100  &  \nodata  &  \nodata  &  \nodata  &   \nodata  &        WISE\\ 
J1514+4000  & 228.520730  &  40.013724  &   2.1226  &    46.80  &     9.74  &     -1.04  &        SDSS\\ 
J1546+1500  & 236.641541  &  15.010193  &  \nodata  &  \nodata  &  \nodata  &   \nodata  &        WISE\\ 
J1603+1809  & 240.828155  &  18.151432  &   3.2460  &  \nodata  &  \nodata  &   \nodata  &        SDSS\\ 
J1609+4306  & 242.441175  &  43.106462  &  \nodata  &  \nodata  &  \nodata  &   \nodata  &        WISE\\ 
J2109-0644  & 317.320264  &  -6.743461  &   1.0812  &    46.38  &     9.05  &     -0.77  &        SDSS\\ 
\enddata
\tablecomments{Column 1: Source name.  Columns 2 and 3: Source right ascension and declination. Column 4: Spectroscopic redshift from SDSS DR14 from \citet{paris+18}.  Column 5: Bolometric quasar luminosity. Column 6: Virial supermassive black hole mass estimate from SDSS spectroscopy from \citet{shen+11}.  Column 7:  Eddington ratio. Column 8: AGN type.  Sources are classified as AGN on the basis of optical spectroscopy from SDSS (\citep{paris+18} and/or infrared colors based on data from WISE \citep{assef+18}.}
\end{deluxetable*}

%%%%%%%%%%%%%%%%%%%%%%%%%%%%%%%%%%%%%%%%%%%%%%%
\begin{deluxetable*}{ccccccccc}[t!]
\tablecaption{FIRST and VLASS Properties \label{tab:FIRST_VLASS}}
\tablecolumns{9}
%\tablenum{1}
\tablewidth{0pt}
\tablehead{
\colhead{Source} & \colhead{Date$_{\rm FIRST}$} & \colhead{$\sigma_{\rm FIRST}$} & \colhead{$F_{\rm FIRST}$} & \colhead{Date$_{\rm VLASS}$} & \colhead{$\sigma_{\rm VLASS}$} & \colhead{$F_{\rm VLASS}$} & \colhead{$\log(L_{\rm VLASS})$}\\
\colhead{} & \colhead{} & \colhead{(mJy~beam$^{-1}$)} & \colhead{(mJy~beam$^{-1}$)} & \colhead{} & \colhead{(mJy~beam$^{-1}$)} & \colhead{(mJy~beam$^{-1}$)} & \colhead{(erg~s$^{-1}$)}\\
\colhead{(1)} & \colhead{(2)} & \colhead{(3)} & \colhead{(4)} & \colhead{(5)} & \colhead{(6)} & \colhead{(7)} & \colhead{(8)} 
}
\startdata
J0742+2704	&	1995 Nov 10	&	0.166	&	0.53	&	2019 Apr 10	&	0.146	& 9.19 &    	41.68 \\
J0751+3154	&	1995 Oct 23	&	0.148	&	$<$ 0.44	&	2019 Apr 13	&	0.148	& 3.00 &	42.36 \\
J0800+3124	&	1994 Jun 04	&	0.123	&	0.38	&	2019 Apr 13	&	0.125	& 4.54 &	42.58 \\
J0807+2102	&	1998 Sep	&	0.157	&	$<$ 0.47	&	2019 Apr 14	&	0.137	& 3.69 &	42.26 \\
J0832+2302	&	1995 Dec 28	&	0.157	&	0.48	&	2019 Apr 13	&	0.130	& 3.08 &	41.64 \\
J0950+5128	&	1997 Apr 29	&	0.148	&	$<$ 0.44	&	2019 Apr 18	&	0.126	& 8.77 &	40.57 \\
J1023+2219	&	1996 Jan	&	0.205	&	0.65	&	2019 Apr 18	&	0.145	& 3.50 &	\nodata \\
J1037-0736	&	2002 Jun	&	0.136	&	$<$ 0.41	&	2019 Apr 16	&	0.151	& 12.95 &	\nodata \\
J1112+2809	&	1995 Nov 04	&	0.156	&	0.49	&	2019 Apr 13	&	0.120	& 3.19 &	\nodata \\
J1157+3142	&	1994 Jun 06	&	0.601	&	2.02	&	2019 Apr 14	&	0.135	& 8.58 &	42.03 \\
J1204+1918	&	1999 Nov	&	0.149	&	0.48	&	2019 Mar 21	&	0.145	& 5.09 &	42.84 \\
J1208+4741	&	1997 Apr 05	&	0.157	&	$<$ 0.47	&	2019 Apr 15	&	0.117	& 3.20 &	41.82 \\
J1246+1805	&	1999 Nov 24	&	0.147	&	$<$ 0.44	&	2019 Apr 12	&	0.122	& 7.09 &	\nodata \\
J1254-0606	&	2001 Apr	&	0.144	&	0.46	&	2019 Mar 07	&	0.247	& 8.76 &	\nodata \\
J1333-0349	&	1998 Sep	&	0.14	&	$<$ 0.42	&	2019 Mar 24	&	0.175	& 3.09 &	\nodata \\
J1347+4505	&	1997 Mar 03	&	0.151	&	$<$ 0.45	&	2019 Mar 19	&	0.129	& 3.58 &	\nodata \\
J1357-0329	&	1998 Sep	&	0.156	&	0.49	&	2019 Mar 24	&	0.170	& 3.11 &	\nodata \\
J1413+0257	&	1998 Jul	&	0.142	&	0.47	&	2019 Apr 02	&	0.229	& 6.91 &	\nodata \\
J1447+0512	&	2000 Feb	&	0.16	&	0.49	&	2019 Mar 12	&	0.152	& 4.30 &	42.45 \\
J1507-0549	&	2001 Apr	&	0.139	&	$<$ 0.42	&	2019 Mar 07	&	0.159	& 4.73 &	\nodata \\
J1512-0533	&	2001 Apr	&	0.233	&	$<$ 0.70	&	2019 Mar 07	&	0.200	& 3.25 &	\nodata \\
J1514+4000	&	1994 Aug 21	&	0.141	&	$<$ 0.42	&	2019 Mar 28	&	0.164	& 3.49 &	42.57 \\
J1546+1500	&	1999 Dec	&	0.144	&	$<$ 0.43	&	2019 Apr 11	&	0.122	& 4.69 &	\nodata \\
J1603+1809	&	1999 Nov	&	0.148	&	$<$ 0.44	&	2019 Apr 12	&	0.116	& 3.61 &	43.03 \\
J1609+4306	&	1994 Jul 24	&	0.124	&	0.45	&	2019 May 04	&	0.126	& 4.53 &	\nodata \\
J2109-0644	&	1997 Feb 28	&	0.153	&	$<$ 0.46	&	2019 Apr 05	&	0.170	& 18.30 &	42.57 \\
\enddata
\tablecomments{Column 1: Source name. Column 2: Date of FIRST observation. For some sources, only the year and month are provided in the header of the FIRST image. Column 3: Local 1$\sigma$ rms noise in FIRST.  Column 4:  Peak flux density at 1.4~GHz from FIRST.  Newly identified sources in VLASS that have faint FIRST counterparts with peak flux densities $>3\sigma_{\rm FIRST}$ are reported as detections. All other sources are reported as FIRST upper limits.  We note that none of the sources in this table meet the selection criteria of the FIRST catalog \citep{helfand+15}. Column 5: Date of VLASS observation.  Column 6: Local 1$\sigma$ rms noise in VLASS.  Column 7: Peak flux density at 3~GHz from VLASS.  Column 9: Radio luminosity at 3~GHz from VLASS.}
\end{deluxetable*}

%%%%%%%%%%%%%%%%%%%%%%%%%%%%%%%%%%%%%%%%%%%%%%%
\begin{figure*}[t!]
\centering
\includegraphics[clip=true, trim=3.5cm 0cm 5.7cm 0cm, width=\textwidth]{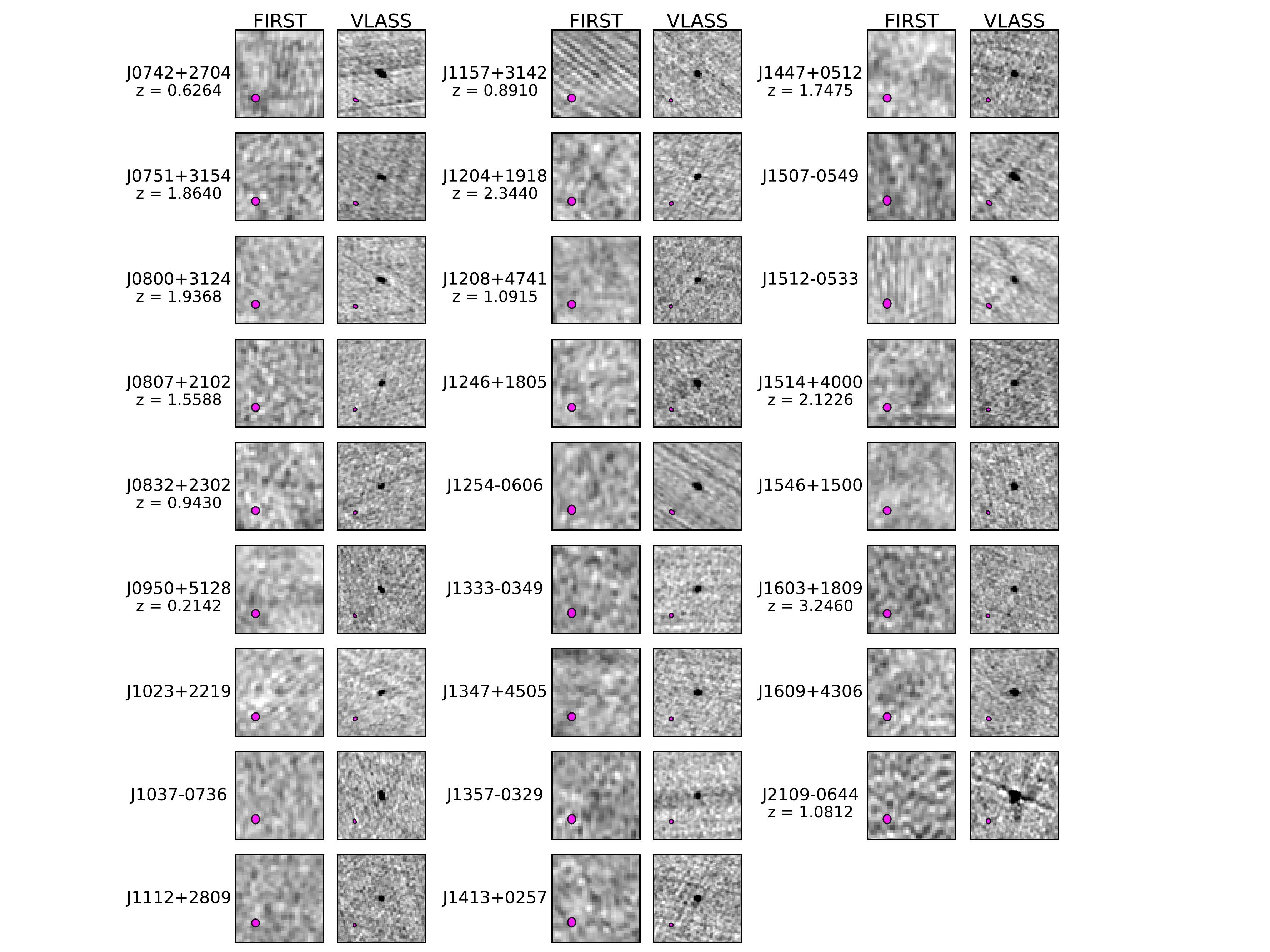}
\caption{Cut-out images (30$^{\prime \prime} \times 30^{\prime \prime}$) from FIRST (1.4~GHz; 1993-2011) and VLASS Epoch~1 (3~GHz; 2017--2019) of our sample.  The synthesized beams (with typical major axis extents of approximately 5$^{\prime \prime}$ and 2.5$^{\prime \prime}$ for FIRST and VLASS, respectively) are shown as magenta ellipses in the lower left corner of each image. 
%highlighting their variability on timescales of $\sim1-2$ decades.
\\}
\label{fig:VLASS_FIRST_compare}
\end{figure*}
%%%%%%%%%%%%%%%%%%%%%%%%%%%%%%%%%%%%%%%%%%%%%%

To select AGN with variable radio emission, we considered both optical and infrared AGN selection techniques to capture both the unobscured and obscured populations.  In the optical, we used the compilation of spectroscopically verified quasars from the Sloan Digital Sky Survey (SDSS; \citealt{york+00}) DR14 \citep{paris+18} and found 52 candidate radio-variable quasars within $<1^{\prime \prime}$ of the VLASS position.  Obscured AGN that have recently brightened in the radio were identified on the basis of their infrared colors using data from the Widefield Infrared Survey Explorer (WISE; \citealt{wright+10}).  We used the "R90" diagnostic criteria from \citet{assef+18}, which were designed to identify AGN with 90\% reliability.  A total of 144 obscured AGN with variable radio emission were identified using the R90 catalog, of which 29 sources also met our optical selection criteria, making a total of 167 AGN associated with our manually-vetted FIRST/VLASS sample. 
%Thus, the 167 unique AGN possibly exhibiting radio variability are detected at a rate of 1 source per 20 deg$^2$.   %Additionally, we note that 29 sources met both our optical and infrared AGN diagnostic criteria.  The detection rate of candidate 

%Due to the sparse density of sources with flux densities above 1~mJy in the radio sky, these spatial coincidences provide very strong evidence that the transient candidates are physically associated with the AGN. In particular, a random point within the footprint from which our transients are drawn has a 1 in 80,000 (1 in 40,000) chance of being located within 1$^{\prime \prime}$ of an SDSS (R90) AGN. Thus, the probability that any of the 52 (144) candidate transients are matched by chance coincidence with their quasar (WISE AGN) counterparts is approximately 1 in 1500 (1 in 250). 

With the AGN associations confirmed, we selected a subset of sources with peak flux densities brighter than 3~mJy~beam$^{-1}$ in VLASS.  %This selects a sample of promising variable AGN, since the implied spectral index between FIRST and VLASS would be steeper than $\nu^{2.5}$, thereby ruling out steady sources that are optically thick due to synchrotron self absorption 
This selects a more reliable sample radio-variable AGN, since the implied spectral index between FIRST and VLASS would be steeper than $\nu^{2.5}$, thereby ruling out steady sources that are optically thick\footnote{We emphasize that steady, albeit optically-thick, radio sources associated with powerful optical/IR AGN are an extremely interesting source population in their own right.  Such sources may represent young radio AGN with radio spectral energy distributions peaking in the GHz regime (see \citealt{odea+98} for a review).} due to synchrotron self absorption (though free-free absorption is still a possibility).    
%Our final sample of transients from Epoch~1.2 of VLASS associated with an AGN consists of 26 sources. %, 13 of which have spectroscopic redshifts in the range $0.2 \lesssim z \lesssim 3.2$. 
%The final sample of radio-variable AGN considered in this study consists of 26 sources. 
Following the 3~mJy cut, the final sample of radio-variable AGN contains 26 objects, and it is these that we consider in this paper.  We provide an illustration of our selection criteria in Figure~\ref{fig:venn_diagram}.

%%%%%%%%%%%%%%%%%%%%%%%%%%%%%%%%%%%%%%%%%%%%%%%
\subsection{Source Properties}
We summarize the basic properties %(positions, redshifts, and radio survey data) 
of our sample in Tables~\ref{tab:sample} and \ref{tab:FIRST_VLASS}.  Of these 26 sources, 13 are selected from SDSS spectroscopy and the remaining 13 are obscured AGN selected based on the WISE criteria described in Section~\ref{sec:selection_criteria}.  
The 13 optically-selected AGN are classified as broad-line quasars with redshifts in the range $0.2 < z < 3.2$.   \citet{shen+11} report bolometric luminosities of $\log(L_{\rm bol}/{\rm erg \, s^{-1}}) \approx 45.2 - 46.8$ from SDSS spectroscopy, virial SMBH masses in the range $\log(M/{\rm M}_{\odot}) \approx 8-9.7$, and Eddington ratios of 6--20\% (consistent with radiatively efficient SMBH accretion; e.g. \citealt{heckman+14}).  All of our sources would have previously been classified as radio-quiet based on their upper limits in FIRST.  The observed increase in radio flux between FIRST and VLASS suggests they are no longer in a radio-quiet state.  Given the high VLASS radio luminosities ($L_{\rm 3\,GHz} = 10^{40 - 42} \,\, {\rm erg} \,{\rm s}^{-1}$), our sources are now consistent with radio-loud quasars (\citealt{kellermann+16}). 

In Figure~\ref{fig:VLASS_FIRST_compare}, we show image cut-outs from FIRST and VLASS.  The 1$\sigma$ rms sensitivities of the FIRST and VLASS images are typically 0.15~mJy~beam$^{-1}$ and 0.12~mJy~beam$^{-1}$, respectively.  The local rms noise level measured in each image is given in Table~\ref{tab:FIRST_VLASS}.  The quality of the FIRST images is generally quite good.  However, we note that one source in particular (J1157+3142) appears to have unusually poor image quality in the FIRST survey, making its classification as a variable AGN uncertain.  We emphasize that while none of our sources met the formal detection threshold criteria ($F_{\rm peak} >1$~mJy) of the final release of the FIRST source catalog \citep{helfand+15}, 12 sources have faint FIRST emission at the location of the VLASS source at the $3-3.6\sigma$ level (J1204+1918).  We list sources with peak FIRST flux densities $>3\sigma$ as FIRST ``detections" in Table~\ref{tab:FIRST_VLASS}.  

%%%%%%%%%%%%%%%%%%%%%%%%%%%%%%%%%%%%%%%%%%%%%%%
\begin{deluxetable*}{cccccc}[t!]
\tablecaption{Summary of New Multiband VLA Observations  \label{tab:summary}}
\tablecolumns{18}
\tablewidth{0pt}
\tablehead{
\colhead{Source} & \colhead{Observing Date} & \colhead{Configuration} & \colhead{VLA Bands} & \colhead{Flux calibrator} %& \colhead{Notes} 
\\
\colhead{(1)} & \colhead{(2)} & \colhead{(3)} & \colhead{(4)} & \colhead{(5)} %& \colhead{(6)} 
}
\startdata
J0742+2704	& 2019 Oct 03 & A	     &	L, S, C, X, Ku	& 3C147 \\
J0807+2102	& 2019 July 23 & BnA $\rightarrow$ A &	L, S, C, X	    & 3C286 \\
        	& 2019 Sept 19 & A        &	X, Ku, K	    & 3C138 \\
J0832+2302	& 2019 July 23 & BnA $\rightarrow$ A &	L, S, C, X	    & 3C286 \\
        	& 2019 Sept 19 & A        &	X, Ku   	    & 3C138 \\
J0950+5128	& 2019 July 23 & BnA $\rightarrow$ A &	L, S, C, X   	& 3C286 \\
        	& 2019 Sept 19 & A        &	X, Ku   	    & 3C138 \\
J1037-0736	& 2019 Oct 13 & A	     &	L, S, C, X, Ku	& 3C286 \\
%        	& 2019 Oct 20 & A	     &	L, S, C, X, Ku	& 3C286 \\
J1204+1918	& 2019 Oct 11 & A	     &	L, S, C, X, Ku	& 3C286 \\
J1208+4741	& 2019 Oct 11 & A	     &	L, S, C, X, Ku	& 3C286 \\
J1246+1805	& 2019 Oct 11 & A	     &	L, S, C, X, Ku	& 3C286 \\
J1254-0606	& 2019 Nov 01 & A $\rightarrow$ D	  &	L, S, C, X, Ku  & 3C286 \\
J1413+0257	& 2019 Sept 23 & A	     &	L, S, C, X, Ku	& 3C286 \\
J1447+0512	& 2019 Sept 23 & A	     &	L, S, C, X, Ku	& 3C286 \\
J1546+1500	& 2019 Sept 23 & A	     &	L, S, C, X, Ku	& 3C286 \\
J1603+1809	& 2019 Sept 23 & A	     &	L, S, C, X, Ku	& 3C286 \\
J2109-0644	& 2019 Sept 10 & A	     &	L, S, C, X, Ku	& 3C48 \\
\enddata
\tablecomments{Column 1: Source name.  Column 2: Observing date(s).  Column 3: VLA configuration.  Column 4: VLA band, defined as follows: L: 1--2~GHz, S: 2--4~GHz, C: 4--8~GHz, X: 8--12~GHz, Ku: 12--18~GHz, K: 18--26~GHz.  Column 5: Flux density calibrator.  We note that the calibrator 3C48 is currently experiencing an ongoing flaring event.  As a result, the absolute flux uncertainty of sources calibrated against this source will be increased.  We therefore assume a conservative flux uncertainty of 20\% for all sources using 3C48 as the primary flux calibrator.  All other flux calibrators are expected to provide an absolute uncertainty of 3\% \cite{perley+13}.  %Column 6: Notes.  
%{\color{red}[Add notes  correlator malfunction during J1037-0736 observation, and explain move-time data.]}
}
\end{deluxetable*}
%%%%%%%%%%%%%%%%%%%%%%%%%%%%%%%%%%%%%%%%%%%%%%%

The VLASS QuickLook image cut-outs shown in Figure~\ref{fig:VLASS_FIRST_compare} highlight the radio variability of our sources on timescales of $\sim1-2$ decades as well as their compactness on $\sim$arcsecond scales.   
Some of the VLASS QuickLook images suffer from strong imaging artifacts, such as prominent sidelobes due to insufficient cleaning and/or the presence of residual phase errors.  This is due to the rough nature of the QuickLook images\footnote{See \citet{lacy+20} for a detailed description of VLASS and the limitations of the QuickLook image products.}, which we emphasize are not the final VLASS survey products.  Despite the known limitations of the QuickLook images, we find the peak VLASS flux densities of our compact sources to typically be consistent with our independent VLA follow-up imaging at $S$-band (see Section~\ref{sec:variability_timescale} for a discussion of source variability at $S$ band). Thus, our study supports the viability of the use of VLASS QuickLook images for science as long as users take the known limitations into account. 

%%%%%%%%%%%%%%%%%%%%%%%%%%%%%%%%%%%%%%%%%%%%%%%
%%%%%%%%%%%%%%%%%%%%%%%%%%%%%%%%%%%%%%%%%%%%%%%
%%%%%%%%%%%%%%%%%%%%%%%%%%%%%%%%%%%%%%%%%%%%%%%
\section{VLA Follow-up Data}
\label{sec:data}
We performed high-resolution, multiband VLA observations for 14/26 candidate variable AGN through project 19A-422 (PI: Gregg Hallinan).  These observations took place from July 23 -- October 13, 2019, primarily during the VLA A configuration.  For each source, the observations at $L$ (1--2~GHz), $S$ (2--4~GHz), $C$ (4--8~GHz), and $X$ (8--12~GHz) band were performed within a single scheduling block (SB) no longer than 2.75~hr, and we therefore refer to these observations as ``quasi-simultaneous" in nature.  

In addition to the 1--12~GHz observations, the majority of our sources also have quasi-simultaneous data at \emph{Ku} (12--18~GHz) band.  Because a preliminary analysis of our first three pilot sources (J0807+2102, J0832+2302, and J0950+5128) showed rising fluxes at high frequencies, we modified our observing strategy to add higher frequency $Ku$-band observations (12--18~GHz) for the remaining sample, and returned later to our three pilot sources on September 19, 2019 to repeat the $X$-band observations and add $Ku$ band. We discuss the X-band variability on $\sim$weeks timescale for these three sources in Section~\ref{sec:variability_timescale}.  Finally, for the pilot source J0807+2102 with the most optically-thick spectrum, we added a $K$-band (18-26 GHz) observation. The dates and frequencies of all our VLA observations are given in Table~\ref{tab:summary}.  
% The first 3 ``pilot" sources (J0807+2102, J0832+2302, and J0950+5128) were observed on July 23, 2019 as part of our multi-band follow-up campaign, at which point VLA data were only obtained from 1--12~GHz.  
 %After processing these data and constructing preliminary radio SEDs, we found higher-frequency spectral peaks ($>10$~GHz) than originally expected.  
%Preliminary data processing and radio SED modelling revealed rising radio spectral shapes over this frequency range, thus motivating the need for observations at higher frequencies.  
% We thus modified our observing strategy for the remainder of our sample to cover the 1--18~GHz range.  For the 3 pilot sources, we obtained \emph{Ku}-band data to extend the frequency coverage up to 18~GHz several weeks (on September 19, 2019) later, along with a second $X$-band epoch to check for any substantial variability that might complicate our radio SED analysis.  We discuss the $X$-band variability on timescales of $\sim$weeks for these sources in Section~\ref{sec:variability_timescale}). In addition to the extension to \emph{Ku} band for the pilot sources, we also obtained data at $K$ band (18--26~GHz) for J0807+2102, the pilot source with the most optically-thick spectral index from the original observations at 1--12~GHz. 
%A summary of our new multi-band VLA observations is provided in Table~\ref{tab:summary}. 

\subsection{Calibration and Imaging}
%We calibrated our data using the scripted CASA VLA calibration pipeline\footnote{\url{https://science.nrao.edu/facilities/vla/data-processing/pipeline/scripted-pipeline}} for CASA 5.3.0.  
We calibrated our data using the scripted VLA calibration pipeline\footnote{\url{https://science.nrao.edu/facilities/vla/data-processing/pipeline/scripted-pipeline}} for the Common Astronomy Software Applications (CASA) package \citep{mcmullin+07} version 5.3.0. 
In order to ensure the use of an optimal calibration solution interval for each band, the pipeline was run separately for each band of each SB. 

Imaging and self-calibration were performed in CASA 5.6.0 following standard heuristics for widefield, broadband VLA data.  We used the TCLEAN task to produce a full field-of-view image at each band and employed the $w$-projection algorithm \citep{cornwell+05} to correct for non-coplanar baselines.  To avoid bandwidth smearing over the large fractional bandwidths of our observations, deconvolution was performed using the MTMFS (multi-term, multi-frequency synthesis) algorithm \citep{rau+11}.  We adopted a Briggs weighting scheme \citep{briggs+95} with a robust parameter between -0.5 and 0.5 to achieve the best compromise among sidelobe levels, resolution, and sensitivity given the quality of the $uv$-coverage and resulting point-spread function (PSF) of each dataset\footnote{As indicated in Table~\ref{tab:summary}, some of our observations took place during transitions between different VLA antenna configurations (BnA $\rightarrow$ A and A $\rightarrow$ D), thus leading to poor PSF quality.  We mitigated this effect by fine-tuning the CASA TCLEAN parameters by hand as needed, in particular the Briggs robust parameter.}.   

\subsection{VLITE}
Commensal low-frequency ($<$1~GHz) data were also recorded during our VLA follow-up campaign.  This commensal system, known as the VLA Low-band Ionosphere and Transient Experiment (VLITE; \citealt{clarke+16,polisensky+16}), records data at 340~MHz simultaneously with regular VLA observing programs\footnote{VLITE data are recorded simultaneously with nearly all VLA observations, including VLASS.  The VLITE counterpart to VLASS is known as the VLITE Commensal Sky Survey (VCSS).  As of the first epoch of VLASS observations, VCSS images reach typical rms noises of 3~mJy~beam$^{-1}$ and have angular resolutions of 12--25 arcseconds \citep{lacy+20}.}.  
%VLITE accumulates more than 6200 hours of data per year from the VLA, making a vast -- yet largely untapped -- resource for enhancing the legacy value and scientific impact of VLA studies.
%VLITE is capable of continuously accessing a 64~MHz bandwidth from the 236--492 MHz VLA Low-band system.  The VLITE system includes a dedicated DiFX software correlator, automated data processing pipelines, and an SQL database.  VLITE science operations with 10 VLA antennas began on November 25, 2014, and the system has since been expanded to 16 and 18 antennas.  Data are recorded simultaneously with nearly all VLA observations, including the VLA Sky Survey\footnote{The VLITE counterpart to VLASS is known as the VLITE Commensal Sky Survey (VCSS).  As of the first epoch of VLASS observations, VCSS images reach typical rms noises of 3~mJy~beam$^{-1}$ and have angular resolutions of 12--25 arcseconds (Lacy et al., submitted).} (VLASS).  
%Notable exceptions when VLITE data are not recorded include NRAO TAC-approved VLA P-band projects and observations of moving targets (e.g., solar system objects) at any band.  
%VLITE science operations with 10 VLA antennas began on November 25, 2014, and the system has since been expanded to 16 and 18 antennas in 2017 and 2019, respectively.  
%All VLITE data products, including images processed by an automated pipeline, are currently hosted by the U.S. Naval Research Laboratory. 
During our VLA observations, commensal VLITE data with a maximum of 18 antennas were recorded.  

VLITE data are automatically calibrated and imaged through an analysis pipeline based on the Obit \citep{cotton+08} data reduction package.  The VLITE pipeline images from our snapshot observations typically achieve a 1$\sigma$ depth of 3~mJy~beam~$^{-1}$ and a maximum angular resolution of $\approx5^{\prime \prime}$.  
We visually inspected all VLITE imaging pipeline products and found that none of our sources was detected, as expected given their curved or flat radio SEDs. We discuss constraints on the radio SEDs and underlying emission mechanisms using the upper limits from VLITE in Section~\ref{sec:SEDs}. 
%We visually inspected all VLITE imaging pipeline products associated with our sources.  None of our sources are detected in the commensal VLITE data, as would be expected for sources with flat or curved radio spectral shapes. 
%We further discuss constraints on the radio SEDs and underlying emission mechanisms using upper limits from VLITE in Section~\ref{sec:SEDs}. 

%%%%%%%%%%%%%%%%%%%%%%%%%%%%%%%%%%%%%%%%%%%%%%%
%%%%%%%%%%%%%%%%%%%%%%%%%%%%%%%%%%%%%%%%%%%%%%%
%%%%%%%%%%%%%%%%%%%%%%%%%%%%%%%%%%%%%%%%%%%%%%%
\section{Results}
\label{sec:results}

%%%%%%%%%%%%%%%%%%%%%%%%%%%%%%%%%%%%%%%%%%%%%%%
\begin{figure*}[t!]
\centering
\includegraphics[clip=true, trim=0cm 0.45cm 0cm 0cm, width=0.78\textwidth]{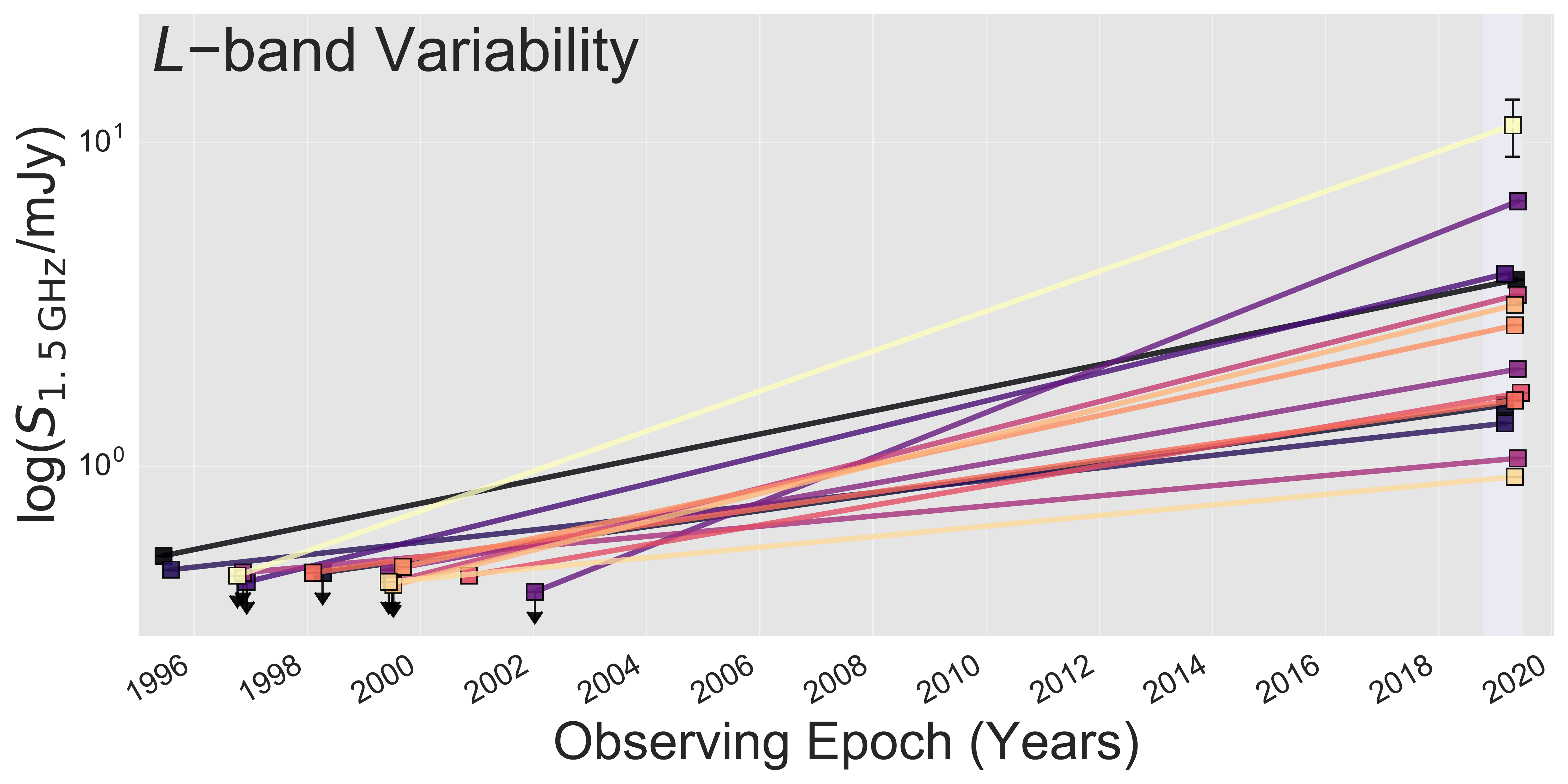}
\includegraphics[clip=true, trim=0cm 0.45cm 0cm 0cm, width=0.78\textwidth]{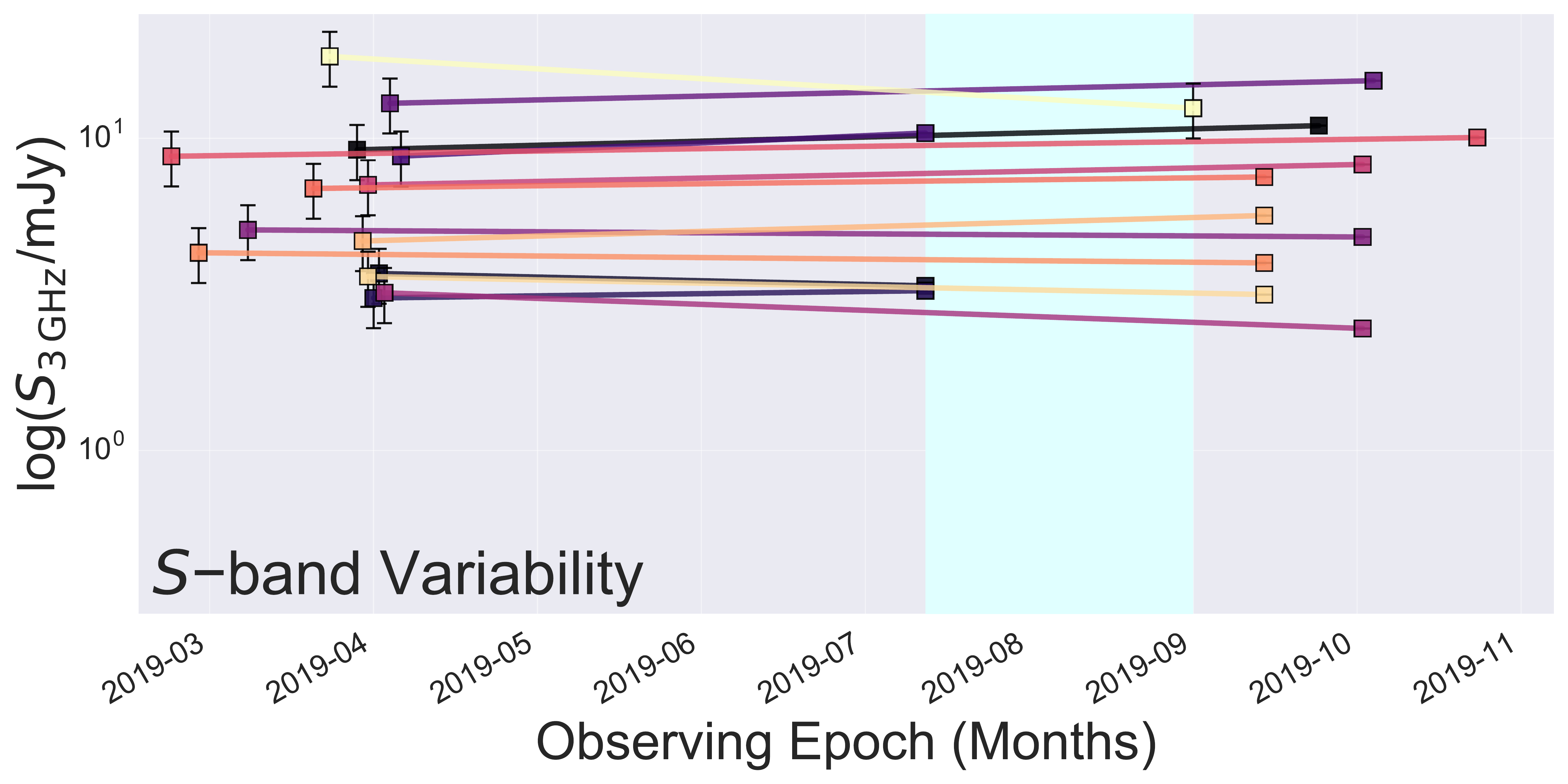}
\includegraphics[clip=true, trim=0cm 0.45cm 0cm 0cm, width=0.78\textwidth]{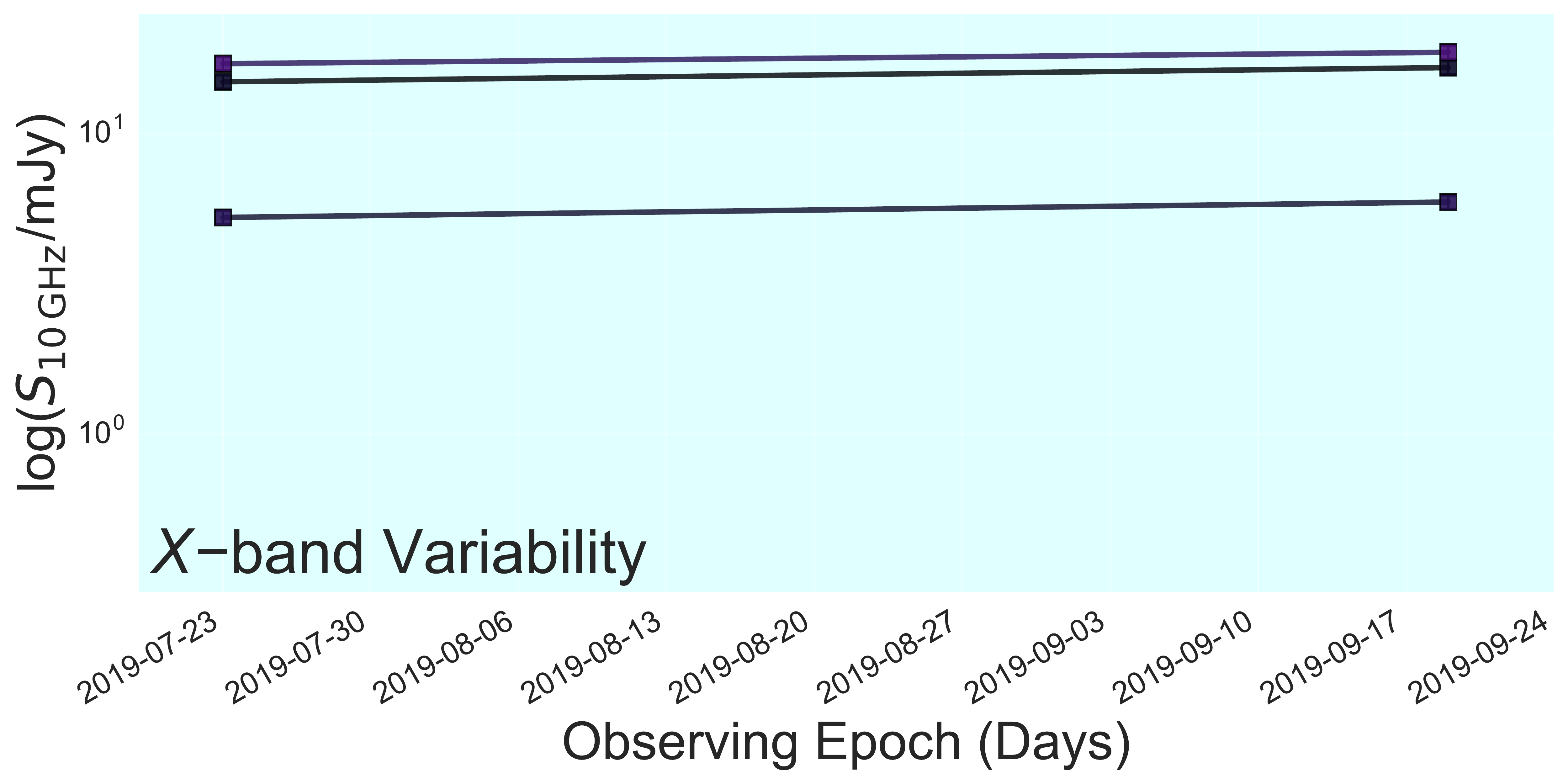}
\caption{Radio flux variability at $L$, $S$, and $X$ band.  Two observing epochs are shown for each band. At $L$ band, we show the flux measurements or upper limits from FIRST (1993--2011) and the flux densities from our 2019 JVLA follow-up observations.  At $S$ band, we show the flux densities from VLASS Epoch~1.2, which were observed in early 2019, and the $S$-band flux densities from our follow-up observations taken a few months later.  All of the VLASS $S$-band measurements are shown with a conservative 20\% flux uncertainty \citep{lacy+20}.  The large uncertainty (20\%) in the flux of a single object (J2109-0644) is apparent at both $L$ and $S$ band in our 2019 observations; we discuss limitations of the absolute flux scale calibration for this source in Section~\ref{sec:data}.  At $X$ band, only 3 sources (J0807+2102, J0832+2302, and J0950+5128), which have two epochs of $X$-band observations, are shown (see Sections~\ref{sec:data} and \ref{sec:variability_timescale}).
\\}
\label{fig:timeseries}
\end{figure*}
%%%%%%%%%%%%%%%%%%%%%%%%%%%%%%%%%%%%%%%%%%%%%%

%%%%%%%%%%%%%%%%%%%%%%%%%%%%%%%%%%%%%%%%%%%%%%%
\begin{figure*}[t!]
\centering
\includegraphics[clip=true, trim=0.5cm 0.5cm 0cm 0cm, width=0.495\textwidth]{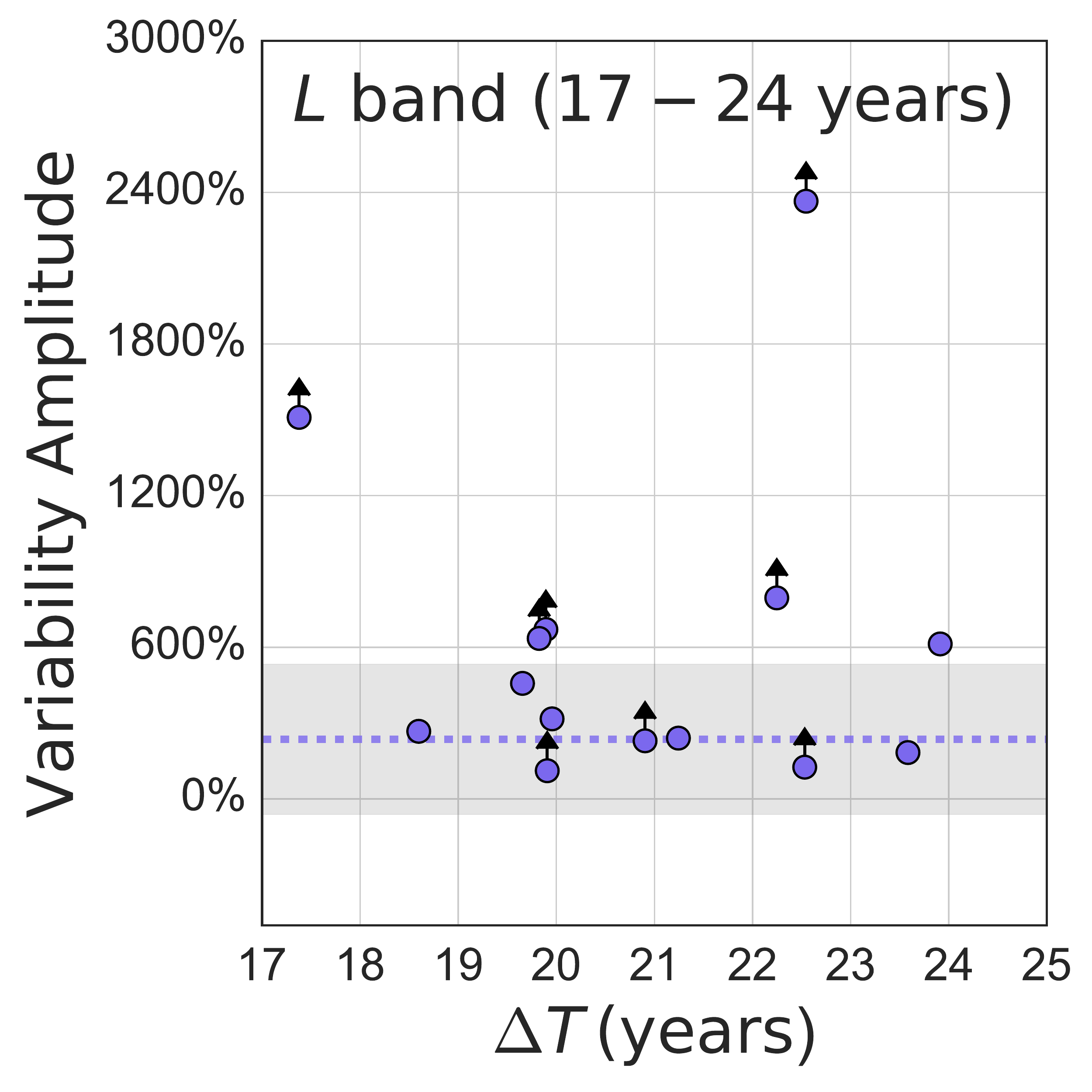}
\includegraphics[clip=true, trim=0cm 0.5cm 0.5cm 0cm, width=0.495\textwidth]{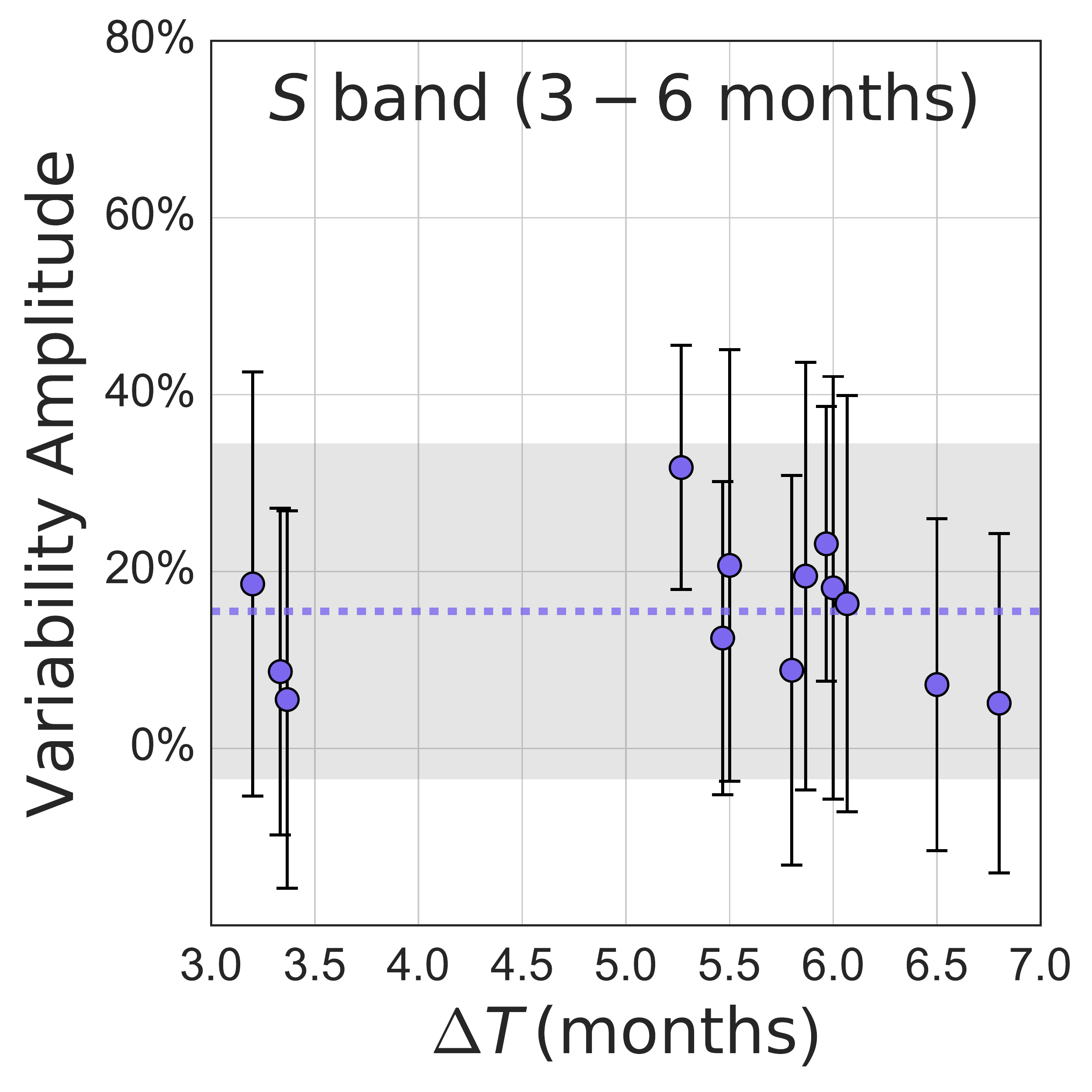}
\caption{Variability amplitude as a function of time from the dual-epoch observations at $L$ and $S$ band shown in Figure~\ref{fig:timeseries}.  
In the left panel, the variability amplitude is shown as the percent change in flux  between FIRST and our 1.5 GHz follow-up observations, or  $(F_{\rm FIRST}-F_{1.5\,\rm GHz})/F_{\rm FIRST}$. In the right panel, the variability amplitude is the percent change in flux between VLASS and our 3 GHz follow-up observations, or  $(F_{\rm VLASS}-F_{3\,\rm GHz})/F_{\rm VLASS}$.  The dashed lines indicate the median variability amplitude at each band and the gray shaded regions denote the weighted 1$\sigma$ standard deviations.  The large errors on the $S$-band variability amplitudes are dominated by the conservative 20\% uncertainty in the flux densities measured from the VLASS Epoch~1 QuickLook images \citep{lacy+20}.   
\\}
\label{fig:variability}
\end{figure*}
\begin{deluxetable*}{ccccccccccc}[t!]
\tablecaption{Radio Spectral Modeling Parameters \label{tab:sizes}}
\tablecolumns{10}
%\tablenum{1}
\tablewidth{0pt}
\tablehead{
\colhead{Source} & \colhead{$\theta_{\rm max}$} & \colhead{Linear Size} & \colhead{$\nu_{\rm peak}$} & \colhead{$\nu_{\rm peak,\,rest}$} & \colhead{$\chi_{\rm red,\,PL}^2$} & \colhead{$\chi_{\rm red,\,CPL}^2$} & \colhead{$q$} & \colhead{$\alpha_{\rm thick}$} &  \colhead{$\alpha_{\rm thin}$} \\
\colhead{} & \colhead{(arcsec)} & \colhead{(pc)} & \colhead{(GHz)} & \colhead{(GHz)}\\
\colhead{(1)} & \colhead{(2)} & \colhead{(3)} & \colhead{(4)}  & \colhead{(5)}  & \colhead{(6)} & \colhead{(7)} & \colhead{(8)} & \colhead{(9)} & \colhead{(10)}
}
\startdata
J0742+2704 &  $<$0.17 &  $<$1193 &   8.6 &      14.0 &    107.75 &       9.23 &  $-0.65\pm0.09$ &  1.59$\pm$0.10 & $-0.62\pm0.17$\\ 
J0807+2102 &  $<$0.11 &   $<$956 &  21.9 &      56.2 &     67.53 &      35.97 &  $-0.36\pm0.16$ &  1.25$\pm$0.11 &        \nodata\\ 
J0832+2302 &  $<$0.17 &  $<$1379 &  11.4 &      22.1 &     39.49 &       3.96 &  $-0.40\pm0.06$ &  1.36$\pm$0.11 &        \nodata\\ 
J0950+5128 &  $<$0.16 &   $<$575 &  13.2 &      16.0 &     34.53 &       5.78 &  $-0.34\pm0.07$ &  1.43$\pm$0.10 &        \nodata\\ 
J1037-0736 &  $<$0.17 &  \nodata &   6.2 &   \nodata &     74.88 &       5.74 &  $-0.54\pm0.07$ &  1.35$\pm$0.11 & $-0.44\pm0.17$\\ 
J1204+1918 &  $<$0.16 &  $<$1341 &   4.6 &      15.5 &    132.82 &       3.99 &  $-0.75\pm0.06$ &  1.27$\pm$0.10 & $-1.16\pm0.17$\\ 
J1208+4741 &  $<$0.12 &  $<$1006 &   5.6 &      11.8 &     89.17 &       6.44 &  $-0.58\pm0.08$ &  1.32$\pm$0.11 & $-0.61\pm0.16$\\ 
J1246+1805 &  $<$0.13 &  \nodata &   9.0 &   \nodata &     46.30 &       2.52 &  $-0.40\pm0.05$ &  1.28$\pm$0.10 & $-0.09\pm0.17$\\ 
J1254-0606 &  $<$0.15 &  \nodata &   6.6 &   \nodata &    236.24 &       2.25 &  $-1.09\pm0.04$ &  2.57$\pm$0.10 & $-1.16\pm0.17$\\ 
J1413+0257 &  $<$0.21 &  \nodata &   6.4 &   \nodata &    215.22 &       1.96 &  $-1.02\pm0.04$ &  2.20$\pm$0.10 & $-1.22\pm0.17$\\ 
J1447+0512 &  $<$0.18 &  $<$1562 &   2.6 &       7.3 &     43.75 &      29.84 &  $-0.34\pm0.17$ &        \nodata & $-0.45\pm0.17$\\ 
J1546+1500 &  $<$0.16 &  \nodata &   4.8 &   \nodata &     43.58 &      11.91 &  $-0.38\pm0.10$ &  0.84$\pm$0.10 & $-0.29\pm0.17$\\ 
J1603+1809 &  $<$0.17 &  $<$1308 &   8.1 &      34.6 &    109.40 &       0.23 &  $-0.66\pm0.01$ &  1.76$\pm$0.10 & $-0.47\pm0.17$\\ 
J2109-0644 &  $<$0.17 &  $<$1423 &   6.5 &      13.5 &     13.74 &       9.94 &  $-0.18\pm0.09$ &  0.15$\pm$0.11 & $-0.41\pm0.16$\\ 
\enddata
\tablecomments{Column 1: Source name.  Column 2: Angular size upper limit from the highest resolution VLA data available.  Column 3: Linear size upper limit for sources with known redshifts.  Column 4: Spectral peak (or ``turnover") frequency ($\nu_{\rm peak}$) from a fit to a curved power-law model.  Column 5: Same as Column 4, but in the rest frame for sources with known redshifts. 
Column 6: Reduced chi squared value for the fit to a simple power-law model.   
Column 7: Reduced chi squared value for the fit to a curved power-law model. 
Column 8: Spectral curvature parameter, $q$, from a fit to a curved power-law model. 
Column 9: The optically-thick spectral index, $\alpha_{\rm thick}$, estimated by a power-law fit to the two lowest frequency VLA bands ($L$ and $S$ band).    
The uncertainties were calculated using standard propagation of errors.  We required $\nu_{\rm peak}>$4~GHz to estimate $\alpha_{\rm thick}$. 
Column 10: Estimates of the optically-thin spectral index, $\alpha_{\rm thin}$ (and the corresponding uncertainties), from a power-law fit to the two highest frequency VLA bands (either $X$ and \emph{Ku} band or \emph{Ku} and $K$ band) above $\nu_{\rm peak}$.  We required quasi-simultaneous VLA measurements in at least two bands above $\nu_{\rm peak}$ to estimate $\alpha_{\rm thin}$. %{\bf \color{red}[To do: add peak CPL model flux]}
}
\end{deluxetable*}
%%%%%%%%%%%%%%%%%%%%%%%%%%%%%%%%%%%%%%%%%%%%%%%

%%%%%%%%%%%%%%%%%%%%%%%%%%%%%%%%%%%%%%%%%%%%%%%
\subsection{Radio flux densities and Morphologies}
Source flux and shape measurements for the 
%WIDAR 
VLA data above 1~GHz were made using the CASA IMFIT task and are reported in Table~\ref{tab:JVLA}.  For the majority of sources, the flux errors include a standard 3\% uncertainty in the absolute flux scale \citep{perley+17}, added in quadrature with the flux error reported by IMFIT.  Due to scheduling limitations, one of our sources (J2109$-$0644) was observed using 3C48 as the flux calibrator. Because 3C48 is currently flaring, we conservatively report a flux uncertainty of 20\% for each measurement of J2109$-$0644 following current NRAO guidelines\footnote{\url{https://science.nrao.edu/facilities/vla/docs/manuals/oss/performance/fdscale}}.

Our sources are characterized by compact ($\lesssim$0.1$^{\prime \prime}$) radio continuum emission over the full frequency range of our VLA study, which provides a maximum angular resolution of $\theta_{\rm max} \lesssim 0.1^{\prime \prime}$ (for the  \emph{Ku}-band/A-configuration data).  This corresponds to an upper limit on the intrinsic linear source sizes that is $<1$~kpc.   

%%%%%%%%%%%%%%%%%%%%%%%%%%%%%%%%%%%%%%%%%%%%%%%
\subsection{Constraints on Variability}
\label{sec:variability_timescale}
We illustrate the flux variations of our sample at $L$ and $S$ band over two epochs probing two different timescales in   Figures~\ref{fig:timeseries} and \ref{fig:variability}.  We also show the variability at $X$-band for the 3 sources (J0807+2102, J0832+2302, and J0950+5128) in our sample with a gap of several weeks between observations from 1--12~GHz ($L$, $S$, $C$, and $X$ band) and 12--18~GHz (\emph{Ku} band).   Tables~\ref{tab:FIRST_VLASS} and \ref{tab:summary} provide additional details on the observing dates of the data from FIRST, VLASS, and our new multiband JVLA follow-up observations.   

At $L$-band, the two epochs shown in the top panel of  Figure~\ref{fig:timeseries} are from the FIRST survey (with observing dates ranging from 1995 to 2002 for our sample) and our 2019 JVLA follow-up study.  The $L$-band source flux densities from our new JVLA observations range from $\sim1-11$~mJy.  Thus, compared to the FIRST flux measurements made 17 to 24 years earlier, our sources have $L$-band flux densities that have increased dramatically by factors of $\sim$2--25.  
The left panel of Figure~\ref{fig:variability} shows the variability amplitude at $L$ band between FIRST and our 2019 observations for each source.  The weighted median value is $\sim$200\%.  However, we emphasize that since half of our sources remain undetected below the 3$\sigma$ level in FIRST, this is likely to be a lower limit.  Thus, the true $L$-band variability amplitude on timescales of $\sim$1--2 decades could be even higher.  

At $S$-band, the two observing epochs shown in Figure~\ref{fig:timeseries} span a much narrower range of time ($\sim$months) compared to the variability timescale constraints at $L$ band ($\sim$decades).  The $S$-band flux densities between our JVLA follow-up observations and VLASS are approximately constant 
over timescales of 3--7 months. This is clearly illustrated in the right panel of Figure~\ref{fig:variability}, which shows 
the $S$-band variability amplitude has a median value of $\sim$15\%.   
Thus, our follow-up JVLA $S$-band data are typically in good agreement within the current $\sim$20\% flux uncertainties of VLASS.  
We note that the most substantial outlier in the right panel of Figure~\ref{fig:variability} is the source J2109-0644, which as discussed  in Section~\ref{sec:data} may have unusually large absolute flux errors due to the variability of the flux calibrator. 

At $X$-band, the two observing epochs shown in Figure~\ref{fig:timeseries} for J0807+2102, J0832+2302, and J0950+5128 span an even narrower range of time ($\sim$weeks) compared to the variability timescale constraints at $L$ and $S$ band.  The $X$-band flux densities are roughly constant, although all 3 sources exhibit a slight increase in flux of $\sim$10\% over a period of 58 days.  Although an increase in flux density of this magnitude could be due to the evolution of a nascent jet (see Section~\ref{sec:origin}), the presence of residual errors in the absolute flux scale of similar magnitude cannot be entirely ruled out.  This is because the first epoch of $X$-band observations was taken during the move from the hybrid BnA configuration to the standard A array, which lead to poor PSF quality that ultimately posed challenges for self calibration and deconvolution (see Section~\ref{sec:data}).  Nevertheless, we conclude that the $X$-band flux densities are in close enough agreement to justify the inclusion of the observations above 12~GHz in our radio SED analysis.% for J0807+2102, J0832+2302, and J0950+5128.  

Although our time domain assessment is currently crude and would benefit from higher-cadence monitoring in the future, we conclude that our sources are characterized by large amplitude (2--25X) variability at $L$ band on timescales 1--2 decades, but maintain roughly constant flux densities over timescales of a few months at $S$ band.  Thus, the typical variability timescale, $\tau_{\rm var}$, likely falls in the range of $3-7$ months $< \tau_{\rm var} < 17-24$ years.  We discuss implications of these constraints on the origin of the radio variability in Section~\ref{sec:origin}.  

%%%%%%%%%%%%%%%%%%%%%%%%%%%%%%%%%%%%%%%%%%%%%%%
\begin{figure*}
\centering

\includegraphics[clip=true, trim=0cm 0.25cm 0cm 0cm, width=0.25\textwidth]{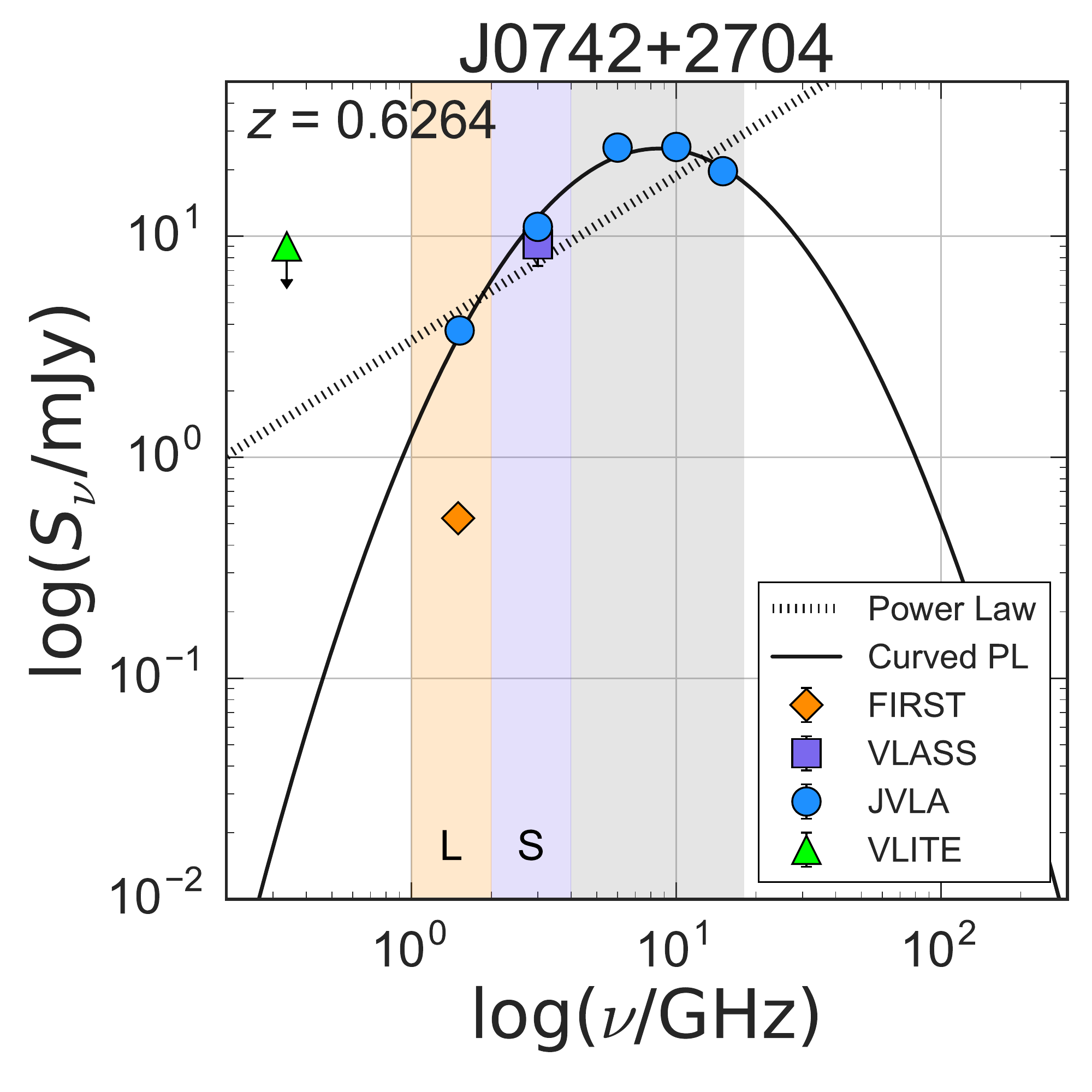}
\includegraphics[clip=true, trim=0cm 0.25cm 0cm 0cm, width=0.25\textwidth]{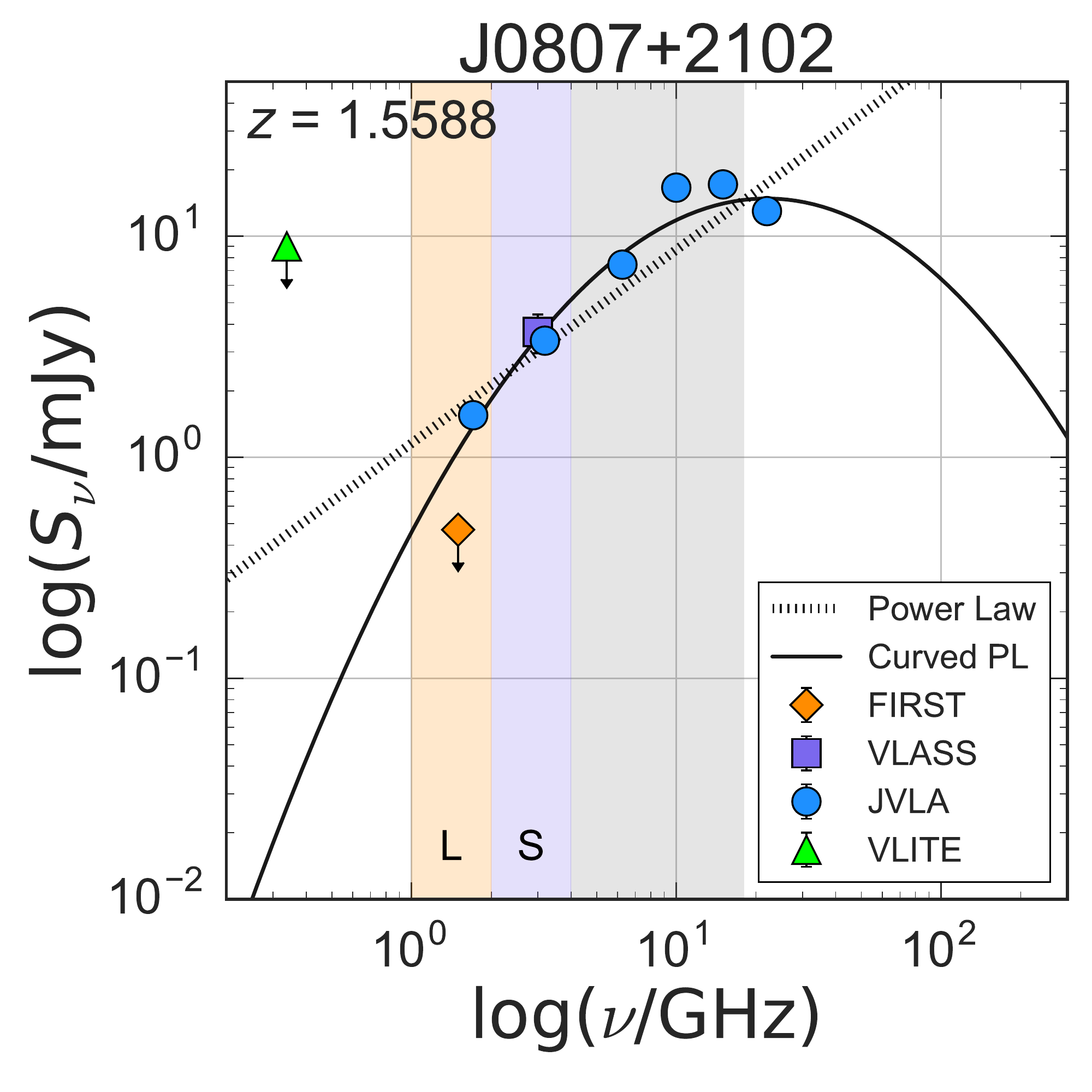}
\includegraphics[clip=true, trim=0cm 0.25cm 0cm 0cm, width=0.25\textwidth]{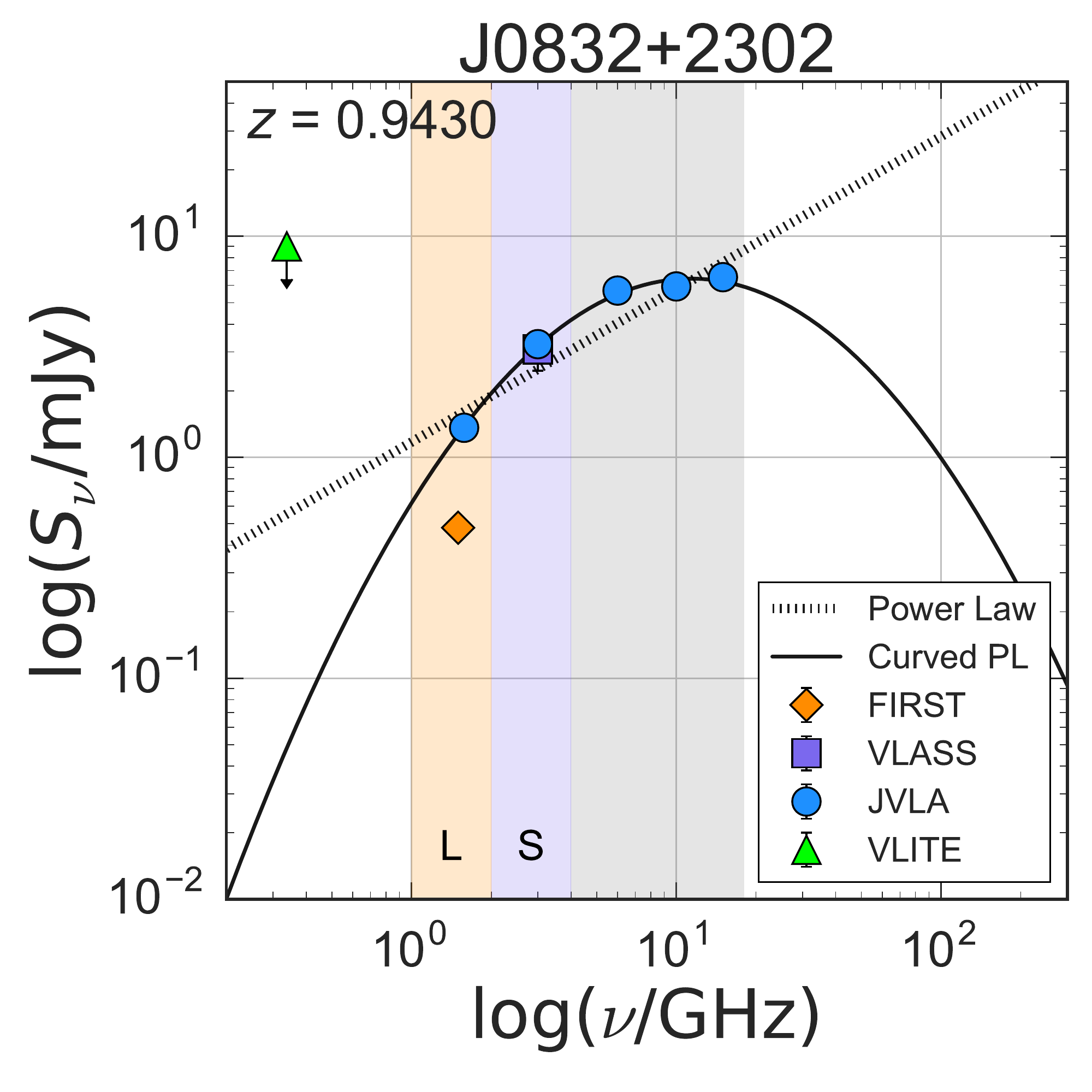}

\includegraphics[clip=true, trim=0cm 0.25cm 0cm 0cm, width=0.25\textwidth]{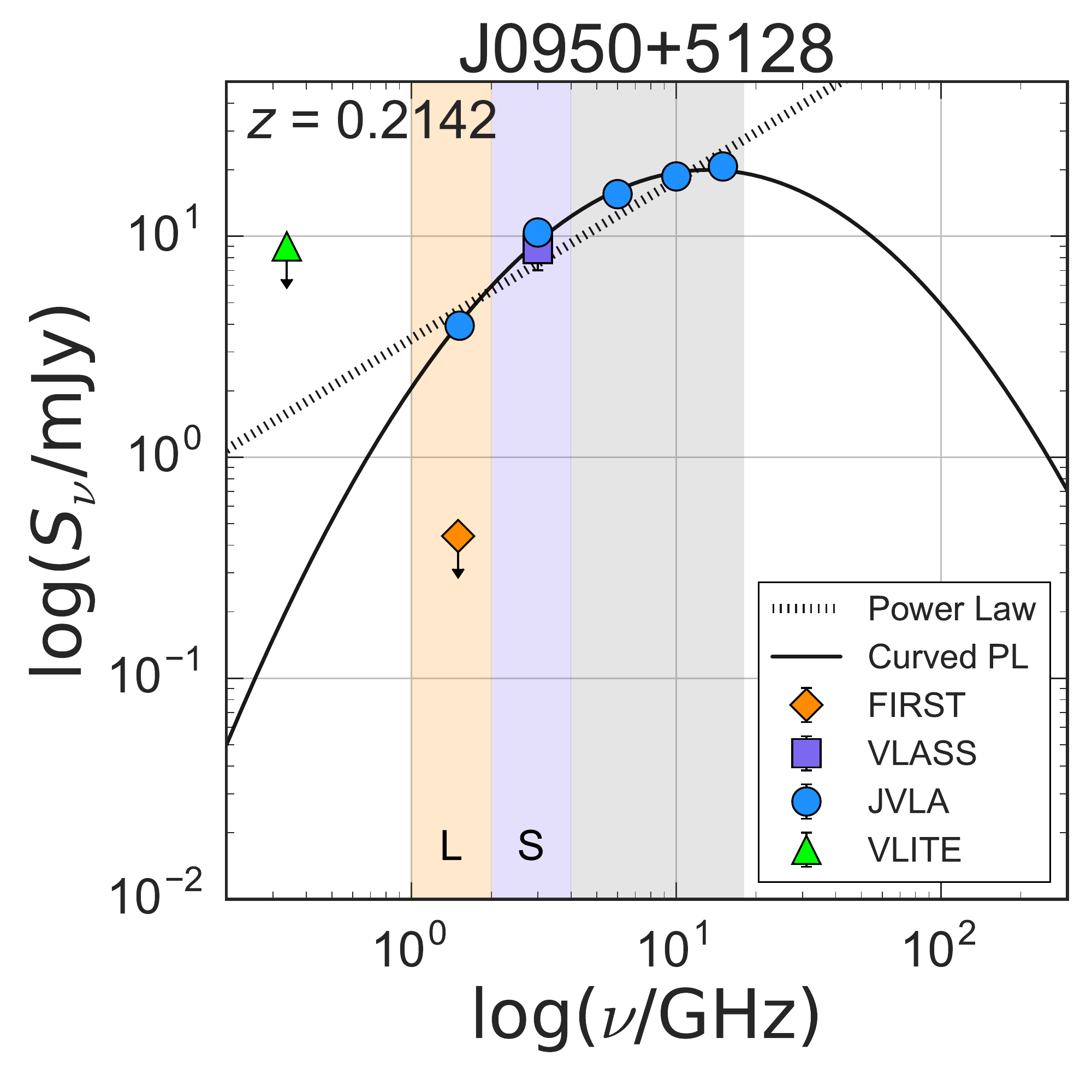}
\includegraphics[clip=true, trim=0cm 0.25cm 0cm 0cm, width=0.25\textwidth]{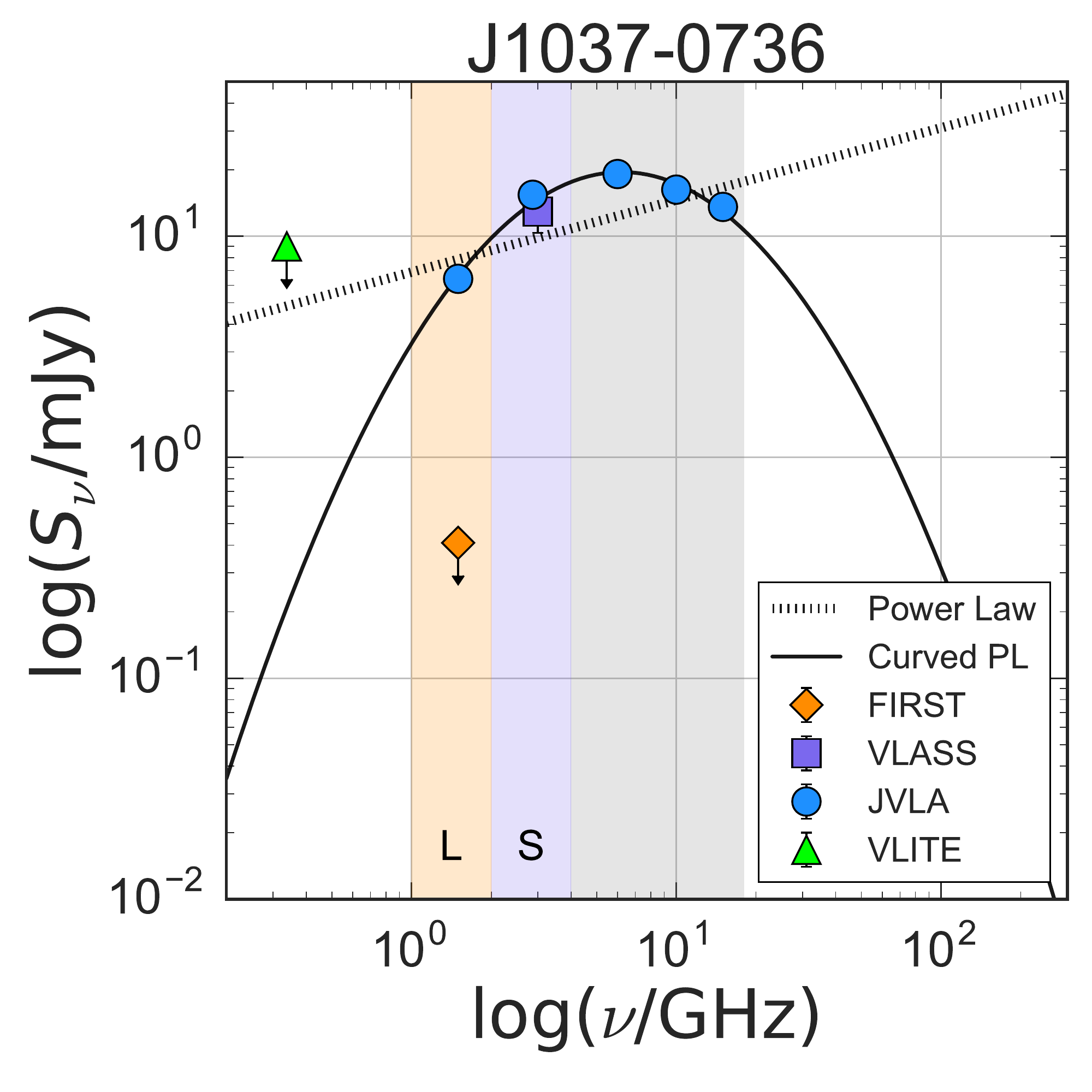}
\includegraphics[clip=true, trim=0cm 0.25cm 0cm 0cm, width=0.25\textwidth]{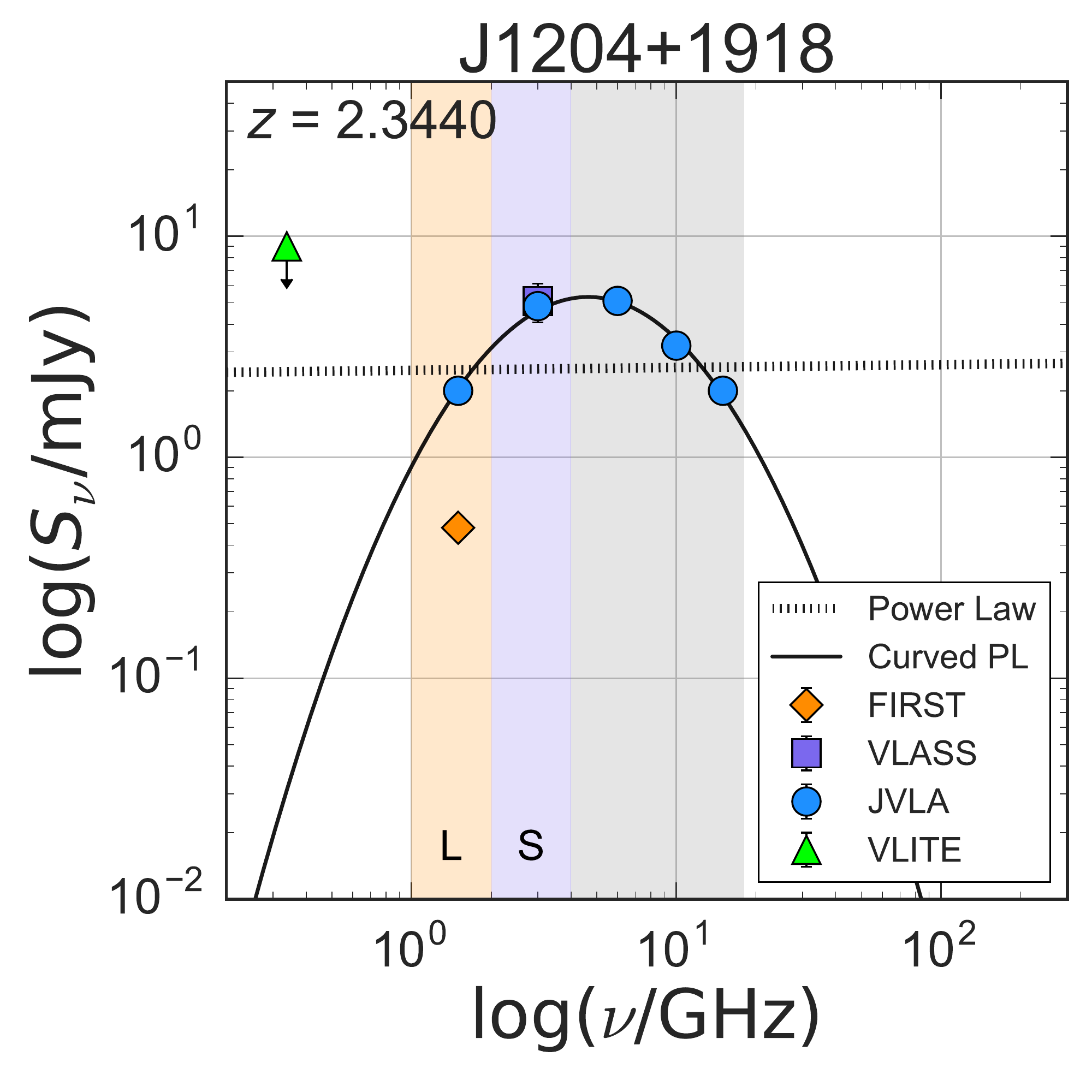}

\includegraphics[clip=true, trim=0cm 0.25cm 0cm 0cm, width=0.25\textwidth]{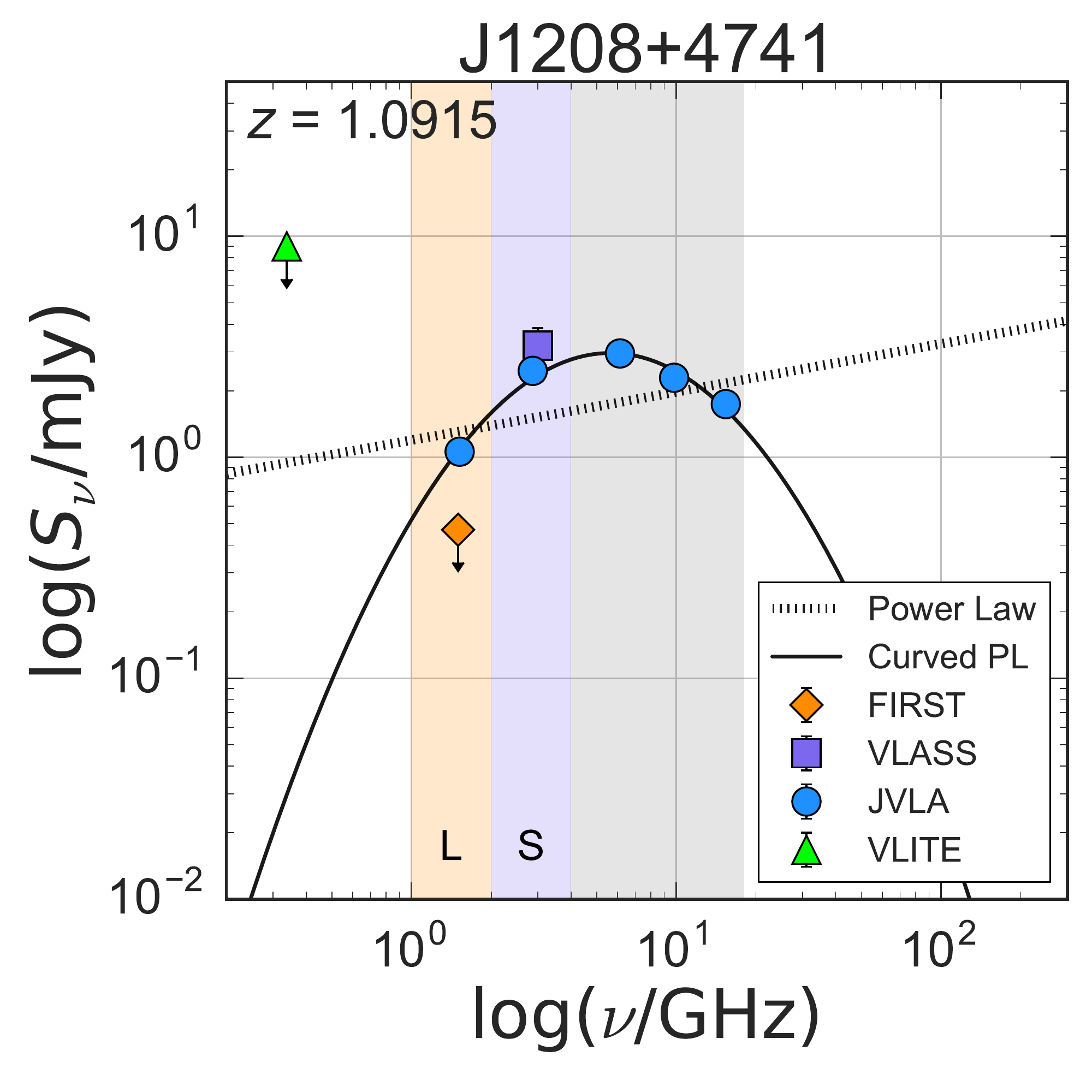}
\includegraphics[clip=true, trim=0cm 0.25cm 0cm 0cm, width=0.25\textwidth]{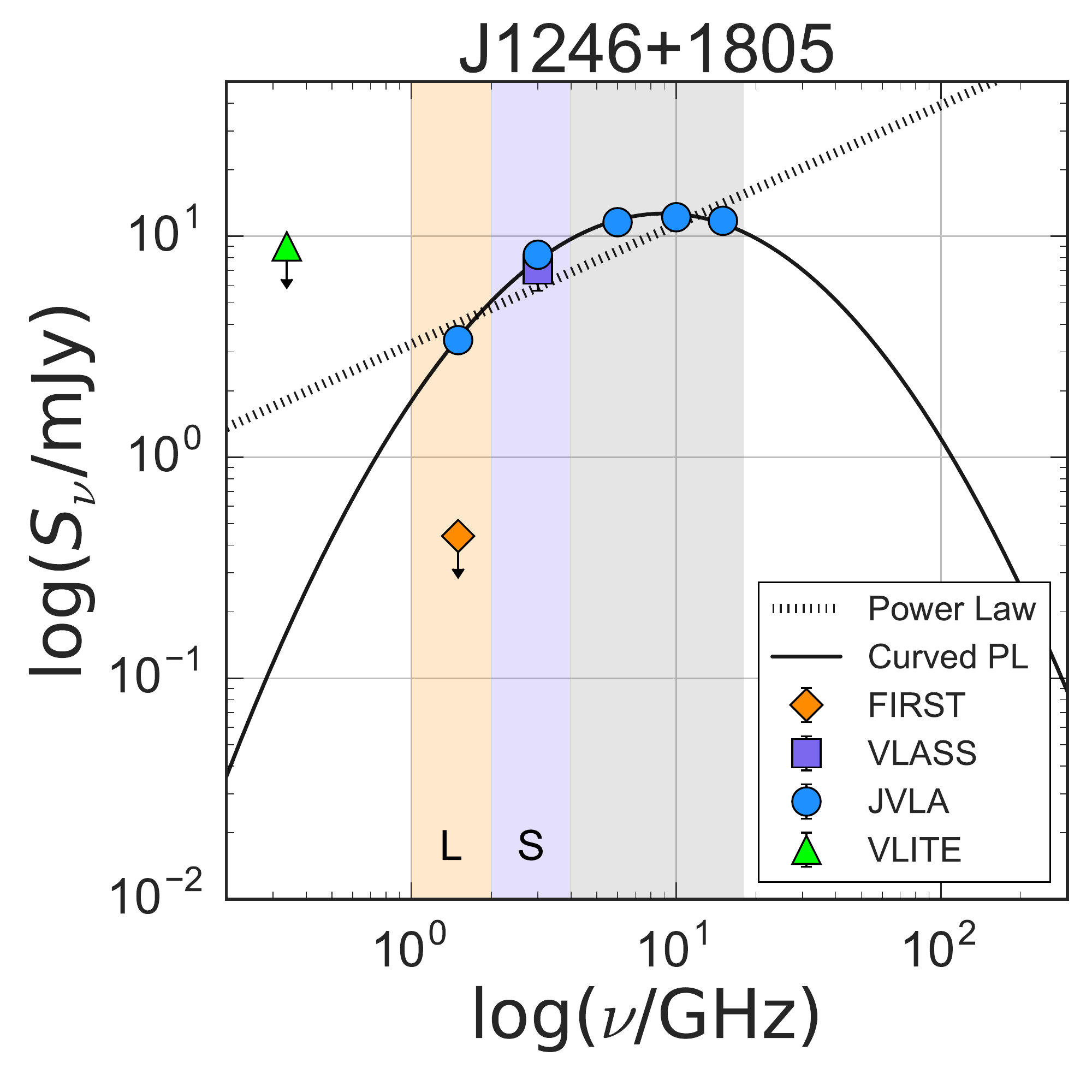}
\includegraphics[clip=true, trim=0cm 0.25cm 0cm 0cm, width=0.25\textwidth]{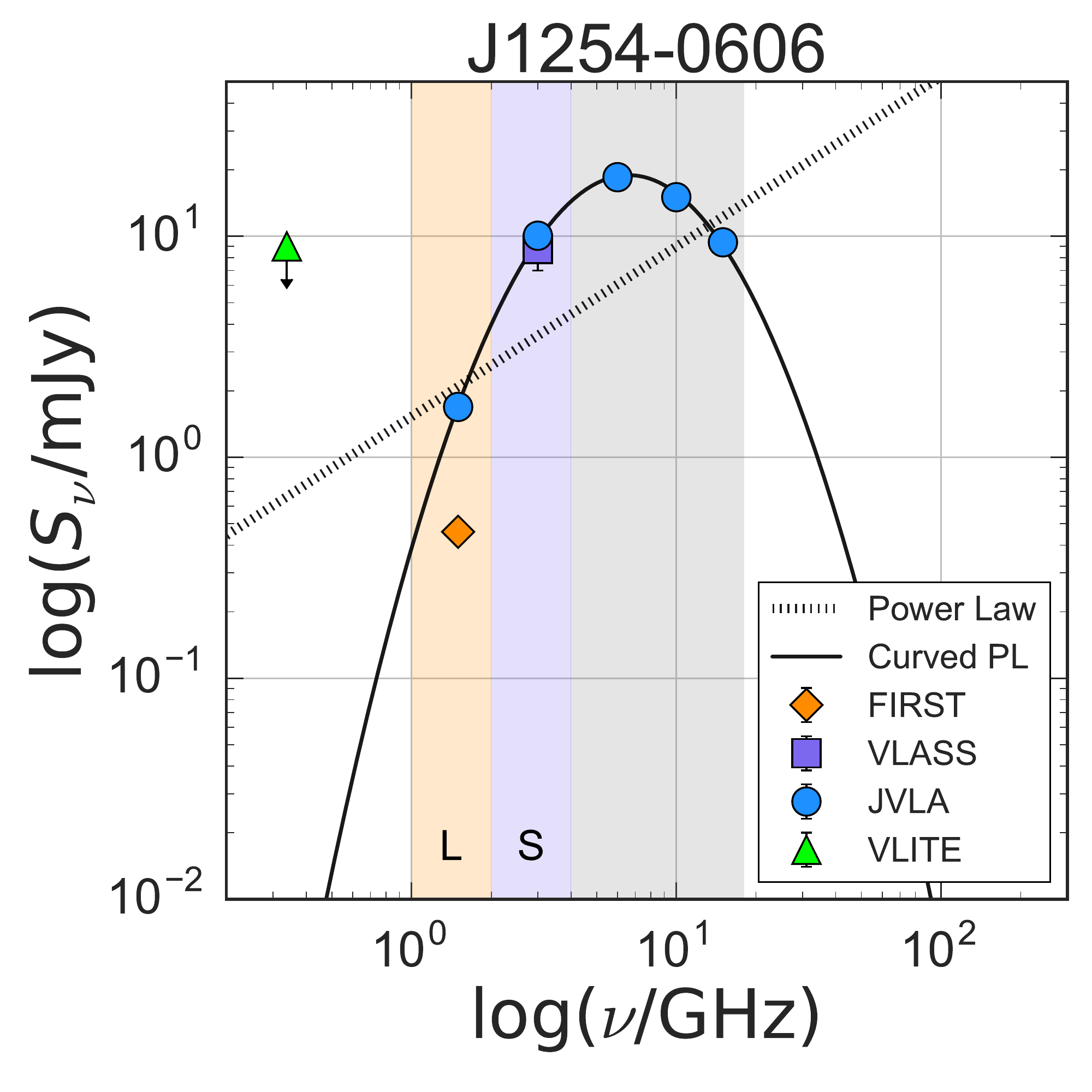}

\includegraphics[clip=true, trim=0cm 0.25cm 0cm 0cm, width=0.25\textwidth]{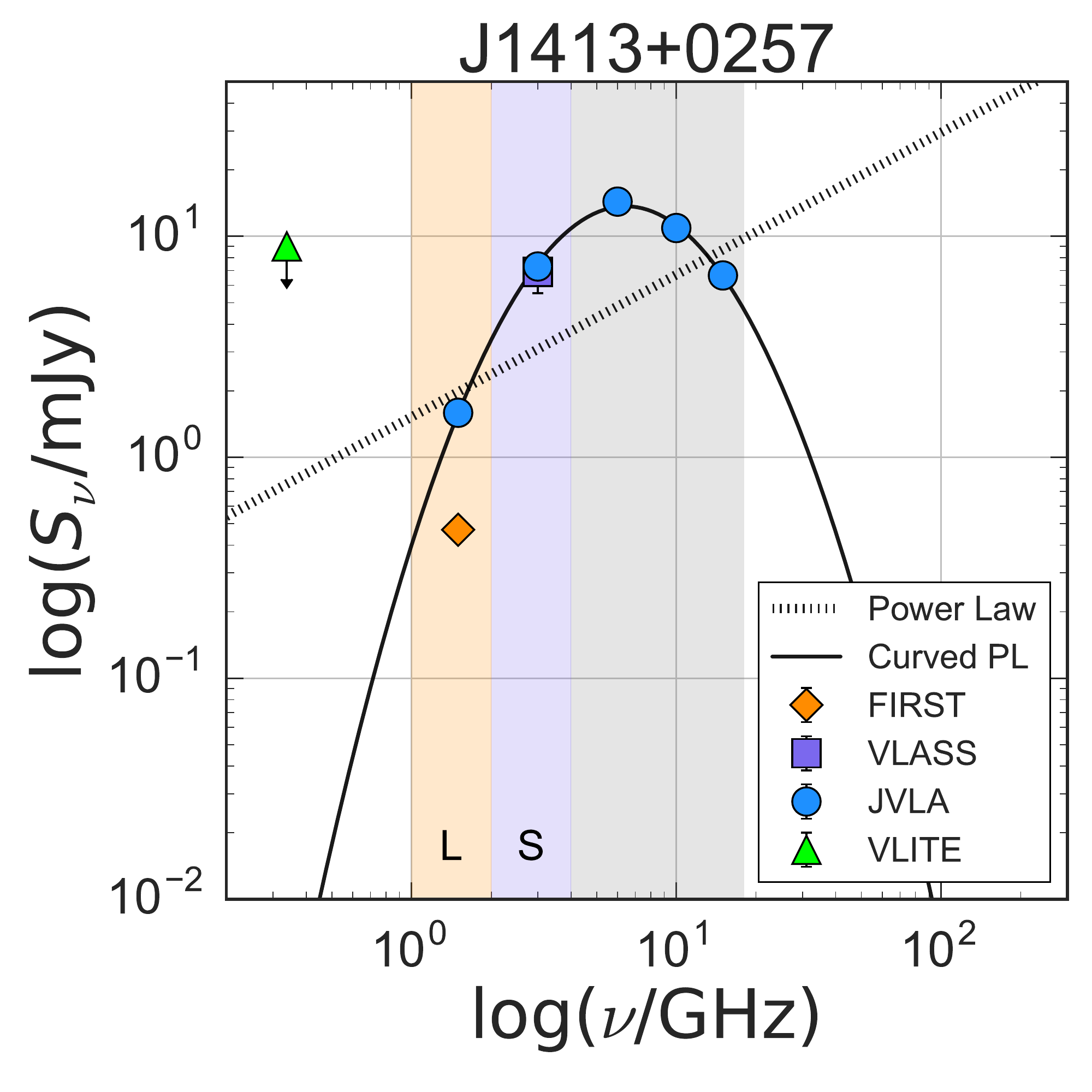}
\includegraphics[clip=true, trim=0cm 0.25cm 0cm 0cm, width=0.25\textwidth]{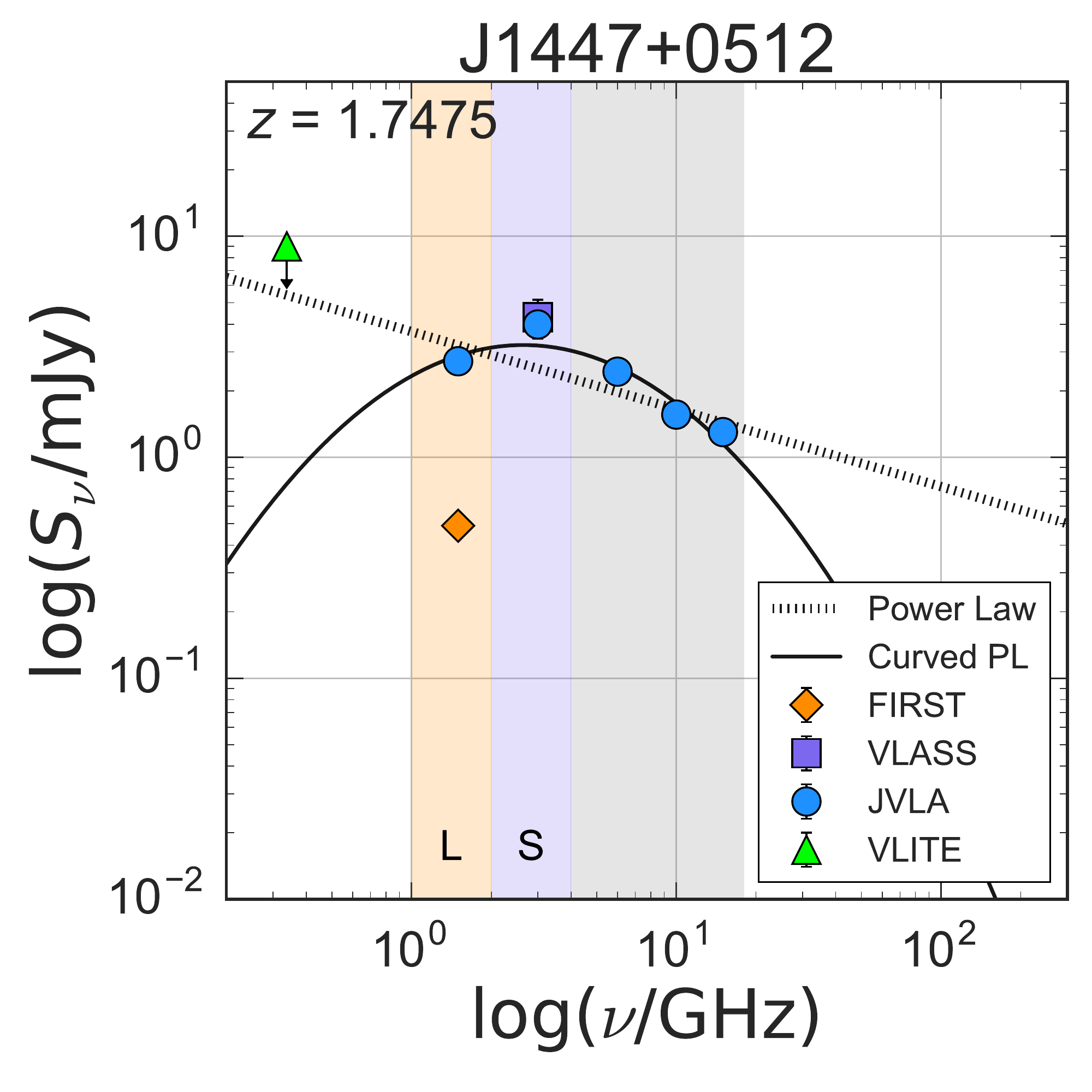}
\includegraphics[clip=true, trim=0cm 0.25cm 0cm 0cm, width=0.25\textwidth]{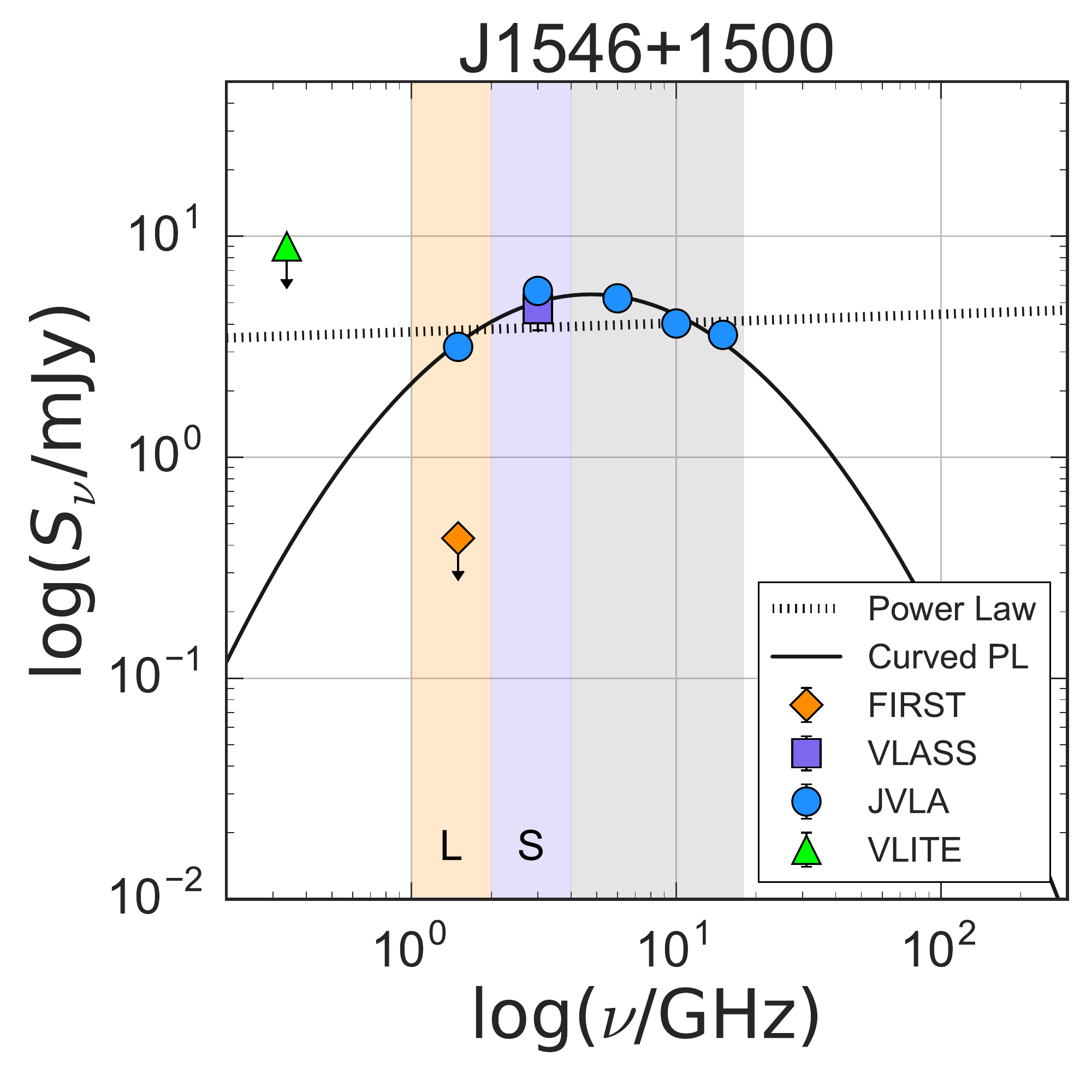}

\includegraphics[clip=true, trim=0cm 0cm 0cm 0cm, width=0.25\textwidth]{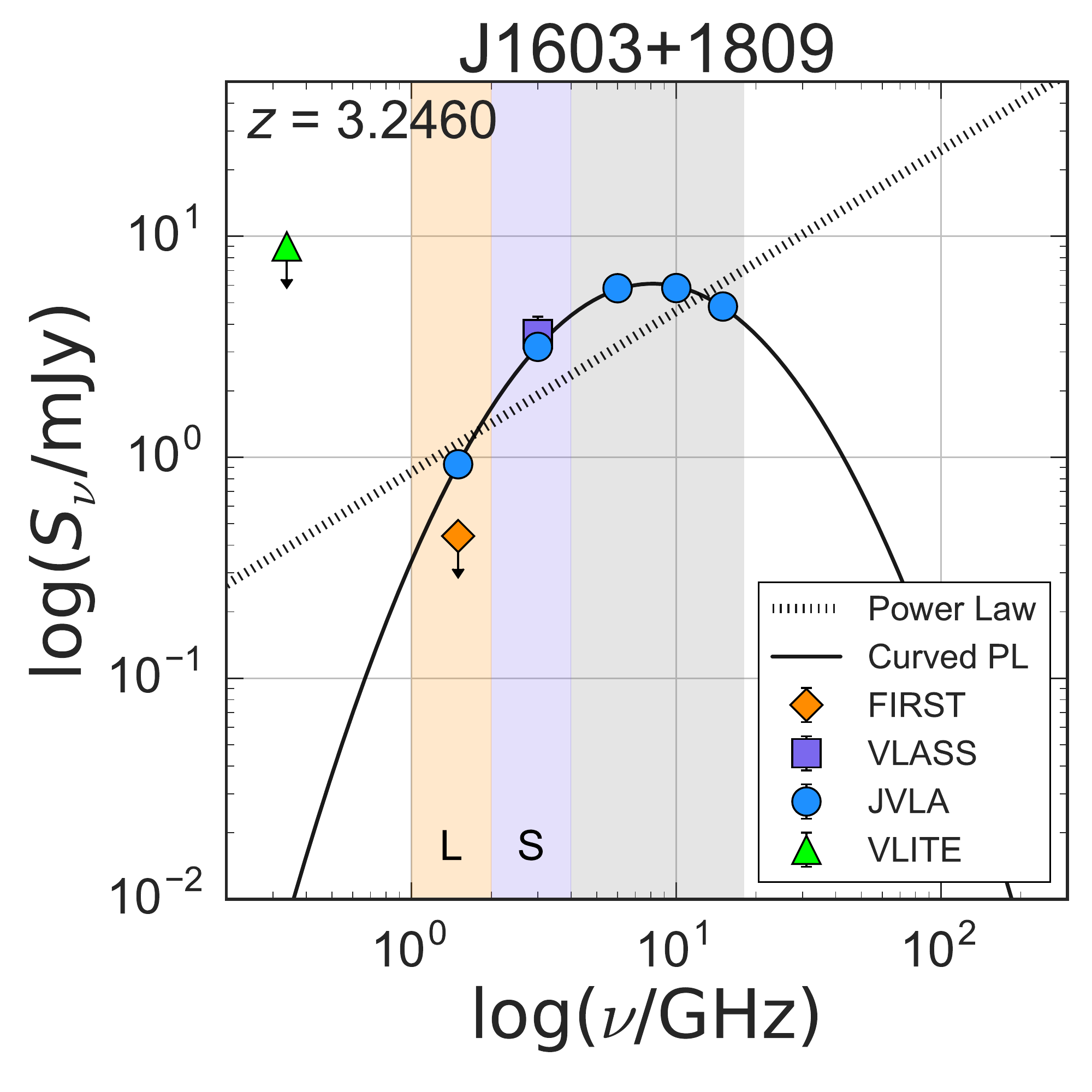}
\includegraphics[clip=true, trim=0cm 0cm 0cm 0cm, width=0.25\textwidth]{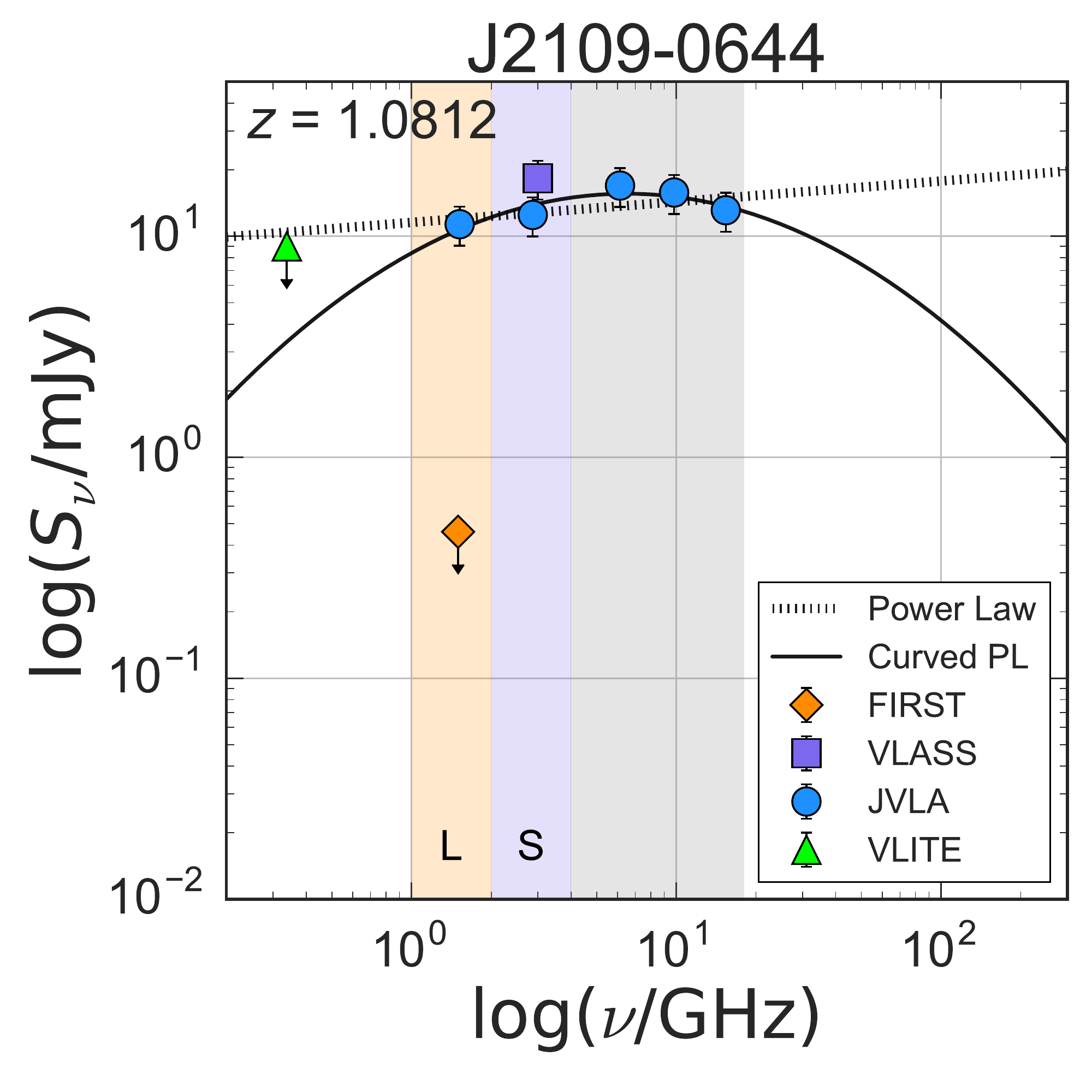}

\caption{%Broadband radio SEDs showing our quasi-simultaneous multi-band JVLA imaging (purple circles) as well as the VLASS detection (orange square), FIRST upper limit or 3$\sigma$ detection, and VLITE upper limit.  Two models based on the quasi-simultaneous JVLA data (purple circles) are shown: 1) a standard non-thermal power-model and 2) a curved power-law model.
Broadband radio SEDs showing our quasi-simultaneous multiband JVLA imaging (blue circles), the VLASS detection (purple squares), FIRST upper limit or 3$\sigma$ detection (orange diamonds), and VLITE upper limit (green triangles).  For each source, two models based on the quasi-simultaneous JVLA data are shown: a standard non-thermal power-model (dotted line) and a curved power-law model (solid line).
\label{fig:SEDs}}
\end{figure*}

%%%%%%%%%%%%%%%%%%%%%%%%%%%%%%%%%%%%%%%%%%%%%%%
\subsection{Quasi-simultaneous Radio SEDs}
\label{sec:SEDs}
 In Figure~\ref{fig:SEDs}, we show the radio SEDs of the 14 sources from our sample with quasi-simultaneous VLA observations from 1--18~GHz. 
 %as well as the upper limits at 340~MHz from the commensal VLITE data.  
 %The presence of varying degrees of spectral curvature in the radio SED shapes is easily discernible by eye.  
 The radio SEDs have peaked spectral shapes with varying degrees of curvature easily discernible by eye.  This indicates the presence of an underlying absorption mechanism, such as synchrotron self absorption (SSA) or free-free absorption (FFA), as is commonly associated with compact radio sources  (\citealt{callingham+17,collier+18}). 
 
%%%%%%%%%%%%%%%%%%%%%%%%%%%%%%%%%%%%%%%%%%%%%%%
\subsubsection{Radio SED Modeling} 
%\label{sec:SEDs}
To quantify the location of the spectral peak and degree of curvature, we performed least squares fits to the quasi-simultaneous VLA data above 1~GHz using two basic synchrotron emission models: 
 
 \begin{enumerate}
 \item A standard non-thermal power-law model defined by $S_{\nu} = a \nu^{\alpha}$, where $S_{\nu}$ is the flux at frequency $\nu$, $a$ represents the amplitude, and $\alpha$ is the spectral index.  
 
 \item A curved power-law model defined by $S_{\nu} = a \nu^{\alpha}e^{q(\ln \nu)^2}$, where $q$ represents the degree of spectral curvature (the breadth of the peak of the radio SED).  The $q$ parameter is defined by $\nu_{\rm peak} = e^{-\alpha/2q}$, where $\nu_{\rm peak}$ is the turnover frequency.  %We note that while this model provides a convenient parametrization for the spectral curvature, it is agnostic  
 Significant spectral curvature is typically defined as $|q| \geq 0.2$  \citep{duffy+12, callingham+17}.
  \end{enumerate}

We note that our use of the curved power-law model is phenomenological in nature.  More physically motivated spectral curvature models, such as SSA or FFA (or models with contributions from multiple electron populations; \citealt{tingay+15}),  
%require more than 3 free parameters, and are thus not well constrained by the number of bands (typically 5; see Table~\ref{tab:summary}) available for each source presented in this paper. 
require more than 3 free parameters, and thus have only a few degrees of freedom given the  number of bands (typically 5; see Table~\ref{tab:summary}) available for each source. 
% In a future study, we will increase the number of spectral data points by dividing each VLA band into multiple frequency segments, independently imaging them, and measuring the flux densities.    

We summarize our spectral modeling analysis in Table~\ref{tab:sizes}.  Based on the reduced chi-square values provided in this table, a curved power-law model provides a better fit compared to a simple power-law model for all sources. 
%\footnote{We emphasize again that the curved power-law model used in our study is not physically motivated, thus some sources have relatively large curved power-law reduced chi-square values.}.  
%Our sources have median spectral turnovers estimated from the curved power-law fits of $\nu_{\rm peak} = 6.6$~GHz (2.5--22.7~GHz).  
The observed spectral turnover frequencies for our sources range from 2.5--22.7~GHz, with a median value of $\nu_{\rm peak} = 6.6$~GHz.  
For sources with measured redshifts, the corresponding rest-frame turnover frequencies lie in the range of $\nu_{\rm peak,\,rest} =  6.8-58.1$~GHz.  
Likewise, the values of $q$ from the curved power-law fits span the range $0.18 < |q| < 1.09$ confirming that essentially all are strongly curved. The exception is J2109-0644, which is only marginally below the formal limit of $|q| > 0.2$, and the VLITE upper limit at 340~MHz supports a curved, not flat, radio SED. 
%We also consider the curvature parameter, $q$, estimated from the curved power-law model fits.  
%This parameter has a range of $0.18<|q|<1.09$ for our sample.  Thus, the majority of our sources satisfy the standard literature criterion for substantial spectral curvature ($|q| \geq 0.2$).  The single exception is J2109$-$0644, for which $|q| = 0.18\pm0.09$.  Given the uncertainty in the $q$ value, a curved radio SED shape is still plausible for J2109$-$0644. In addition, the upper limit from VLITE at 340~MHz for this source argues against a flat radio SED. 

In Table~\ref{tab:sizes}, we also provide the optically-thick and optically-thin spectral index values ($\alpha_{\rm thick}$ and $\alpha_{\rm thin}$) below and above the turnover frequency, respectively.  We estimated $\alpha_{\rm thick}$ by fitting a power-law model to the quasi-simultaneous flux densities at the two lowest frequency VLA bands ($L$ and $S$ band) below the turnover frequency.  Estimates for $\alpha_{\rm thick}$ are only provided for sources with $\nu_{\rm peak}>4$~GHz (spectral peaks above the VLA $S$ band).  For $\alpha_{\rm thin}$, we performed a power-law fit to the quasi-simultaneous flux densities at the two highest frequency VLA bands (either $X$ and \emph{Ku} band or \emph{Ku} and $K$ band\footnote{$K$-band data were obtained for one source, J0807+2102, in order to constrain the radio SED shape on the optically-thin side of $\nu_{\rm peak}$ (which is above $X$-band for this source).}  above the turnover frequency.  We required the quasi-simultaneous VLA data to include at least two independent bands above the turnover frequency to estimate $\alpha_{\rm thin}$. The uncertainties in $\alpha_{\rm thick}$ and $\alpha_{\rm thin}$ provided in Table~\ref{tab:sizes} were calculated using standard propagation of errors.  

Based on the radio spectral modeling analysis described here, we conclude that the radio SEDs of our sources are best represented by curved power-law models (as opposed to flat power-law models lacking curvature), thus supporting the presence of significant spectral curvature.      
In a future study, we will 
%increase the number of spectral data points by dividing each VLA band into multiple frequency segments, independently imaging them, and measuring the flux densities.    
incorporate new data from forthcoming VLA and Giant Metre-wave Radio Telescope (GMRT) observations spanning a broader frequency range (the full contiguous frequency range of the VLA from 1--50~GHz and deep measurements below 1~GHz using the VLA and GMRT).  
%This will enable a more rigorous radio SED modeling analysis that will allow us  to distinguish between different physical models for the underlying absorption. 
This will enable a more rigorous radio SED modeling analysis that will allow us to test whether SSA of FFA are responsible for the observed spectral curvature (\citealt{mhaskey+19a, mhaskey+19b}).  

%%%%%%%%%%%%%%%%%%%%%%%%%%%%%%%%%%%%%%%%%%%%%%%
\subsubsection{Radio Color-color Diagram} 
\begin{figure}
\centering
\includegraphics[clip=true, trim=4cm 0.25cm 3.5cm 0.25cm, width=0.5
\textwidth]{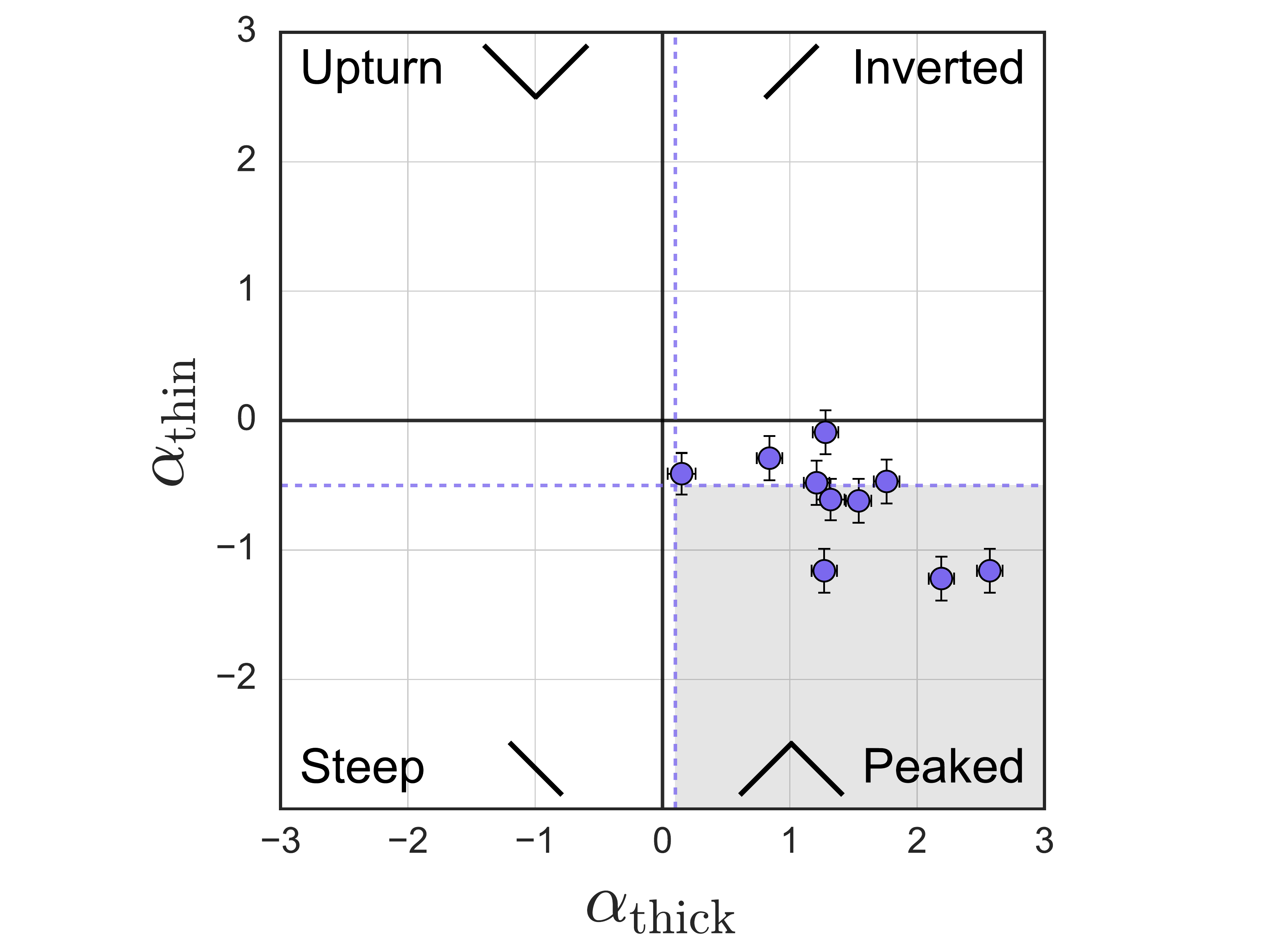}
\caption{Radio color-color diagram for sources with well-defined optically-thick and optically-thin spectral indices ($\alpha_{\rm thick}$ and $\alpha_{\rm thin}$, respectively).  Details on the calculation of $\alpha_{\rm thick}$ and $\alpha_{\rm thin}$ are provided in Section~\ref{sec:SEDs} and Table~\ref{tab:sizes}.  
%The dashed horizontal line at $\alpha_{\rm thin}=-0.5$ denotes the limit below which sources are considered to have significant spectral curvature.  Sources to the right of the dashed vertical line at  $\alpha_{\rm thick}=0.1$ are more likely to have substantial curvature.
The gray shaded region highlights the selection criteria for peaked-spectrum radio sources from \citet{callingham+17} of $\alpha_{\rm thick}>0.1$ and $\alpha_{\rm thin}<-0.5$.
%sources to the right of the dashed vertical line at  $\alpha_{\rm thick}=0.1$ are more likely to have substantial curvature. 
%The gray-shaded region highlights optically-thick spectral indices above the canonical limit of $\alpha = 2.5$ for synchrotron self absorption. 
\\}
\label{fig:radio_colors}
\end{figure}

In Figure~\ref{fig:radio_colors}, we show our sources on a radio color-color diagram ($\alpha_{\rm thick}$ vs. $\alpha_{\rm thin}$).  All of our sources reside in the second quadrant of this figure, which characterizes peaked radio spectral shapes indicative of an underlying absorption mechanism commonly associated with source compactness due to youth and/or confinement by an external medium (\citealt{odea+98,odea+20,orienti+16}).  Most of our sources also meet the more strict selection criteria for peaked-spectrum radio sources from \citet{callingham+17} of $\alpha_{\rm thick}>0.1$ and $\alpha_{\rm thin}<-0.5$).  

%%%%%%%%%%%%%%%%%%%%%%%%%%%%%%%%%%%%%%%%%%%%%%%
%%%%%%%%%%%%%%%%%%%%%%%%%%%%%%%%%%%%%%%%%%%%%%

%%%%%%%%%%%%%%%%%%%%%%%%%%%%%%%%%%%%%%%%%%%%%%%
\begin{figure*}
\centering
\includegraphics[clip=true, trim=0cm 0cm -0.75cm 0cm, width=0.475\textwidth]{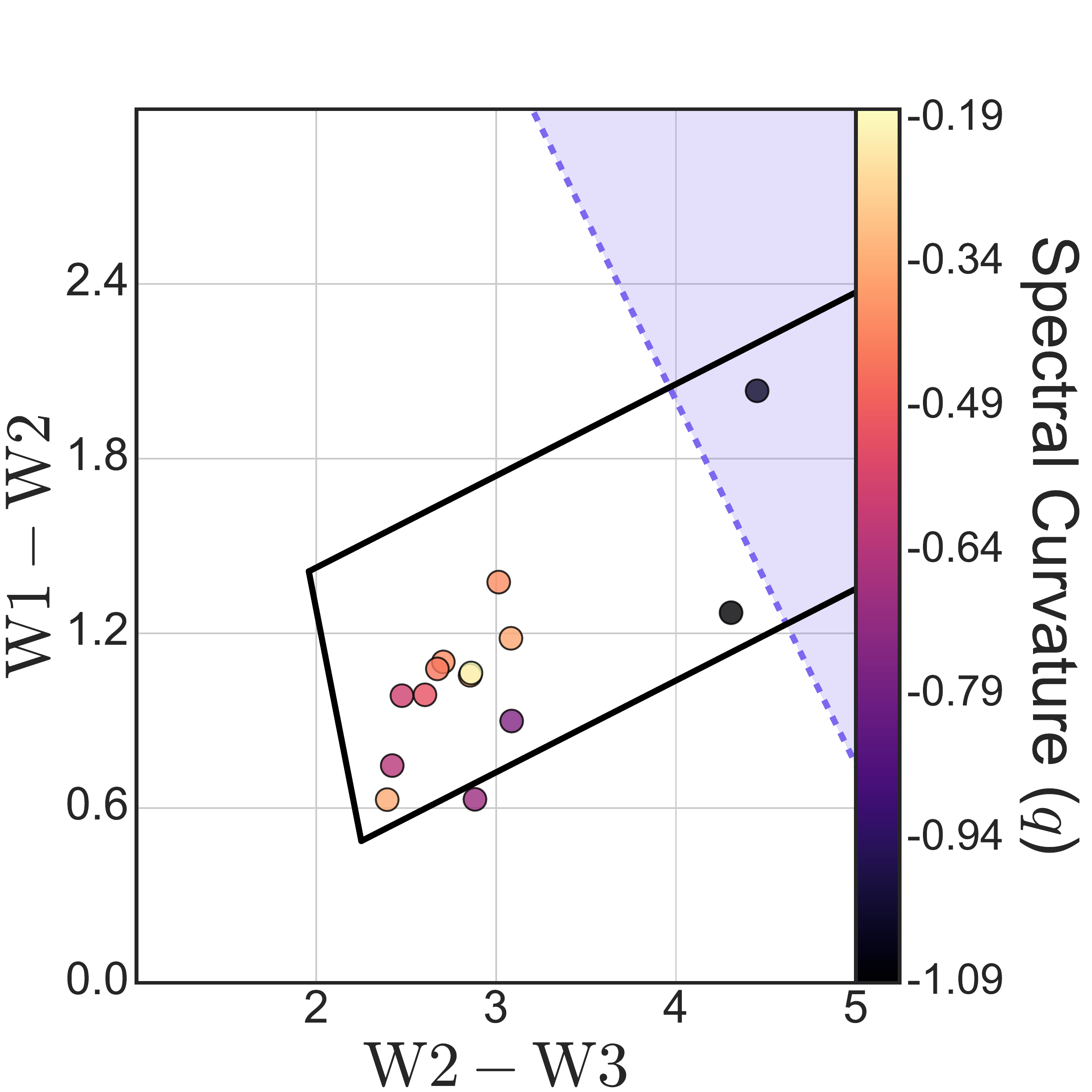}
\includegraphics[clip=true, trim=-0.75cm 0cm 0cm 0cm, width=0.475\textwidth]{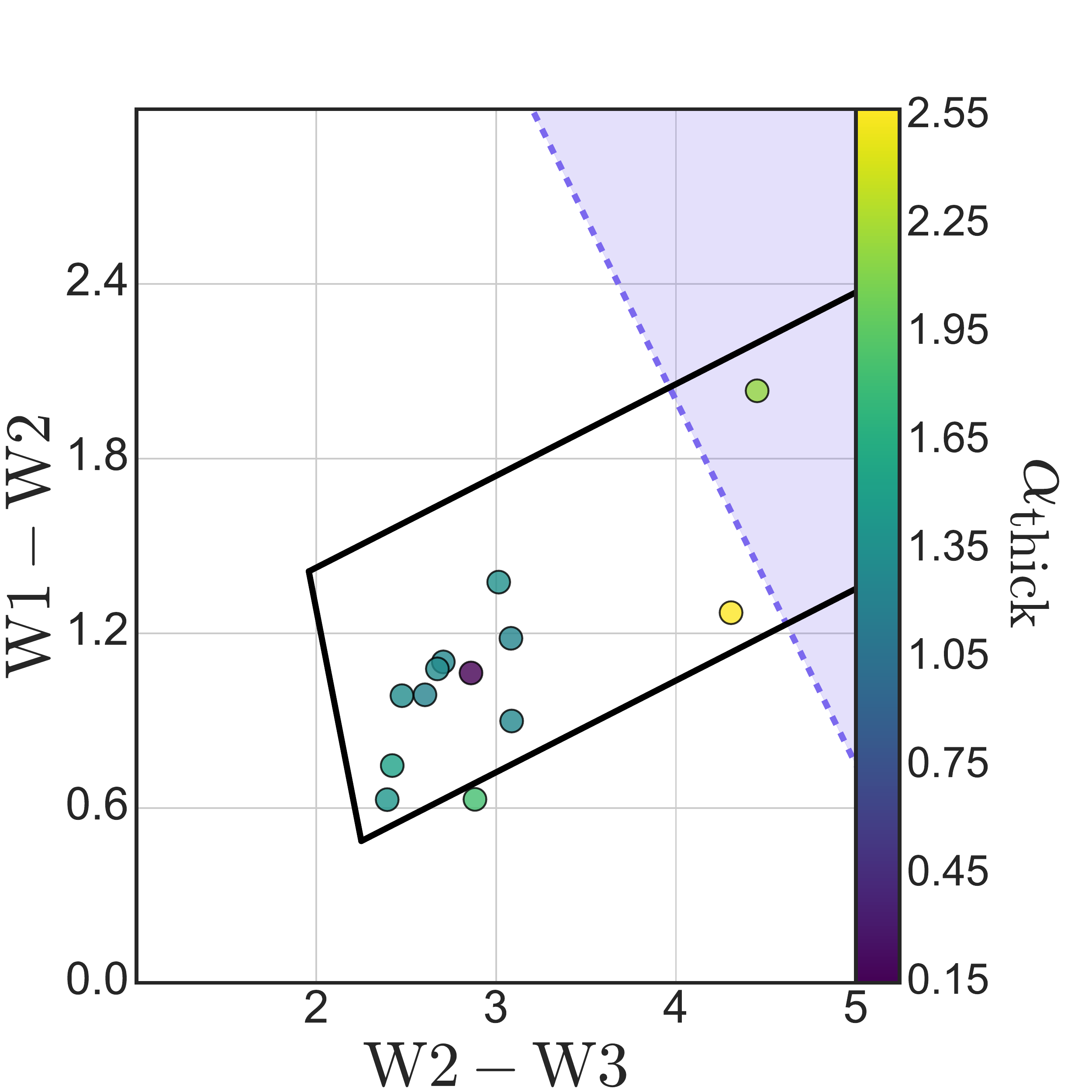}
\caption{{\it WISE} infrared color-color diagrams (W1 = 3.4$\mu$m, W2 = 4.6$\mu$m, and W3 = 12$\mu$m) in Vega magnitudes. 
%The dashed horizontal line represents the AGN selection criteria of $W1-W2 > -0.8$ \citep{stern+12}, which provides a completeness of $\sim$75\% \citep{assef+18}.  
The black wedge defines the color space for luminous AGN in {\it WISE} defined by \citet{mateos+12}.  
The shaded region bounded by the dashed diagonal line, defined as $(W1 - W2) + 1.25 (W2 - W3) > 7$, identifies extremely red sources likely to be heavily obscured, luminous AGN \citep{lonsdale+15,patil+20}.  The points in the left- and right-hand plots are colorized by the radio spectral curvature parameter ($q$) and the optically-thick radio spectral index ($\alpha_{\rm thick}$), respectively. 
\\}
\label{fig:WISE_colors}
\end{figure*}
%%%%%%%%%%%%%%%%%%%%%%%%%%%%%%%%%%%%%%%%%%%%%%

% Most of our sources have optically-thick spectral index constraints that are consistent with SSA.  
% %However, for  J1254-0606, $\alpha_{\rm thick}=2.57 \pm 0.10$, which exceeds the canonical limit for synchrotron self absorption ($\alpha_{\rm thick}=2.5$). 
% In the most extreme case in sample (J1254-0606),  $\alpha_{\rm thick}=2.57 \pm 0.10$.  This source could therefore be an example of FFA (\citealt{callingham+15,collier+18,mhaskey+19a,mhaskey+19b, keim+19}, although we point out that idealized single-population SSA cannot be ruled out given the current uncertainty in $\alpha_{\rm thick}$.
% % extreme value it might also be an example of FFA
% %This suggests that FFA may contribute to the curved spectral shape of this source (\citealt{callingham+15,collier+18,mhaskey+19a,mhaskey+19b, keim+19}).
% %The presence of FFA  would be of particular interest in the context of galaxy evolution since simulations have shown that it may arise from jet-ISM interactions  \citep{bicknell+97,bicknell+18}. 
Our sources have optically-thick spectral index constraints that are consistent with SSA, though improved spectral coverage at low frequencies ($<1$~GHz) will ultimately be needed (\citealt{callingham+15,collier+18,mhaskey+19a,mhaskey+19b, keim+19}).  The identification of FFA associated with any of the newly radio-loud AGN in our sample would be of particular interest in the context of galaxy evolution  since simulations have shown that it may arise from jet-ISM interactions \citep{bicknell+97,bicknell+18}. 
Such jet-ISM interactions may subsequently influence the ambient star formation rate and/or efficiency, as argued by recent observational studies in low-redshift galaxies (\citealt{morganti+13, nyland+13, mukherjee+18, husemann+19, jarvis+19, ruffa+19}). 
%The possibility of FFA in J1254$-$0606 will be further tested once additional radio data covering a broader frequency range are available.  
% From Bicknell et al. 1997:" In our model, the strong radiative shocks which precede the advantage of the lobe into the ISM create an ionized shell of shocked ISM capable of free-free absorbing low-frequency radio emission, thereby causing the peak in the radio spec- trum. "
% Distinguishing between these absorption mechanisms is important since it provides clues about the physical conditions in the vicinity of the jet.
% FFA of the radio emission via ambient thermal gas (by obtaining a robust upper limit that effectively rules out SSA) 
% Future high-resolution radio observations spanning a broader range of frequencies will be needed to robustly constrain the underlying absorption mechanisms (synchrotron self absorption or free-free absorption). 
 
%We also show the spectral curvature parameter, $q$, as a function of $\alpha_{\rm thick}$ in the right panel of Figure~\ref{fig:radio_colors}. This figure illustrates the expected  connection between the degree of spectral curvature and $\alpha_{\rm thick}$, with the most sharply-peaked source exhibiting the steepest optically-thick spectral index.  

%%%%%%%%%%%%%%%%%%%%%%%%%%%%%%%%%%%%%%%%%%%%%%%
\begin{deluxetable*}{ccccccccccccc}[t!]
\tablecaption{Infrared Source Properties \label{tab:IR}}
\tablecolumns{12}
%\tablenum{1}
\tablewidth{0pt}
\tablehead{
\colhead{Source} & \colhead{W1} & \colhead{W2} & \colhead{W3} & \colhead{W4} & \colhead{$\log(L_{6\mu\rm m,\,rest})$} & \\
\colhead{} & \colhead{(mag)} & \colhead{(mag)} & \colhead{(mag)} & \colhead{(mag)} & \colhead{(erg~s$^{-1}$)}\\
\colhead{(1)} & \colhead{(2)} & \colhead{(3)} & \colhead{(4)} & \colhead{(5)} & \colhead{(6)} 
}
\startdata
J0742+2704  &  14.42$\pm$0.03  &  13.68$\pm$0.04  &  11.25$\pm$0.15  &         \nodata  &              45.05\\ 
J0751+3154  &  15.98$\pm$0.06  &  14.86$\pm$0.07  &         \nodata  &         \nodata  &              46.24\\ 
J0800+3124  &  15.91$\pm$0.05  &  14.70$\pm$0.07  &  12.26$\pm$0.41  &         \nodata  &              46.30\\ 
J0807+2102  &  15.91$\pm$0.06  &  14.72$\pm$0.07  &  11.64$\pm$0.25  &         \nodata  &              46.14\\ 
J0832+2302  &  13.84$\pm$0.03  &  12.46$\pm$0.03  &   9.45$\pm$0.04  &   7.21$\pm$0.11  &              46.07\\ 
J0950+5128  &  13.43$\pm$0.02  &  12.80$\pm$0.03  &  10.40$\pm$0.08  &   8.22$\pm$0.25  &              44.13\\ 
J1023+2219  &  15.85$\pm$0.05  &  14.77$\pm$0.07  &  11.32$\pm$0.20  &   8.47$\pm$0.34  &            \nodata\\ 
J1037-0736  &  14.82$\pm$0.03  &  13.83$\pm$0.04  &  11.22$\pm$0.16  &         \nodata  &            \nodata\\ 
J1112+2809  &  16.33$\pm$0.07  &  15.40$\pm$0.10  &         \nodata  &         \nodata  &            \nodata\\ 
J1157+3142  &  14.98$\pm$0.03  &  13.82$\pm$0.04  &  11.26$\pm$0.16  &   8.62$\pm$0.41  &              45.40\\ 
J1204+1918  &  16.07$\pm$0.06  &  15.17$\pm$0.09  &  12.08$\pm$0.33  &   8.85$\pm$0.41  &              46.54\\ 
J1208+4741  &  16.16$\pm$0.06  &  15.18$\pm$0.07  &  12.70$\pm$0.43  &         \nodata  &              45.26\\ 
J1246+1805  &  16.53$\pm$0.07  &  15.43$\pm$0.10  &  12.72$\pm$0.47  &         \nodata  &            \nodata\\ 
J1254-0606  &  15.92$\pm$0.05  &  14.65$\pm$0.06  &  10.35$\pm$0.06  &   7.83$\pm$0.16  &            \nodata\\ 
J1333-0349  &  15.34$\pm$0.04  &  14.35$\pm$0.05  &  11.99$\pm$0.25  &         \nodata  &            \nodata\\ 
J1347+4505  &  15.22$\pm$0.04  &  13.87$\pm$0.03  &  10.82$\pm$0.09  &         \nodata  &            \nodata\\ 
J1357-0329  &  13.87$\pm$0.03  &  12.77$\pm$0.03  &   9.87$\pm$0.05  &   7.39$\pm$0.10  &            \nodata\\ 
J1413+0257  &  18.14$\pm$0.25  &  16.11$\pm$0.17  &  11.65$\pm$0.17  &   8.94$\pm$0.32  &            \nodata\\ 
J1447+0512  &  16.20$\pm$0.06  &  15.14$\pm$0.07  &         \nodata  &         \nodata  &              46.03\\ 
J1507-0549  &  15.97$\pm$0.06  &  15.12$\pm$0.10  &  12.04$\pm$0.29  &         \nodata  &            \nodata\\ 
J1512-0533  &  16.66$\pm$0.08  &  15.19$\pm$0.10  &  12.43$\pm$0.39  &   9.10$\pm$0.47  &            \nodata\\ 
J1514+4000  &  16.22$\pm$0.05  &  15.54$\pm$0.08  &  12.13$\pm$0.19  &   9.53$\pm$0.52  &              46.21\\ 
J1546+1500  &  15.97$\pm$0.05  &  14.90$\pm$0.07  &         \nodata  &         \nodata  &            \nodata\\ 
J1603+1809  &  16.10$\pm$0.05  &  15.47$\pm$0.09  &  12.59$\pm$0.48  &         \nodata  &              47.05\\ 
J1609+4306  &  16.38$\pm$0.05  &  15.41$\pm$0.08  &  13.12$\pm$0.43  &   9.62$\pm$0.48  &            \nodata\\ 
J2109-0644  &  15.14$\pm$0.04  &  14.07$\pm$0.04  &  11.21$\pm$0.17  &   8.22$\pm$0.31  &              45.71\\ 
\enddata
\tablecomments{Column 1: Source name.  Columns 2-5: WISE W1 (3.4$\mu$m), W2 (4.6$\mu$m), W3 (12$\mu$m), and W4 (22$\mu$m) band magnitudes and errors from the AllWISE source catalog \citep{cutri+13}.  Column 6: Estimated rest-frame 6$\mu$m luminosity extrapolated from a power-law fit to the AllWISE photometry.  
}
\end{deluxetable*}

%%%%%%%%%%%%%%%%%%%%%%%%%%%%%%%%%%%%%%%%%%%%%%
\subsubsection{Infrared Color-Color Diagram} 
In Figure~\ref{fig:WISE_colors}, we show the distribution of the 14 sources from our multiband VLA follow-up campaign in {\it WISE} infrared color space ($W2-W3$ vs. $W1-W2$).  A summary of the infrared properties of our sources is provided in Table~\ref{tab:IR}.  All but one these sources meet the infrared color selection criteria for obscured quasars defined by \citet{mateos+12}.  The single exception is J1603+1809, which is identified as a Type 1 quasar in SDSS, but does not meet the {\it WISE} AGN selection criteria of the  \citet{assef+18} R90 catalog.  

We also highlight the region occupied by extremely red and powerful AGN defined by $(W1 - W2) + 1.25 (W2 - W3) > 7$ \citep{lonsdale+15} in Figure~\ref{fig:WISE_colors}.   
%Sources with such extreme mid-infrared colors are likely to be heavily obscured, ultra-luminous ($\log(L_{\rm bol}/{\rm L}_\odot) \sim 11.7-14.2$) quasars with redshifts in the range $0.4 < z < 3.0$ \citep{patil+20}.  
Sources with such extreme mid-infrared colors are believed to be heavily obscured, ultra-luminous quasars.      
Of our 14 sources %with multi-band JVLA follow-up data, 
one source, J1413+0257, meets the \citet{lonsdale+15} selection criteria.  Although this source currently lacks redshift information, \citet{patil+20} reported redshifts in the range $0.4 < z < 3.0$ (with a median value of $z \sim 1.5$) and luminosities of $\log(L_{\rm bol}/{\rm L}_\odot) \sim 11.7-14.2$ for a sample of quasars in the \citet{lonsdale+15} color space also harboring compact radio sources.

Finally, we use Figure~\ref{fig:WISE_colors} to investigate possible relationships between the {\it WISE} colors and two key parameters from our radio spectral modeling: the spectral curvature parameter, $q$, and the optically-thick spectral index, $\alpha_{\rm thick}$.  Qualitatively, Figure~\ref{fig:WISE_colors} suggests an association between both a higher degree of spectral curvature (a narrower spectral peak) and $\alpha_{\rm thick}$ values for redder sources.  

Previous studies have reported evidence for a connection between quasar reddening and radio jet properties possibly arising from hierarchical SMBH-galaxy co-evolution (\citealt{georgakakis+12, glikman+12, klindt+19, patil+20, fawcett+20, rosario+20}).  In such a scenario, a quasar may transition to a radio-loud phase following a gas-rich merger and subsequent change in SMBH accretion state and/or spin conducive to jet formation.  In a future study, we will investigate this possibility in more detail by constraining the origin of the reddening in our sample through optical and infrared SED modeling.    
Ultimately, observations of a larger sample will be needed to more firmly establish the relationship between the infrared colors and radio SED properties in young radio quasars.
% NOTE: possible key connection between jet-driven kpc-scale outflows and SMBH-galaxy co-evolution

%%%%%%%%%%%%%%%%%%%%%%%%%%%%%%%%%%%%%%%%%%%%%%%
%%%%%%%%%%%%%%%%%%%%%%%%%%%%%%%%%%%%%%%%%%%%%%%
%%%%%%%%%%%%%%%%%%%%%%%%%%%%%%%%%%%%%%%%%%%%%%%
\section{Origin of the Radio Variability}
\label{sec:origin}
In this section, we consider different scenarios for the origin of the radio variability in our sample based on the current constraints on source   luminosities, radio spectral shapes, and observed variability properties (amplitude and timescale).  
%One of the primary source characteristics of our sources that bears on the origin of the variability is the variability's timescale. 
Regarding the variability properties, we emphasize that our cadence is currently very sparse, with only three observations (FIRST, VLASS, and our multiband follow-up) spaced by roughly six months and two decades. Although this is clearly insufficient to infer the full character of the variability, we assume that the simple flux ratios provide a provisional measure of the variability on each timescale (see Section~\ref{sec:variability_timescale}). Future observations will yield a richer cadence that will further clarify the variability on different timescales.

% stanimirovic+18 - ionized ISM

%%%%%%%%%%%%%%%%%%%%%%%%%%%%%%%%%%%%%%%%%%%%%%%
\subsection{Plasma Propagation Effects}
Radio waves traveling through an inhomogeneous plasma may be modulated by scattering or lensing phenomena.  This may lead to variations in the observed properties of compact sources, such as flat-spectrum radio AGN.  Physical mechanisms responsible for such propagation effects include interplanetary scintillation (IPS; \citealt{morgan+18}, and references therein),  interstellar scintillation (ISS; \citealt{jauncey+16}, and references therein), and extreme scattering events (ESEs; \citealt{fiedler+87}).

IPS arises from turbulence in the solar wind plasma and leads to rapid flux variability on timescales of roughly seconds in low-frequency ($\sim$100~MHz) radio observations.  Although IPS has proven to be a powerful tool for identifying compact radio AGN in low-frequency surveys (e.g. \citealt{chhetri+18}), it is not compatible with the variability timescales and amplitudes observed in our sample above 1~GHz.  

ISS is caused by the refraction or diffraction of radio waves emitted by a microarcsecond-scale source as they pass through fluctuations in the plasma density and/or magnetic field of the ionized ISM in our Galaxy \citep{stanimirovic+18}.  
This leads to variable radio emission on timescales of hours to days with a strong dependence on galactic latitude.
%and modulations with periods of precisely one year

%The observed variability of compact radio sources due to ISS is dependent on the Galactic free electron density along a particular line-of-sight.  
Observational manifestations of ISS (i.e. variability amplitude and timescale) depend on the Galactic free electron density along a particular line of sight as well as the observing frequency.  
The distribution of our sources in Galactic latitude is approximately flat (Figure~\ref{fig:footprint}), arguing against ISS as the main driver of the observed radio variability.   
As a more quantitative test of an ISS variability origin for each source, we computed the critical frequency,\footnote{The critical frequency was estimated using the online calculator available at \url{https://www.nrl.navy.mil/rsd/RORF/ne2001/} based on the Cordes-Lazio model for the Galactic distribution of free electrons \citep{cordes+02,cordes+03}.} 
$\nu_0$,  
%(where the modulation due to scintillation is the strongest)
below which the modulation due to scintillation is in the strong regime.\footnote{For an in-depth review of optics and ISS theory, we refer readers to \citet{narayan+92}}  At the positions of our sources we find $\nu_0$ = 7.1--11~GHz.  
For observations at $\nu$ = 1.5~GHz, we are therefore well within the strong scattering regime %and the scattering strength, $\xi$ = ($\nu$/$\nu_0$)$^{17/10}$, ranges from 16 to 33. Since $\xi >>$ 1, we are well within the strong %scattering limit 
and the equations from Section~3.2 of \citet{walker+98} apply. 
For point sources in this limit that are experiencing refractive scintillation, the expected flux modulation at 1.5~GHz is m = ($\nu_0$/$\nu$)$^{17/30}$ $\sim$ 30--40\%. This modulation is expected to occur on a timescale of $t_r \sim 2(\nu_0/\nu)^{11/5}$ hours $\sim$ 2--7 days. This stands in sharp contrast to the observed modulations of $\gtrsim$100\% to $\gtrsim$2500\% 
occurring on observed timescales of months to decades for our sources, thus ruling out refractive ISS as the origin of the radio variability in our sample. 

Diffractive ISS, which is associated with interference, leads to variability over an even narrower bandwidth and shorter timescale, making it considerably less plausible than refractive ISS.  The fluctuations (of order unity) last for only $\sim$0.1 -- 0.3 hours.  Furthermore, at 1.5~GHz, any diffractive ISS would be decorrelated over a bandwidth of $\delta \nu \sim 1-5$~MHz. This modulation would thus be washed out by averaging over a single VLA frequency channel.

%note to self - the index of refraction of a plasma lens is < 1.  plasma lenses are therefore divergent, and result in de-magnification of the background source.
Another plasma lensing phenomenon known to produce radio variability is that of ESEs (\citealt{fiedler+87}).  In this case, refractive defocusing by the transverse passage of a discrete, high-density plasma lens in front of a compact radio source leads to a reduction of flux with a characteristic U-shaped light curve (\citealt{clegg+98, bannister+16}).  The variability amplitude (a reduction in emission of $\lesssim50$\%) and timescale (weeks to months) for an ESE are typically larger in magnitude compared to ISS.  At least one case of an ESE with a threefold modulation in flux in the centimeter-wave radio regime is known \citep{bannister+16}.  This overlaps with the lower range in variability amplitude of our sample.  However, we note that the timescale for the high-amplitude radio variability reported in \citet{bannister+16} is a few months.  This stands in sharp contrast to our sources, which vary on decades-long timescales but are steady over periods of a few months.  

While we cannot totally rule out contributions to the observed radio variability by serendipitous propagation effects, we find such scenarios to be unlikely given inconsistencies with the high variability amplitudes and long timescales observed in our sample.  

%{\bf \color{red}[Mention extreme scattering events and symmetric achromatic variability]}
%Other possible extrinsic variability scenarios worth considering are extreme scattering events and symmetric achromatic variability

%%%%%%%%%%%%%%%%%%%%%%%%%%%%%%%%%%%%%%%%%%%%%%%
\subsection{Supernovae and Gamma-Ray Bursts}
Variable radio emission may be associated with supernovae and gamma-ray bursts (GRBs), arising from collimated outflows of high-mass supernovae and compact object mergers (neutron stars and stellar-mass black holes) \citep{weiler+02,woosley+06,berger+14}. However, the high luminosities of our sources (Table~\ref{tab:FIRST_VLASS}) rule out the possibility of a radio supernova afterglow \citep{palliyaguru+19} as a progenitor scenario. Regarding the possibility of radio variability associated with a GRB origin, we note that our sources are on par with, or more luminous than, on-axis GRBs, which have peak luminosities of $\sim 10^{31}$~erg~s$^{-1}$~Hz$^{-1}$ \citep{chandra+12}. However, the typical variability timescale of radio emission associated with GRBs is around 1--2 weeks \citep{pietka+15}. This is considerably shorter than the variability timescale constraints for our sources ($>$ a few months), thus ruling-out the possibility of a GRB progenitor.

%%%%%%%%%%%%%%%%%%%%%%%%%%%%%%%%%%%%%%%%%%%%%%%
\subsection{Tidal Disruption Events}
Tidal disruption events (TDEs; e.g., \citealt{komossa+15}, and references therein) occur when a star passes within the tidal radius of a SMBH and is ripped apart and partially accreted onto the SMBH.  This leads to a multi-wavelength flare, which in some cases includes the production of radio continuum emission associated with non-thermal jets or thermal outflows  (\citealt{vanvelzen+11,vanvelzen+16,anderson+20}).  
The most radio luminous known TDE, Swift J1644+57, peaked at $\sim$10$^{32}$ ~erg~s$^{-1}$~Hz$^{-1}$ \citep{eftekhari+18}, which is roughly in line with the radio luminosities of our sources.  
In addition, recent studies have suggested that TDEs may be responsible for triggering a significant fraction of the changing-look AGN population, particularly at $z>2$ \citep{padmanabhan+20}.  
%However, TDEs only arise from SMBH with masses $\lesssim10^8$~M$_{\odot}$.  Stars that approach more massive SMBH are "swallowed whole" \citep{kesden+12}, and thus do not lead to the production of an electromagnetic flare).  Our sources tend to have more massive SMBHs (Table~\ref{tab:sample}), and we therefore conclude that TDEs are unlikely progenitors for the majority of our sources.  
Because more massive black holes have weaker tidal fields near their event horizons, TDEs for main sequence stars become increasingly rare above a critical mass of about ($10^8$ M$_{\odot}$, \citealt{hills+75}).  Main sequence stars that approach SMBHs above this limiting mass are ``swallowed whole" (\citealt{macleod+12}), and thus do not lead to the production of an electromagnetic flare\footnote{We note that rapidly rotating stars, including those on the main sequence, could conceivably be disrupted by SMBHs with $M_{\rm SMBH}$ up to $\sim 7 \times 10^8$ M$_{\odot}$ \citep{kesden+12}.}.  As shown in Table~\ref{tab:sample}, the available SMBH mass estimates of our sources typically exceed $10^8$ M$_{\odot}$, suggesting that TDEs are unlikely progenitors for the majority of our sources.

On the other hand, evolved stars with large diffuse envelopes, such as red giants, are in principle susceptible to tidal disruption by a SMBH with $M_{\rm SMBH} > 10^8$ M$_{\odot}$.  However, such events are expected to be rare due to the relatively short duration of the red giant phase compared to main sequence lifetimes.  Recent studies have also argued that TDEs of giant stars by SMBHs with masses above $10^8$ M$_{\odot}$ may be less luminous than TDEs associated with lower-mass SMBHs (\citealt{bonnerot+16}).  
%We also note that rapidly rotating stars, including those on the main sequence, could conceivably be disrupted by SMBHs with $M_{\rm SMBH}$ up to $\sim 7 \times 10^8$ M$_{\odot}$ \citep{kesden+12}.  %However, to our knowledge, observational evidence for such a scenario is currently lacking.    
 
Although TDEs may provide a plausible mechanism for driving the extreme radio variability in some of our sources, the current literature consensus is that TDEs with powerful relativistic jets are rare (e.g., \citealt{vanvelzen+18}), composing only a few percent of the known TDE population\footnote{For an alternative perspective on the radio detection rates of TDEs, we refer readers to \citet{dai+20}.} (\citealt{alexander+20}, and references therein).  Thus, based on the high radio luminosities and SMBH masses of our sample, as well as the low space density of radio-loud TDEs, we conclude that the radio variability in our sources is most likely not associated with TDEs. 

%\section{Plausible Intrinsic Variability Scenarios}
%Based on the radio SED shapes, variability timescale and amplitude constraints, and high radio luminosities ($L_{\rm 3\,GHz} = 10^{40 - 42} \,\, {\rm erg} \,{\rm s}^{-1}$) of our sources,   
%we conclude that variability due to extrinsic propagation effects or transient phenomena (including on-axis GRBs, radio supernovae afterglows, and tidal disruption events is unlikely.  
%In summary, extrinsic variability, as well as intrinsic variability associated with radio supernova afterglows, on-axis GRBs, and TDEs, can be ruled-out. We therefore favor an intrinsic AGN variability scenario potentially driven by shocks along the jet, jet reorientation, or young jets that have been recently launched.  

%%%%%%%%%%%%%%%%%%%%%%%%%%%%%%%%%%%%%%%%%%%%%%%
%\subsection{AGN Flaring}
\subsection{Intrinsic AGN Variability}
While the large-scale lobes of radio galaxies are believed to remain steady over Myr to Gyr timescales \citep{blandford+19}, intrinsic radio variability on human timescales (days to years) is common among AGN with compact ($<1$~kpc) jets such as blazars \citep{lister+01}, unbeamed radio quasars/galaxies with young jets \citep{torniainen+05, orienti+20, kunert-bajraszewska+20}, the cores of FRI/FRII \citep{fanaroff+74} radio galaxies \citep{chatterjee+11, macagni+20}, and low-luminosity AGN \citep{mundell+09, brunthaler+05}.  Although the physics of  the intrinsic variability of compact AGN jets remains an unsolved problem, plausible mechanisms include the propagation of shocks along the jet \citep{marscher+85} and changes in the SMBH accretion properties such as accretion disk instabilities \citep{czerny+09, janiuk+11} or accretion state changes \citep{koay+16, wolowska+17}.  In addition to variability mechanisms directly related to accretion, radio variability may also arise from the jet itself owing to jet reorientation and the evolution of a young, expanding jet (e.g. \citealt{kunert-bajraszewska+20,an+20}).  We focus on accretion-driven radio variability in this section and discuss jet reorientation and youth in Sections~\ref{sec:reorientation} and \ref{sec:young}. 
%, the propagation of shocks along the jet \citep{marscher+85}, %interactions between the jet and the ambient ISM \citep{mundell+09}, jet reorientation \citep{}, and the evolution of a young, recently-triggered jet \citp{}.    

Each radio variability scenario described above is characterized by differences in variability amplitude level,  as well as temporal and spectral evolution.  Thus, distinguishing amongst them is best accomplished through multiepoch, broadband radio studies.  
For instance, low-amplitude ($\sim$tens of percent) radio flares occurring on timescales of days to months are popularly attributed to shock propagation. 
The low variability amplitudes and short timescales typically associated with this flaring mode contrast with the properties of our sources, which were selected to exhibit large (100\% to $>$2500\%) flux increases at $\sim$GHz frequencies over decadal timescales and were later found to be steady over timescales of a few months.  

Furthermore, the radio luminosities of our sources are typical of flares from AGN with bright, persistent counterparts (such as the $\sim$2~Jy, 6 $\times 10^{33}$ ~erg~s$^{-1}$~Hz$^{-1}$ blazar J0851+202; \citealt{pietka+15}).  Blazar flares typically represent less than a twofold change in quiescent flux, which we emphasize is considerably lower than the typical variability amplitude observed in our sample (more than an order of magnitude in the most extreme cases).  

\subsection{Jet Reorientation}
\label{sec:reorientation}
For jets aligned at small angles to our line of sight, special relativistic effects cause the source to be beamed, which in the time domain leads to an apparent increase in the variability amplitude and a decrease in the variability timescale \citep{lister+01}.  A rapid reorientation of a compact jet toward our line of sight during the $\sim$20 years between FIRST and VLASS Epoch~1 would lead to an apparent brightening of an intrinsically low-luminosity radio source due to an increase in the Doppler factor.  

Potential underlying causes for jet reorientation that may lead to variability  on human timescales include 
%magneto-hydrodynamic (MHD) instabilities \citep{bodo+13}, 
helical magnetic fields, flaring blazars, jet-ISM interactions, and SMBH orbital motion.  Confirming or refuting the possibility of helical magnetic fields requires multiepoch observations with milliarcsecond-scale resolution to identify key signatures such as periodic changes in the position angle of the jet (\citealt{britzen+17}).  We discuss the the remaining possibilities in this section.  

\subsubsection{Blazar Contamination}
Blazars represent the brightest and most well-studied class of relativistically beamed objects and have characteristic flat radio spectral shapes and double-peaked multiwavelength SEDs from the radio to gamma-energy regimes \citep{fossati+98, meyer+11, bottcher+19}.  Unlike blazars, the majority of our sources (with the exception of J2109-0644; see Section~\ref{sec:SEDs}) have peaked radio spectra.  However, we note that flaring blazars, including those hosted by quasars, are known to exhibit temporarily peaked radio spectra on timescales of weeks to months \citep{tinti+05, torniainen+05, orienti+10, fromm+15}.  The high-energy (X-ray and gamma-ray) properties of our sources, and hence their multiwavelength SEDs, are not currently known.  

%Although we cannot entirely rule out the possibility of blazar contamination in our sample, the lack of substantial variability on timescales of a few months (Section~\ref{sec:variability_timescale}) argues against it being a dominant factor.  
Besides the radio SED shapes, another argument  against blazar variability as the origin of the observed radio brightening in our sources is the lack of substantial variability on timescales of a few months (Section~\ref{sec:variability_timescale}).  
We plan to constrain the Doppler factors of our sources by measuring their parsec-scale brightness temperatures and morphologies in an upcoming VLBA study. 

%Finally, important evidence against the hypothesis that the observed transient radio emission is due to blazar variability is obtained from the NEOWISE data 
Important evidence against the blazer hypothesis is also obtained from the NEOWISE data \citep{mainzer+11}. These data comprise about 5~years of monitoring observations at 3.4 and 4.6 $\mu$m. With one exception (J1413+0257), all sources in our sample have sufficient detections in NEOWISE to allow the construction of a light curve. None of the sources in our sample display the erratic and large-amplitude variability that blazars (or  flat-spectrum radio quasars) typically display at mid-IR wavelengths \citep{anjum+20}. 

We have also investigated the optical variability by comparing SDSS imaging observations (a single observation, obtained between 2000 and 2005) and the more recent multiepoch Zwicky Transient Survey (ZTF; \citealt{graham+19})  observations, obtained between 2018 and 2020 (DR3). Since the ZTF catalog contains only PSF photometry, we restrict our sample to sources that are detected as point-like in SDSS imaging observations, leaving 14 quasars with detections in both SDSS and ZTF. The ZTF light curves of these sources are unremarkable. Furthermore, we find no evidence for a persistent flux increase or decrease between the SDSS and ZTF epoch. The mean magnitude difference between ZTF and SDSS observations is $-0.08$~mag with a standard deviation of 0.35~mag, which is common for the $\approx 15$ year time difference between SDSS and ZTF \citep{macleod+12}.  

%Although it may seem plausible that our sample could suffer from substantial blazar contamination, we argue that this is not likely to be the case.  
%Our sources have considerably lower peak radio flux densities ($\sim$10~mJy) than blazar samples, which typically contain sources with Jansky-level peak flux densities. 

% %While jet precession due to accretion disk instabilities is unlikely given its long variability timescale ($\sim10^{3-4}$~yr), 
  
%%%%%%%%%%%%%%%%%%%%%%%%%%%%%%%%%%%%%%%%%%%%%%%
\subsubsection{Interaction with a Dense ISM}
Radio variability may also arise from jet deflection \citep{an+20} or interaction with a dense ambient medium  (\citealt{middelberg+07,kunert-bajraszewska+10,an+13,mukherjee+16, siemiginowska+16, williams+20, lister+20}).  In such scenarios, brightening in the radio may be caused by an increased Doppler factor and/or the production of shocks associated with the jet plowing into the ISM. Such interactions have been observed out to high redshift ($z>5$; \citealt{an+20}). In addition to flux variability, evidence for jet-ISM interactions on parsec scales typically includes jet asymmetries, evidence for jet deceleration, and enhanced polarization at the edge of the jet.  

While we emphasize that there is currently no evidence for the presence of jet-ISM interactions in our sample, at least one of our sources has extremely red {\it WISE} colors in the hyperluminous quasar regime.  Recent studies have argued that the combination of radio compactness and mid-infrared colors may be associated with subgalactic jets propagating through a dense ISM \citep{patil+20}.  Such jets have the potential to influence galaxy evolution, perhaps by altering star formation rates/efficiencies or through ``self-regulation" of the SMBH accretion rate.  Quantifying the amount of energy transferred by subgalactic jets to a dense ambient ISM in the form of feedback, particularly at $z \sim 1-3$, is a key goal of studies with next-generation telescopes \citep{nyland+18}.  

%%%%%%%%%%%%%%%%%%%%%%%%%%%%%%%%%%%%%%%%%%%%%%%
\subsubsection{SMBH Binary Orbital Motion}
Orbital motion associated with an SMBH binary system (\citealt{palenzuela+10, kaplan+11, schnittman+13, kulkarni+16}) could  conceivably lead to periodic radio variability.  However, such systems should be quite rare: \citet{holgado+18} predict that no more than 1/100 blazars host SMBH binaries with periods $< 1$~yr.  
%However, the SMBH merger rate, as well as the precise nature of their electromagnetic counterparts (including radio jets), remain observationally unconstrained at this time.  
Merging SMBHs with smaller separations would be even rarer than this, though such systems may produce radio variability by either the disruption of an existing jet or the formation of a new jet launched by the circumbinary accretion disk (e.g. \citealt{komossa+20}).  

Simulations of SMBH mergers predict boosts in radio jet luminosity that are expected to have maximum magnitudes and durations for near-equal-mass binaries of high-mass SMBHs ($10^9 - 10^{10} M_{\odot}$; \citealt{khan+18}).  An improved understanding of the physics of jet formation associated with SMBH binary mergers awaits future multimessenger studies incorporating constraints from both multiepoch radio surveys \citep{murphy+13} and pulsar timing arrays \citep{burke-spolaor+19}.  

%Our current VLA data cannot confirm or rule out the possibility of jet reorientation as the primary (or perhaps contributing) mechanism responsible for the observed radio variability in our sample.  Long-term monitoring of the radio light curves, parsec-scale imaging, and full-polarimetric\footnote{Although VLASS is a full-polarization radio survey, only Stokes {\it I} image products are available at this time \citep{lacy+20}.} studies (\citealt{an+20}) will be necessary to test this possibility.       
%{\bf \emph{Identifying free-free absorption via multi-band radio SED modeling would serve as a direct probe of this process}} (jets plowing into a dense circumnuclear medium).

%%%%%%%%%%%%%%%%%%%%%%%%%%%%%%%%%%%%%%%%%%%%%%%
% Magnetically choked accretion flow and radio loudness - Sikora et al. 2013: https://iopscience.iop.org/article/10.1088/2041-8205/764/2/L24/pdf

%%%%%%%%%%%%%%%%%%%%%%%%%%%%%%%%%%%%%%%%%%%%%%%
\begin{figure*}[t!]
\centering
\includegraphics[clip=true, trim=0cm 0.5cm 0cm 0cm, width=\textwidth]{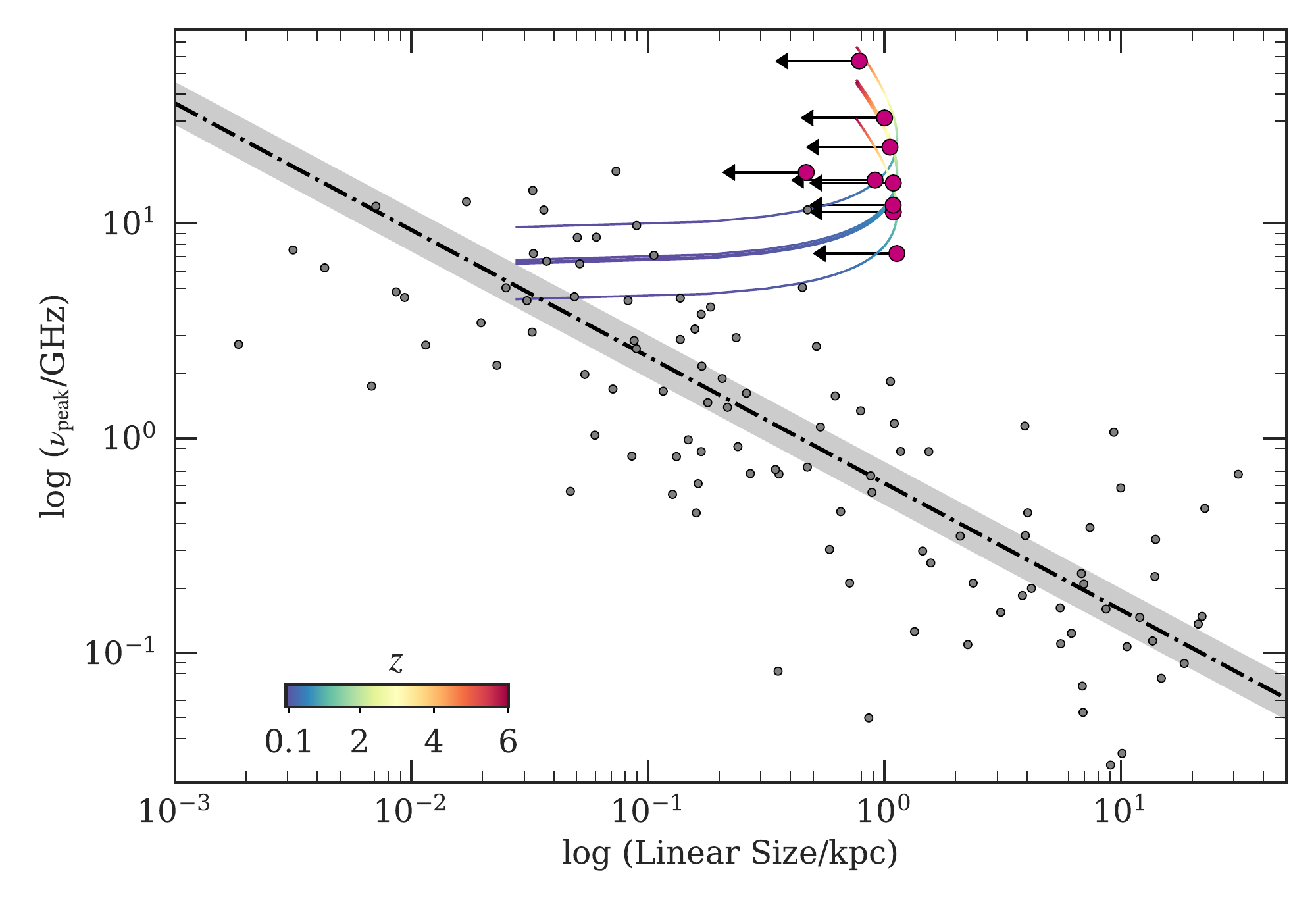}
\caption{Spectral turnover as a function of linear size.  The small gray circles are drawn from literature measurements of young radio AGN compiled by \citet{jeyakumar+16}.  The black dot-dashed line shows the empirical fit to the turnover-size relation from \citet{odea+98}, and the dark gray shaded region indicates the uncertainty in the relation.  Upper limits on the sizes of sources with quasi-simultaneous, multiband VLA follow-up from our sample that have spectroscopic redshifts available are shown by the large purple circles.  For sources lacking redshifts, the linear size upper limits are shown over a range of possible redshifts as indicated by the rainbow-colored arcs. 
\\}
\label{fig:LS_turnover}
\end{figure*}
%%%%%%%%%%%%%%%%%%%%%%%%%%%%%%%%%%%%%%%%%%%%%%

%%%%%%%%%%%%%%%%%%%%%%%%%%%%%%%%%%%%%%%%%%%%%%%
\subsection{Young Radio Jets}
\label{sec:young}
Young radio AGN, such as gigahertz-peaked spectrum (GPS) sources, are characterized by compact morphologies and inverted radio SEDs below their turnover (peak) frequencies, which are typically in the GHz regime 
\citep{odea+98}, consistent with the morphologies and radio SEDs of the majority of our sources. 
After a jet is launched, models predict a rapid increase in luminosity ($P_{\rm radio} \sim t^{2/5}$) as the dominant energy-loss mechanism transitions from adiabatic to synchrotron losses \citep{an+12}, making the identification of young radio AGN in VLASS that have emerged in the time since the FIRST survey (17--24~yr for our sample) plausible.  For a nascent radio jet that has been triggered within the past 20~yr, the model of \citet{an+12} suggests an increase in radio luminosity of $>300$\%.  
The identification of such young radio AGN is not unprecedented; the youngest known sources have kinematic ages as low as 20~yr \citep{gugliucci+05}.  
%{\bf \emph{Monitoring the evolution of the radio SED turnover of variable compact radio AGN is crucial for testing this possibility.}}

Based on the currently available radio continuum constraints, which include large variability amplitudes (with flux increases from 100\% to $>$2500\%) at 1.5~GHz compared to FIRST over timescales of decades, steady flux densities on timescales less than a few months at 3~GHz compared to VLASS, source size constraints $<0.1^{\prime \prime}$ ($\lesssim 1$~kpc), and curved radio SEDs peaking at $\sim$5--10~GHz, we find the radio properties of our sources to be consistent with young and compact radio AGN.  
 
If the jet youth scenario for our sources is supported by higher-cadence, multiband follow-up data, there are exciting prospects for forthcoming radio surveys.  Wide-area, broadband, synoptic radio surveys above a few GHz conducted over timescales of years to decades would be particularly well tuned for identifying large statistical samples of AGN with recently launched jets.

In Figure~\ref{fig:LS_turnover}, we plot our sources on the turnover-size relation, along with a sample of young radio AGN from the literature.  This relationship is believed to arise from the evolving radio spectral shapes of expanding young radio sources.  
%As the jets of a young and compact radio AGN expand, they lose energy, thereby causing the peak (or spectral turnover) of the relativistic electron energy distribution to shift to lower frequencies.  
As a young and compact radio source expands, the opacity due to SSA or FFA decreases, which causes the spectral peak (or turnover) to shift to lower frequencies (e.g. \citealt{bicknell+97, odea+98, tingay+03}).       
Since we only have upper limits on the linear sizes of our sources, we cannot yet directly confirm whether they follow the turnover-size relation.  However, if we assume that our sources do follow the relation, we can obtain a rough estimate of their sizes.  

As an example, we consider J0807+2102 at $z = 1.5588$.  The observed spectral turnover of this source is 21.9~GHz, which corresponds to a rest-frame turnover frequency of 56.2~GHz.  Based on Figure~\ref{fig:LS_turnover}, this implies a source size of 1--10~pc.  Assuming a conservative value for the jet advance speed of $0.1c$, %{\bf \color{red}[Citations?]}, 
we obtain a rough estimate of the source age of $\sim$30--300~yr.  Thus, the young jet model passes this important consistency check.   
%We emphasize that the identification of sources with ages of only a few decades is not unprecedented.  Young radio AGN with dynamical ages as low as 20~yr have been reported in the literature based on proper motion studies \citep{gugliucci+05}.  
%{\bf For sources with known redshifts, we provide a summary of the rough source age estimates from the turnover-size relation in Table~\ref{tab:ages}.}

Very long baseline interferometric (VLBI) studies with sub-milliarcsecond resolution will ultimately be needed to test whether our sources are indeed compact on parsec scales, as expected for young jets.  Nevertheless, the radio SED shapes and source size limits from the VLA data presented in this study are consistent with radio variability arising from the evolving radio SEDs of jets that have recently been triggered. 

\section{Discussion}
\label{sec:discussion}
Based on the radio properties of our sources, in particular their high luminosities, size constraints, variability amplitudes/timescales, and sky coordinates, we rule out extrinsic variability related to plasma propagation as a radio variability scenario.  We also conclude that intrinsic variability associated with supernovae, GRBs, or TDEs is unlikely.  We therefore favor an intrinsic AGN variability scenario in which the observed radio brightening is driven by a change in the SMBH accretion rate, jet reorientation, or the formation of newborn jets. 

We emphasize that multiple mechanisms may  contribute to the observed radio variability in a given source.  A source could simultaneously be young, mildly Doppler boosted, and flaring %in response to a change in the accretion state. 
owing to the propagation of shocks along the jet.  
Since our source flux densities are typically steady on timescales of a few months but vary substantially (100\% to $>$2500\%) over timescales of two decades, we find jet youth and/or jet reorientation to be the most plausible radio variability scenarios.  

% %%%%%%%%%%%%%%%%%%%%%%%%%%%%%%%%%%%%%%%%%%%%%%%
\subsection{Accretion State Change}
% As described in Section~\ref{sec:selection_criteria}, our source selection includes a mix of SDSS quasars \citep{paris+18}, all of which in our sample are Type I, broad-line quasars, and WISE AGN \citep{assef+18}.  When available, the Eddington ratio estimates for the SDSS quasars are in the range of $L_{\rm Edd} \sim 6-20$\%, consistent with the high-excitation AGN population characterized by efficient SMBH accretion with ($L_{\rm Edd} > 1$\%; \citealt{heckman+14}).  

% For the obscured AGN, the WISE color-color plots shown in Figure~\ref{fig:WISE_colors} demonstrate that our sources have mid-infrared colors consistent with the quasar population.  Thus, all of our sources are high-excitation AGN with a new onset of radio-loud activity featuring young, compact jets.  
The large observed changes in radio flux may be associated with AGN transitioning between a radio-quiet state to one in which synchrotron-emitting radio jets are present. Extreme radio variability is typical of Galactic radio sources such as X-ray binaries \citep[e.g.][]{mirabel+99}. In these sources, the short timescales associated with black holes with masses $\sim 1-10\, M_{\odot}$ mean that the change in state from a radio-quiet mode associated with soft X-ray emission to a mode with a hard X-ray spectrum and radio jets can happen on timescales of minutes. Given the $\sim 10^7-10^9\, M_{\odot}$ SMBH masses that are typical of AGN, similar transitions may occur \citep{maccarone+03,falcke+04,nipoti+05,kording+06}, but the corresponding timescales may be longer. 

Previous attempts have been made to identify radio sources whose jets may have recently switched off \citep[e.g.][]{marecki+11}, but such objects still have radio emission from their lobes (though such sources are interesting in their own right in the context of restarted jets).  
An alternative approach is to identify AGN with a new onset of radio activity potentially associated with jets that have recently switched on.  By their very nature these objects are expected to be relatively rare, but by surveying the radio emission from a large number of AGN in two or more well-separated epochs, it may be possible to find objects that are candidates for AGN undergoing this transition.

%%%%%%%%%%%%%%%%%%%%%%%%%%%%%%%%%%%%%%%%%%%%%%%
\subsection{Comparison to Previous Studies} 
Although rare, previous studies have identified quasars with high-amplitude radio variability over timescales of years to decades 
\citep{devries+04, barvainis+05, prandoni+10, bell+15}.  In the Caltech NRAO Stripe 82 Survey (CNSS) pilot survey, \citet{mooley+16} reported the detection of a single radio-loud type 2 quasar at $z = 1.65$ (VTC233002-002736) that lacked any detectable emission in the FIRST survey. Follow-up observations revealed a nearly ten-fold flux increase at 1.5 GHz over a 15 year period and a curved radio SED, very similar in nature to our sources. 
Another example of such behavior in the CNSS, found in the $z = 0.94$ quasar CNSS 013815+00, has been reported recently by \citet{kunert-bajraszewska+20}. A newborn expanding radio jet is responsible for GPS-like characteristics of the radio spectrum of 013815+00. In addition, the transition from the radio-quiet to the radio-loud phase in 013815+00 coincides with changes in its accretion disk luminosity. Thus, the burst of radio activity in 013815+00 is interpreted as a result of an enhancement in the SMBH accretion rate.  We note that only 013815+00 is identified in the \citet{paris+18} SDSS quasar catalog, but both of the CNSS quasars are classified as AGN in the mid-infrared by \citet{assef+18} and  would therefore meet the AGN and radio selection criteria of our sample (see Section~\ref{sec:sample}). 

We note that only one source matching our selection criteria was identified in the 50~deg$^2$ pilot CNSS footprint\footnote{There are 3 additional variable radio sources identified as AGN in Table~2 of \citet{mooley+16} that meet our VLASS and FIRST selection criteria ($S_{\rm VLASS}>3$~mJy and a non-detection in FIRST).  However, none of these sources are identified in the optical or infrared AGN surveys (\citealt{paris+18} and/or \citealt{assef+18}) as required for our sample.}.  This is loosely consistent (within a factor of a few) with our observed detection rate so far over 3440~deg$^2$ of the VLASS-FIRST footprint of one confirmed optical or infrared AGN with newly radio-loud activity per $\sim 20 \pm 13$~deg$^2$.  
A more formal assessment of the areal density of AGN that have transitioned from radio-quiet to radio-loud on decades-long timescales that incorporates the entirety of VLASS Epoch~1 will be presented in a future study. 

%%%%%%%%%%%%%%%%%%%%%%%%%%%%%%%%%%%%%%%%%%%%%%%
\subsection{Implications for Galaxy Evolution}
The large-scale radio jets and lobes launched by some active galactic nuclei (AGN) are believed to have long lifetimes and duty cycles on the order of millions of years.  The properties of such ``classical" radio AGN contrast sharply with those of radio jets featuring compact (subgalactic) extents, younger ages (decades to thousands of years), and shorter lifetimes that have previously been identified \citep{Lahteenmaki+18, kunert-bajraszewska+20}. 

In the case of re-orienting jets, the change in jet direction may facilitate the transfer of energy over a large volume of the ISM of the host galaxy, thus potentially influencing ambient ISM conditions \citep{gaibler+11, mukherjee+16}.  However, the prevalence and basic physical properties of re-orienting jets, regardless of origin (see Section~\ref{sec:reorientation}), have not been well constrained owing to inherent observational challenges (the combined requirements of high angular resolution imaging and high-cadence monitoring).  If re-orienting compact jets are common, particularly at $z > 1$, they may contribute substantially to SMBH-galaxy co-evolution via the regulation of star formation from jet-ISM feedback or a disruption in SMBH growth in response to the launching of a jet. 
%A similar conclusion was reached by \citet{vanvelzen+15}, who found that the fraction of powerful jets that grow beyond $\sim 100$~kpc scales decreases with redshift. 

%We also consider the impact of intermittent, albeit short-lived episodes of radio-loud activity associated with powerful AGN. 
Intermittent, albeit short-lived, episodes of radio-loud activity associated with powerful AGN may also be important in the context of galaxy evolution.  
Numerous studies over the past few decades have used a variety of arguments (such as the overrepresentation of compact jetted AGN in flux-limited surveys) in favor of the existence of a large population of short-lived and/or rapidly retriggered compact radio AGN (\citealt{czerny+09, mooley+16, jarvis+19, patil+20, shabala+20}).  
As a rough assessment of whether the rate of newly radio-loud AGN identified in VLASS so far is consistent with this possibility, we compare the areal densities of the overall radio-loud AGN population based on constraints from FIRST with our sample.  
\citet{ivezic+02} found 1154 SDSS quasars detected by FIRST over 774~deg$^2$, corresponding to an areal density of radio-loud quasars of $\sim$1~deg$^{-2}$. If these quasars have a lifetime of $\sim$10$^6$~yr, we would only expect a very small fraction, $\sim$1 in 10$^5$, to be within their first decade of being identified as radio-loud. This translates to an areal density of $\sim$10$^{-5}$ deg$^{-2}$ compared to our detection rate of 13 newly radio-loud type 1 quasars in 3,440~deg$^2$, i.e. $\sim 4\times10^{-3}$~deg$^{-2}$, consistent with a typical period of occurrence of $\sim$10$^5$~yr.  

We speculate that frequent episodes of short-lived AGN jets that do not necessarily grow to large scales could be associated with higher-efficiency jet-driven feedback into the hosts of high-$z$ galaxies.  
A similar conclusion was reached by \citet{vanvelzen+15}, who found that the fraction of powerful jets that grow beyond $\sim 100$~kpc scales decreases with redshift.  Future studies investigating the ISM content and conditions of the hosts of newly radio-loud AGN will be important for placing quantitative constraints on the energetic impact of feedback from compact jets at cosmic noon.
%\citet{vanvelzen+15} found that the fraction of powerful jets that grow beyond $\sim 100$~kpc scales decreases with redshift \citet{vanvelzen+15}, possibility associated with a higher efficiency of jet-induced feedback into the hosts of high-redshift galaxies. 
%Ultimately, a robust quantitative measure of the duty cycles and lifetimes of radio-loud activity, and the dependence on parameters such as redshift, SMBH mass, etc., will require further study.  

%%%%%%%%%%%%%%%%%%%%%%%%%%%%%%%%%%%%%%%%%%%%%%%
%%%%%%%%%%%%%%%%%%%%%%%%%%%%%%%%%%%%%%%%%%%%%%%
%%%%%%%%%%%%%%%%%%%%%%%%%%%%%%%%%%%%%%%%%%%%%%%
\section{Summary}
\label{sec:summary}
As part of an ongoing search for slow radio transients between FIRST and VLASS Epoch~1 covering 3,440~deg$^2$ so far, we have identified a sample of 26 sources with radio variability on decadal timescales associated with known powerful quasars in SDSS and/or WISE.  These sources were previously radio-quiet quasars in FIRST but are now consistent with the radio-loud quasar population following their detection in VLASS Epoch~1.  

To investigate the origin of the radio variability in our sample of newly radio-loud quasars, we performed multiband, quasi-simultaneous VLA follow-up observations of 14 sources.  
All of our sources are characterized by high radio luminosities ($L_{\rm 3\,GHz} = 10^{40 - 42} \,\, {\rm erg} \,{\rm s}^{-1}$)  in the quasar regime, compact ($\lesssim 0.1^{\prime \prime}$) emission, and broadband radio SEDs from 1--18~GHz with significant spectral curvature.  

A comparison between the VLASS images (all of which were observed in 2019 during Epoch~1.2) and our follow-up VLA $S$-band data a few months later revealed good agreement within the current $\sim$20\% flux uncertainties of VLASS, suggesting a typical variability timescale longer than a few months. At $L$-band, the observed variability amplitudes range from 100\% to $>$2500\% in the 17--24~yr since FIRST.  

Based on the radio properties of our sources, including their SED shapes, variability timescale and amplitude constraints, and high radio luminosities, we conclude that variability due to extrinsic propagation effects or transient phenomena (including GRBs, supernovae, and TDEs) is unlikely.  We therefore favor an intrinsic AGN variability scenario for our sample.  

We conclude that our sources are most consistent with powerful quasars hosting compact, possibly young jets, which poses a challenge the generally accepted idea that ``radio-loudness" is a property of the quasar/AGN population that remains fixed on human timescales.  Our study suggests that frequent episodes of short-lived AGN jets that do not necessarily grow to large scales may be common at high redshift.  We speculate that intermittent but powerful jets on subgalactic scales could interact with the interstellar medium leading to feedback that could influence the evolution of galaxies at cosmic noon.

Further multiband follow-up with the VLA, as well as parsec-scale imaging with the Very Long Baseline Array, will be essential for placing tighter constraints on the evolutionary stages of our sources.  Additional follow-up efforts across the electromagnetic spectrum, including optical/infrared observations to determine the basic properties of the host galaxies, studies of the molecular gas content and conditions using millimeter/submillimeter telescopes, and explorations of the accretion physics and large-scale environments from high-energy (e.g. X-ray) data, will be required.  

Ultimately, the completion of the remaining two VLASS epochs, as well as future surveys of the dynamic radio and millimeter sky with the Square Kilometre Array and its pathfinders (e.g., \citealt{bignall+15,murphy+13}), will provide new insights into the life cycles of radio jets.   
In addition to radio surveys, detailed studies of individual objects quantifying the impact of jets on their ambient environments with telescopes such as the Atacama Large Millimeter/Submillimeter Array and the next-generation Very Large Array \citep{nyland+18} will be essential for determining the overall importance of feedback on ISM scales driven by compact jets for galaxy evolution.  
%VAST: An ASKAP Survey for Variables and Slow Transients

%The X-ray emission associated with young radio AGN may be thermal (arising from the accretion disk or shock-heated gas from an interaction with the jet) or non-thermal (inverse-Compton emission produced by the jet) in nature.  We will investigate the origin of the X-ray emission using the scaling relation between the X-ray and radio emission   %as well as the fundamental plane of radio activity 
%that has been established for a wide range of sources, including young radio AGN with X-ray counterparts %\citep{kunert-bajraszewska+14, wojtowicz+20}.  

%%%%%%%%%%%%%%%%%%%%%%%%%%%%%%%%%%%%%%%%%%%%%%%
%%%%%%%%%%%%%% ACKNOWLEDGEMENTS %%%%%%%%%%%%%%%
%%%%%%%%%%%%%%%%%%%%%%%%%%%%%%%%%%%%%%%%%%%%%%%
\acknowledgments
We thank the anonymous referee for providing us with helpful comments that have improved the quality of this work.   
The National Radio Astronomy Observatory is a facility of the National Science Foundation operated under cooperative agreement by Associated Universities, Inc.  
Basic research in radio astronomy at the U.S. Naval Research Laboratory is supported by 6.1 Base Funding. 
%The authors have made use of {\sc Astropy}, a community-developed core {\sc Python} package for Astronomy \citep{astropy+13}.  
%We also used {\sc Montage}, which is funded by the National Science Foundation under Grant Number ACI-1440620, and was previously funded by the National Aeronautics and Space Administration's Earth Science Technology Office, Computation Technologies Project, under Cooperative Agreement Number NCC5-626 between NASA and the California Institute of Technology.  
M.K.-B. acknowledges support from the `National Science Centre, Poland' under grant No. 2017/26/E/ST9/00216. 
% Thank people who discussed project with me?  

\vspace{5mm}
\facilities{VLA}

\software{Astropy \citep{astropy+13}, CASA \citep{mcmullin+07}, Obit \citep{cotton+08}, and Montage \citep{jacob+10,berriman+17}.}

%%%%%%%%%%%%%%%%%%%%%%%%%%%%%%%%%%%%%%%%%%%%%%%%%%%%
%%%%%%%%%%%%%%%%%% BIBLIOGRAPHY %%%%%%%%%%%%%%%%%%%%
%%%%%%%%%%%%%%%%%%%%%%%%%%%%%%%%%%%%%%%%%%%%%%%%%%%%
%\bibliographystyle{aasjournal}
\bibliographystyle{apj}
\bibliography{main}

%%%%%%%%%%%%%%%%%%%%%%%%%%%%%%%%%%%%%%%%%%%%%%%%%%%%
%%%%%%%%%%%%%%%%%%%% APPENDIX %%%%%%%%%%%%%%%%%%%%%%
%%%%%%%%%%%%%%%%%%%%%%%%%%%%%%%%%%%%%%%%%%%%%%%%%%%%
\appendix
\label{appendix}

\section{Source Properties}
\label{sec:source_props}
The basic properties of our sources from our multiband VLA follow-up observations are presented in Table~\ref{tab:JVLA}.  We note that all sources are unresolved at the angular resolution provided in Column 6 of Table~\ref{tab:JVLA}.

\startlongtable
\begin{deluxetable*}{ccccccccc}
\tablecaption{Multiband VLA Follow-up Data \label{tab:JVLA}}
\tablecolumns{7}
\tablewidth{0pt}
\tablehead{
\colhead{Source} & \colhead{Date} & \colhead{Band} & \colhead{$\nu$} & \colhead{$\sigma_{\rm rms}$} & \colhead{$\theta_{\rm maj} \times \theta_{\rm min}$} & \colhead{$S_{\rm peak}$} \\
\colhead{} & \colhead{} & \colhead{} & \colhead{(GHz)} & \colhead{($\mu$mJy~beam$^{-1}$)} & \colhead{($\prime \prime \times \prime \prime$)} & \colhead{(mJy~beam$^{-1}$)} \\
\colhead{(1)} & \colhead{(2)} & \colhead{(3)} & \colhead{(4)} & \colhead{(5)} & \colhead{(6)} & \colhead{(7)} 
}
\startdata
J0742+2704	& 2019 Oct 03 		& L	     &	1.5	& 47 & $2.00 \times 0.89$ & $3.80 \pm 0.04$  \\ 
			& 				& S	     &	3.0	& 25 & $0.95 \times 0.49$ & $10.97 \pm 0.02$   \\
			& 				& C	     &	6.0	& 20 & $0.43 \times 0.25$ & $25.20 \pm 0.04$   \\
			&				& X	     &	10.0	& 20 & $0.27 \times 0.18$ & $25.23 \pm 0.03$  \\
			& 				& Ku	     &	15.0	& 10 & $0.17 \times 0.12$ & $19.70 \pm 0.01$  \\
\hline
J0807+2102	& 2019 July 23 		& L	     &	1.5	&  187 & $4.44 \times 1.11$ & $0.86 \pm 0.03$\\
			& 				& S	     &	3.0	&  87 & $2.31 \times 0.57$ & $3.15 \pm 0.07$\\
			& 				& C	     &	6.0	&  45 & $1.15 \times 0.28$ & $7.83 \pm 0.03$\\
			&				& X	     &	10.0	&  26 & $0.67 \times 0.20$ & $15.10 \pm 0.02$\\
			& 2019 Sept 19	        & X	     &	10.0	&  15 & $0.28 \times 0.19$ & $16.83 \pm 0.02$\\
			& 				& Ku	     &	15.0	&  13 & $0.18 \times 0.13$ & $17.42 \pm 0.01$\\
			& 				& K	     &	22.0	&  28 & $0.11 \times 0.09$ & $13.03 \pm 0.03$\\
\hline
J0832+2302	& 2019 July 23 		& L	     &	1.5	& 70 & $4.63 \times 1.11$ & $1.41 \pm 0.05$\\
			& 				& S	     &	3.0	&  35 & $2.33 \times 0.65$ & $3.25 \pm 0.03$\\
			& 				& C	     &	6.0	&  20 & $1.14 \times 0.34$ & $5.69 \pm 0.01$\\
			&				& X	     &	10.0	&  21 & $0.71 \times 0.21$ & $5.28 \pm 0.02$\\
			& 2019 Sept 19	        & X	     &	10.0	&  13 & $0.26 \times 0.19$ & $5.97 \pm 0.01$\\
			& 				& Ku	     &	15.0	&  21 & $0.17 \times 0.12$ & $6.57 \pm 0.02$ \\
\hline
J0950+5128	& 2019 July 23 		& L	     &	1.5	&  58 & $4.11 \times 1.48$ & $3.93 \pm 0.05$\\
			& 				& S	     &	3.0	&  41 & $2.23 \times 0.68$ & $10.21 \pm 0.02$\\
			& 				& C	     &	6.0	&  19 & $1.10 \times 0.39$ & $15.61 \pm 0.02$\\
			&				& X	     &	10.0	&  23 & $0.65 \times 0.24$ & $17.09 \pm 0.02$\\
			& 2019 Sept 19	        & X	     &	10.0	&  12 & $0.28 \times 0.19$ & $18.69 \pm 0.01$\\
			& 				& Ku	     &	15.0	&  33 & $0.16 \times 0.13$ & $13.12 \pm 0.05$\\
\hline
J1037-0736	& 2019 Oct 13 		& L	     &	1.5	&  69 & $1.79 \times 1.10$ & $6.68 \pm 0.07$\\
			& 				& S	     &	3.0	&  68 & $0.72 \times 0.53$ & $15.48 \pm 0.07$ \\
			& 				& C	     &	6.0	&  22 & $0.34 \times 0.26$ & $19.13 \pm 0.06$\\
			&				& X	     &	10.0	&  16 & $0.24 \times 0.16$ & $16.08 \pm 0.02$\\
			& 				& Ku	     &	15.0	&  14 & $0.17 \times 0.11$ & $13.31 \pm 0.01$\\
\hline
J1204+1918	& 2019 Oct 11   	& L	     &	1.5	&  90 & $1.32 \times 1.14$ & $2.00 \pm 0.01$\\
			& 				& S	     &	3.0	&  26  & $0.70 \times 0.55$ & $4.83 \pm 0.02$\\
			& 				& C	     &	6.0	&  15 & $0.32 \times 0.31$ & $5.10 \pm 0.01$\\
			&				& X	     &	10.0	&  18 & $0.23 \times 0.18$ & $3.20 \pm 0.01$\\
			& 				& Ku	     &	15.0	&  13 & $0.16 \times 0.12$ & $2.00 \pm 0.01$\\
\hline
J1208+4741	& 2019 Oct 11 	        & L	     &	1.5	&  41 & $1.14 \times 0.98$ & $1.06 \pm 0.01$\\
			& 				& S	     &	3.0	&  18 & $0.62 \times 0.49$ & $2.46 \pm 0.01$\\
			& 				& C	     &	6.0	&  13 & $0.29 \times 0.25$ & $2.95 \pm 0.01$\\
			&				& X	     &	10.0	&  15 & $0.18 \times 0.15$ & $2.29 \pm 0.01$\\
			& 				& Ku	     &	15.0	&  13 & $0.12 \times 0.10$ & $1.74 \pm 0.01$\\
\hline
J1246+1805	& 2019 Oct 11 	        & L	     &	1.5	&  130 & $1.45 \times 1.17$ & $3.39 \pm 0.01$\\
			& 				& S	     &	3.0	&  43 & $0.76 \times 0.56$ & $8.25 \pm 0.02$\\
			& 				& C	     &	6.0	&  23 & $0.32 \times 0.27$ & $11.58 \pm 0.02$\\
			&				& X	     &	10.0	&  28 & $0.18 \times 0.18$ & $12.20 \pm 0.01$\\
			& 				& Ku	     &	15.0	&  25 & $0.13 \times 0.11$ & $11.75 \pm 0.03$\\
\hline
% CHECK
J1254-0606	& 2019 Nov 01 		& L	     &	1.5	&  80 & $1.70 \times 0.89$ & $1.69 \pm 0.08$\\
			& 				& S	     &	3.0	&  74 & $0.90 \times 0.55$ & $10.06 \pm 0.03$\\
			& 				& C	     &	6.0	&  37 & $0.33 \times 0.23$ & $18.51 \pm 0.05$\\
			&				& X	     &	10.0	&  28 & $0.22 \times 0.15$ & $15.03 \pm 0.05$\\
			& 				& Ku	     &	15.0	&  26 & $0.15 \times 0.10$ & $9.39 \pm 0.04$\\
\hline
J1413+0257	& 2019 Sept 23 	& L	     &	1.5	&  63 & $1.45 \times 1.04$ & $1.38 \pm 0.06$\\
			& 				& S	     &	3.0	&  23 & $0.87 \times 0.60$ & $7.30 \pm 0.02$\\
			& 				& C	     &	6.0	&  22 & $0.40 \times 0.25$ & $14.58 \pm 0.02$\\
			&				& X	     &	10.0	&  15 & $0.30 \times 0.19$ & $10.92 \pm 0.01$\\
			& 				& Ku	     &	15.0	&  12 & $0.21 \times 0.12$ & $6.68 \pm 0.01$\\
\hline
J1447+0512	& 2019 Sept 23  	& L	     &	1.5	&  48 & $1.55 \times 1.09$ & $2.71 \pm 0.05$\\
			& 				& S	     &	3.0	&  21 & $0.92 \times 0.58$ & $4.02 \pm 0.02$\\
			& 				& C	     &	6.0	&  12 & $0.49 \times 0.30$ & $2.38 \pm 0.01$\\
			&				& X	     &	10.0	&  15 & $0.32 \times 0.19$ & $1.55 \pm 0.01$\\
			& 				& Ku	     &	15.0	&  11 & $0.18 \times 0.13$ & $1.30 \pm 0.01$\\
\hline
J1546+1500	& 2019 Sept 23		& L	     &	1.5	&  39 & $1.41 \times 1.15$ & $3.30 \pm 0.04$\\
			& 				& S	     &	3.0	&  21 & $0.76 \times 0.58$ & $5.61 \pm 0.02$\\
			& 				& C	     &	6.0	&  12 & $0.42 \times 0.30$ & $5.28 \pm 0.01$\\
			&				& X	     &	10.0	&  14 & $0.26 \times 0.19$ & $4.03 \pm 0.01$\\
			& 				& Ku	     &	15.0	&  21 & $0.16 \times 0.10$ & $3.50 \pm 0.02$\\
\hline
J1603+1809	& 2019 Sept 23		& L	     &	1.5	&  40 & $1.22 \times 1.01$ & $0.88 \pm 0.04$\\
			& 				& S	     &	3.0	&  18 & $0.71 \times 0.58$ & $3.16 \pm 0.01$\\
			& 				& C	     &	6.0	&  12 & $0.38 \times 0.30$ & $5.85 \pm 0.01$\\
			&				& X	     &	10.0	&  15 & $0.24 \times 0.19$ & $5.81 \pm 0.01$\\
			& 				& Ku	     &	15.0	&  12 & $0.17 \times 0.12$ & $4.80 \pm 0.01$\\
\hline
J2109-0644	& 2019 Sept 10		& L	     &	1.5	&  58 & $1.43 \times 1.06$ & $11.17 \pm 0.05$\\
			& 				& S	     &	3.0	&  22 & $0.80 \times 0.57$ & $12.51 \pm 0.01$\\
			& 				& C	     &	6.0	&  15 & $0.39 \times 0.29$ &  $16.97 \pm 0.01$\\
			&				& X	     &	10.0	&  10 & $0.25 \times 0.18$ & $15.79 \pm 0.01$\\
			& 				& Ku	     &	15.0	&  11 & $0.17 \times 0.11$ & $13.00 \pm 0.02$\\
\enddata
\tablecomments{Column 1: Source name.  Column 2: Observing date(s).  Column 3: VLA band, defined as follows: L: 1--2~GHz, S: 2--4~GHz, C: 4--8~GHz, X: 8--12~GHz, Ku: 12--18~GHz, K: 18--26~GHz.  Column 4: Central image frequency.  Column 5:  1$\sigma$ rms noise.  Column 6: Clean beam dimensions (major $\times$ minor axis).  Column 7:  Peak flux density.  
}
\end{deluxetable*}
%%%%%%%%%%%%%%%%%%%%%%%%%%%%%%%%%%%%%%%%%%%%%%%

%\listofchanges
\end{document}